\documentclass[aps,prx,twocolumn,footinbib,superscriptaddress]{revtex4-2}
\usepackage{feynmp}

\usepackage{amsmath}
\usepackage{amssymb}
\usepackage{bbm}
\usepackage{braket}
\usepackage{xcolor}
\usepackage{pifont}
\usepackage{slashed}
\usepackage[mathscr]{euscript}
\usepackage[shortlabels]{enumitem}
\usepackage{tikz-feynman}
\usepackage{bm}
\usepackage{dsfont}
\allowdisplaybreaks

\usepackage{graphicx}
\usepackage[colorlinks=true]{hyperref}
\usepackage{comment}

\renewcommand{\approx}{\simeq}

\renewcommand{\Im}{\text{Im}}

\definecolor{wrongultramarine}{rgb}{1,0.5,0}

\newcommand{\rd}{{\rm d}}

\newcommand{\Tr}{{\rm \, Tr\,}}

\newcommand{\calG}{{\mathcal G}}

\newcommand{\calD}{{\mathcal D}}
\newcommand{\calQ}{{\mathcal Q}}
\newcommand{\calH}{{\mathcal H}}

\newcommand{\be}{\begin{equation}}
\newcommand{\ee}{\end{equation}}
\newcommand{\beq}{\begin{eqnarray}}
\newcommand{\eeq}{\end{eqnarray}}
\newcommand{\ba}{\[\begin{aligned}}
\newcommand{\ea}{\end{aligned}\]}
\newcommand{\bal}{\begin{aligned}}
\newcommand{\eal}{\end{aligned}}

\newcommand{\tr}{{\rm tr\,}}

\renewcommand{\Im}{{\rm Im\,}}
\renewcommand{\vec}[1]{{\bf #1}}
\renewcommand{\epsilon}{\varepsilon}
\renewcommand{\dag}{\dagger}

\newcommand{\calO}{\mathcal{O}}

\newcommand{\vn}[1]{{\left|\vec{#1}\right|}}

\newcommand{\rU}{{\rm U}}
\newcommand{\rSU}{{\rm SU}}

\newcommand{\ii}{\mathrm{i}}

\newcommand{\drangle}{\rangle\!\rangle }

\hypersetup{
    bookmarks=true,         
    unicode=false,          
    pdftoolbar=true,        
    pdfmenubar=true,        
    pdffitwindow=false,     
    pdfstartview={FitH},    
    pdfsubject={},   
    pdfcreator={},   
    pdfproducer={}, 
    pdfkeywords={} {} {}, 
    pdfnewwindow=true,      
    colorlinks=true,       
    linkcolor=blue, 
    citecolor=blue,        
    filecolor=magenta,      
    urlcolor=blue           
}

\tikzset{
  mid arrow/.style={postaction={decorate,decoration={
        markings,
        mark=at position .560 with {\arrow[#1]{stealth}}
      }}},
  near arrow/.style={postaction={decorate,decoration={
        markings,
        mark=at position .275 with {\arrow[#1]{stealth}}
      }}},
   far arrow/.style={postaction={decorate,decoration={
        markings,
        mark=at position .800 with {\arrow[#1]{stealth}}
      }}},
      boson/.style={decorate, draw=black,
    decoration={snake,amplitude=1pt, segment length=5pt},
      },
}

\newcommand{\hx}{\hat{X}}
\newcommand{\hxc}{\hat{\mathcal{X}}}
\newcommand{\cl}{\mathsf{cl}}
\newcommand{\q}{\mathsf{q}}

\newcommand{\hqq}{\hat{Q}}
\newcommand{\hy}{\hat{Y}}

\newcommand{\vex}[1]{\bm{\mathrm{#1}}}

\newcommand{\tauh}{\hat{\tau}}
\newcommand{\intl}[1]{\int\limits_{#1}}
\newcommand{\parr}{\partial}
\newcommand{\spp}{{\mathsf{sp}}}
\DeclareMathOperator{\trr}{tr}
\newcommand{\ord}[1]{\bm{\mathit{O}}\left(#1\right)}
\DeclareMathOperator{\diag}{diag}
\newcommand{\pup}[1]{{\scriptscriptstyle{({#1})}}}
\usepackage{color}

\definecolor{OliveGreen}{cmyk}{0.64, 0, 0.95, 0.40}

\begin{document}

\title{Field theory of monitored, interacting fermion dynamics with charge conservation
}

\begin{abstract}
Measurement-induced phase transitions (MIPTs) in monitored quantum dynamics are non-equilibrium phase transitions between quantum-chaotic (volume-law entangled) and entanglement-suppressed, area-law phases. Here we reveal how monitored dynamics are situated within
the framework of general far-from-equilibrium, quantum condensed-matter physics. 
Measurement-induced heating effects scramble the distribution function in generic (interacting) monitored fermion systems, and this enables a simplified symmetry-based description of the dynamics. We 
demonstrate the equivalence of the Keldysh technique with the conventional Statistical-Mechanics Model for circuits,
resulting from a doubled Hilbert-space (Choi-Jamio{\l}kowski) mapping. We illustrate this using the monitored dynamics of interacting fermions with a conserved charge, deriving a unified effective field theory that captures all phases and phase transitions. The non-interacting counterpart in 1D space only has an area-law phase, with no MIPT. This was explained via an effective non-linear sigma model replica field theory
possessing a very large symmetry. We show that other phases and phase transitions emerge when the replica symmetry is reduced by interactions. The reduced symmetry combines a replica permutation symmetry and charge-conservation within each replica. The former and its spontaneous breaking govern the MIPT, which can be recognized via a separatrix in the renormalization group flow. The replica-resolved charge conservation dictates the ``charge-sharpening" transition between two kinds of dynamics, where the global charge information is either hidden or reconstructible from the measurements. The field theory explains why the charge-sharpening transition should occur only in the volume-law phase. Our framework provides a template for other classes of MIPTs and situates these within the arena of non-equilibrium condensed matter physics.

\end{abstract}

\author{Haoyu Guo}
\affiliation{Department of Physics, Cornell University, Ithaca, New York 14853, USA}
\author{Matthew S. Foster}
\affiliation{Department of Physics and Astronomy, Rice University, Houston, Texas 77005, USA}
\author{Chao-Ming Jian }
\email{chao-ming.jian@cornell.edu}
\affiliation{Department of Physics, Cornell University, Ithaca, New York 14853, USA}
\author{Andreas W. W. Ludwig}
\affiliation{Department of Physics, University of California, Santa Barbara, California 93106, USA}

\date{\today}

\maketitle
\tableofcontents

\section{Introduction}
\label{sec:Introduction}

The discovery of measurement-induced non-equilibrium phase transitions~\cite{li2018quantum,li2019measurement,skinner2019measurement,chan2019unitary,fisher2023random,PotterVasseurReview2021}
has nucleated an immense amount of research activity in recent
years, including  Refs.~\cite{zhou2019emergent,cao2019entanglement,VasseurRTN2019,zabalo2020critical,choi_2020,Jian20,Bao20,nahum2020entanglement,lang2020entanglement,gullans2020dynamical,gullans2020scalable,BaoChoiAltman2019,fidkowskiHaahHastingsHowDynamicalMemoriesForget-arXiv2008.10611,lavasani2021measurement,Sang2021LoopModel,Turkeshi2020MIPT2D,Zabalo2021,LiChenLudwigFisher2021,NahumRoySkinnerRuhmanAlltoAll2021,AlbertonBuchholdDiehlFermionPRL2021,MBuchhold2021,ZabaloGullansWilsonVasseurLudwigGopalakrishnanHusePixley,CMJian2022,Agrawal22,Barratt_U1_FT,LearnabilityPhysRevLett.129.200602,li2023cross,MajidyAgrawalGopalakrishnanPotterVasseurHalpern2023,LiVijayPolymer2023entanglement,CMJian2023,MFava2023,Mirlin23,Chahine23,NahumWiese,PiroliLiVasseurNahumControl2023,LiVasseurFisherLudwig,KumarKemalChakrabortyLudwigGopalakrishnanPixleyVasseur,VidalPotterVasseur2024,LovasAgrawalVijayBoundaryDiss2024,Mirlin2024_Above1D,FavaU1,MerrittFidkowski2023,Lumia2024,YoheiAshida2020}.
In the conceptually simplest incarnation, these are quantum circuits with a local qubit or qudit Hilbert space, subjected to unitary time evolution generating many-body chaos and, in the meantime,  ``monitored" or ``measured'' by an external ``environment'' or ``observer.''
It is crucial for the physics of interest that the outcomes of the measurements are not traced over, but instead simply ``recorded'' or ``collected.''
The unitary evolution generates entanglement.
For small measurement rates,
after a long-time evolution
the quantum system
in the steady state possesses volume-law entanglement (characteristic of many-body chaos),
while in the opposite limit frequent measurements `disentangle' the  steady-state wave function, a fact reflected in area-law
entanglement. At a critical measurement rate a novel type of continuous,
non-equilibrium quantum phase transition,
the ``measurement-induced phase transition'' (MIPT)
between phases with volume- and area-law
entanglement is known to occur,
with a critical scaling of the entanglement entropy with subsystem size (which is, for example, a logarithmic scaling in one spatial dimension) \cite{li2018quantum,li2019measurement,skinner2019measurement,chan2019unitary,LiChenLudwigFisher2021,fisher2023random,PotterVasseurReview2021,Jian20,Bao20,Turkeshi2020MIPT2D}.
There are many variations
of this theme
(see, e.g.,
reviews \cite{fisher2023random,PotterVasseurReview2021} for more examples).

One variation on the theme of MIPTs occurs in systems of
{\it non-interacting} fermions subjected to unitary evolution and measurements, both preserving the non-interacting nature in the combined evolution. In this type of system,
a volume law is in general
\footnote{Certainly, special features can enable other types of entanglement scaling in non-interacting fermion systems; see, for example, the spacetime dual a unitary circuit in studied Ref. \cite{Tarun}.   } not possible \cite{fidkowskiHaahHastingsHowDynamicalMemoriesForget-arXiv2008.10611},
but measurement-induced transitions between area-law
and logarithmic entanglement can exist \cite{nahum2020entanglement,Sang2021LoopModel,
AlbertonBuchholdDiehlFermionPRL2021,MBuchhold2021,CMJian2022,CMJian2023,MFava2023,Mirlin23,Chahine23,MerrittFidkowski2023,Mirlin2024_Above1D,FavaU1}. The theoretical description of such non-interacting systems is better understood than that of general qubit/qudit systems, in that the former are typically described by Non-Linear-Sigma Models (NLSMs) invariant under continuous symmetries (rotating replica copies of fermions into each
other) \cite{CMJian2022,CMJian2023,MFava2023,Mirlin23,Chahine23,Mirlin2024_Above1D,FavaU1}.

 While monitored quantum circuits have drawn intense interest from the quantum information community, a key open question has been how to interpret such systems through the condensed-matter lens of far-from-equilibrium, quantum-many body dynamics. Generic quantum circuits can be equivalently cast in terms of interacting, monitored fermions. Theories of far-from-equilibrium, many-fermion systems are typically very complicated, because of the need to keep track of the occupation of all levels (i.e., the fermion distribution function). An extreme example of this complexity is the problem of many-body localization, where an area-law entangled, localized many-body state is frozen out of equilibrium, with a well-defined occupation number (zero or one) for each level \cite{BAA}.
Understanding any new class of far-from-equilibrium quantum many-body phases and phase transitions is a potential milestone for condensed matter physics, given the paucity of existing examples.

Non-equilibrium quantum many-body systems can be formulated quite generally using the Keldysh technique \cite{Kamenev23}. Keldysh field theory describes the time-evolution of the density matrix, which is encoded through twin forward- and backward-time-ordered copies of the theory. If a single copy of the theory possesses an internal symmetry group $G$, then the associated Keldysh \emph{action}
possesses $G$ $\times$ $G$ symmetry, corresponding to independent transformations on the forward- and backward-contours. 
However, this symmetry is ``anomalous,'' 
because the contours are interlinked by the boundary condition of the initial density matrix. The link is precisely the single-particle distribution function \cite{Kamenev23}. 

In this paper, by studying a particular class of interacting, monitored fermion circuits, we reveal how monitored dynamics produce a vastly simplified type of nonequilibrium many-body theory. In particular, we show how the \emph{heating effect} of the measurements scrambles the distribution function, and effectively restores the $G$ $\times$ $G$ symmetry of the 
{\it action}
to a symmetry of the 
{\it full
Keldysh field theory}. 
In other words, measurements eliminate the anomaly by destroying almost all memory of the initial condition. 
As a result, we are able to derive a relatively simple 
effective field theory 
(``statistical mechanics model'') for monitored circuit dynamics from the Keldysh formalism.
The simplicity of the model is tied to the preservation of two copies of the key defining symmetry. (In the case of monitored circuits, $G$ involves transformations between additional ``replica'' copies of the theory on each Keldysh contour, necessary to extract interesting observables such as entanglement.) General consequences of this picture that we articulate here include the absence of linear response to external driving in the long-time dynamics of monitored circuits, and a clarification of which correlation functions describing conserved quantities can remain nonzero despite the complete scrambling of the distribution function. 
Moreover, for the class of monitored, interacting fermions studied in this work, we derive from Keldysh microscopics a unifying effective field theory that can describe all known dynamical phases and phase transitions in the circuit.

The 
focus of the present paper is the dynamics of 
generic (i.e.\ interacting) fermion
systems possessing a conserved charge that
is monitored by charge measurements. 
Related systems of this type in one spatial dimension with a local qudit onsite Hilbert space were studied relatively recently in a set of two seminal papers, Refs.~\cite{Agrawal22,Barratt_U1_FT}.
The interesting  salient result of
these studies
was that, while the dynamics still exhibited a measurement-induced volume-law to area-law entanglement transition
(analogous to what happened in the absence of a conserved charge~\cite{Jian20, Bao20}), the presence of a conserved charge led to additional structure {\it within} the volume-law phase.
A ``charge-sharpening'' transition was identified
separating a ``charge-fuzzy'' region of the volume phase at a low measurement rate, in which measurements do not render a generic initial state a sharp eigenstate of the total charge, from a ``charge-sharpened'' phase at a higher measurement rate where they do. The transition between these two phases was identified to be of Kosterlitz-Thouless type. A statistical mechanics model for the relevant dynamics of the qudit system has been formulated in Refs.~\cite{Agrawal22,Barratt_U1_FT}. The theory for the charge-sharpening transition in these works draws intuition from the 
large-qudit-dimension limit of this statistical mechanics model. However, a unifying theory of entanglement transition and the charge sharpening physics was lacking. Monitored dynamics of interacting fermions with a U(1) charge conservation provides a new angle on the relevant physics, for which we derive a unifying theory. This
will play the role of a Landau-Ginzburg-type theory that facilitates a more comprehensive understanding of the global phase diagram, the universal characters of each phase, and the nature of the
phase transitions via both 
symmetry-centric
considerations and renormalization group analysis.
See Sec.~\ref{sec:survey} for a survey of our results.

\section{Survey of results}
\label{sec:survey}

\subsection{Structure of the effective field theory}

In the present paper, we aim at understanding the monitored dynamics of many-body systems with a conserved charge from an orthogonal perspective, namely by adding short-ranged interactions to monitored systems of non-interacting fermions with charge conservation. This provides a new angle on the physics, independent of earlier qudit-oriented studies,
allowing us to explore the general phase diagram involving measurement rate and interaction strength, as well as general symmetry considerations and the renormalization group (RG) flow in this general parameter space. In particular, we obtain a cohesive view of the phases with different entanglement scaling and charge fuzziness (or sharpness) and their phase transitions from the perspective of a single unifying field theory, together with a comprehensive analysis of its symmetries. In addition to the field-theoretical characterization of the phase diagram, we provide a general symmetry-based insight into why the charge-sharpening transition
should only happen in the presence of volume-law entanglement scaling.
Moreover, and importantly, this orthogonal perspective also allows us
to understand at the same time (as already mentioned in the 
Introduction)
how the physics of monitored dynamics in general is connected with and embedded into the long-standing subject area of general non-equilibrium many-body condensed matter physics~\cite{Kamenev23}.

For the purpose of formulating this orthogonal perspective,
we need to recall that in one spatial dimension, the case we focus on here, a system of charged non-interacting fermions subject to charge-conserving dynamics and local charge measurements only possesses an area-law entangled phase~\cite{Mirlin23,FavaU1}.
There is no transition whatsoever. This can be seen from the particular NLSM (called a Principal Chiral Model [PCM])
describing the corresponding non-interacting fermion dynamics whose field variable is a unitary $R\times R$ matrix~\cite{Mirlin23,FavaU1} and which resides in what is known as symmetry class AIII~\cite{CMJian2022}. This NLSM has unitary U$(R)$ $\times$ U$(R)$
symmetry, where $R$ denotes the number of replicas. In the required limit $R\to 1$ \cite{Jian20,Bao20} relevant for measurements satisfying the Born rule,
Goldstone modes turn out to render a symmetry-broken phase unstable.
This follows as a consequence of the well-known 
\cite{McKaneStone79}
RG equation in this (AIII) symmetry class, and only the phase where the field variable is disordered exists. Such a phase exhibits area-law entanglement entropy scaling.

The unitary matrix-valued field variable of the NLSM naturally factorizes into a phase times an
SU$(R)$
matrix ${\hat X}$. Accordingly, the unitary
U$(R)$ $\times$ U$(R)$
symmetry factorizes into
U$(1)$ $\times$ U$(1)$, simply related to overall charge diffusion,
and the non-Abelian SU$(R)$ $\times$ SU($R$)
symmetry which is
related to the physics of entanglement (and of other observables non-linear in the density matrix)
originating from ${\hat X}$.
It is the latter symmetry that is fundamentally changed by interactions:
By using an exact (minimal coupling)
``gauging trick'' procedure,  
we can determine precisely the field theory into which the NLSM of the non-interacting monitored fermions is transformed when local density interactions are added. 
We also show how the same theory can be derived starting from the {\it microscopic} replicated Keldysh formulation
of the interacting fermion system, subject to weak density measurements.

\begin{figure}[t]
    \centering
    \includegraphics[width=0.52\textwidth]{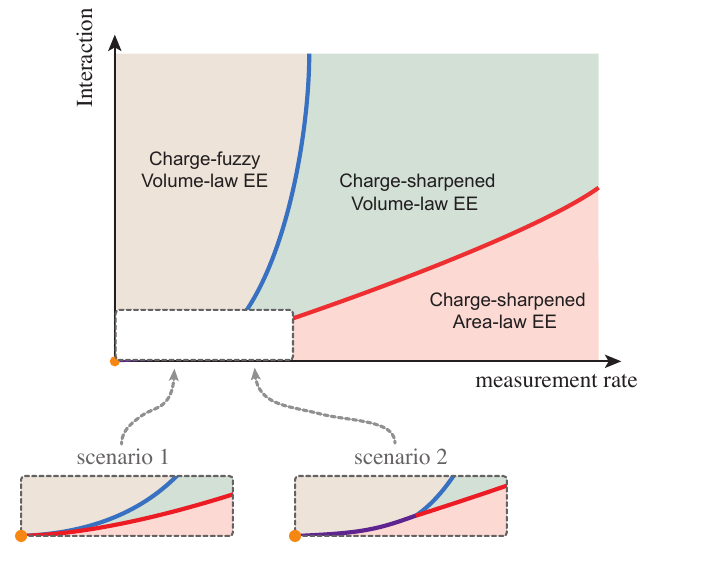}
    \caption{Schematic phase diagram of the  1+1D charge-conserving monitored dynamics of interacting fermions as a function of the measurement rate and interaction. There are three different dynamical phases distinguished by (1) their average entanglement entropy (EE) scaling, i.e., volume law versus area law, and (2) quantum-trajectory-resolved global charge fluctuation, namely ``charge-fuzzy" versus ``charge-sharpened" (Sec. \ref{sec:PhysCons} for details). The blue line indicates the charge-sharpening transition between the charge-fuzzy and charge-sharpened dynamics. The red line represents the entanglement transition between volume-law and area-law EE scaling. Near the origin (orange dot) with vanishing interaction and measurement rate, the phase diagram may exhibit two possible scenarios. In both scenarios, the charge-fuzzy volume-law phase and the charge-sharpened area-law phase extend to the origin (based on our renormalization group study in Sec. \ref{sec:RG}). In scenario 1, these two phases are always separated by an intermediate charge-sharpened volume-law phase. In scenario 2, the charge-sharpening transition and the entanglement transition merge into one single transition depicted as the purple line (see Sec. \ref{sec:PhysCons} for more detailed discussions).
    }
    \label{fig:PhaseDiagram_AIII}
\end{figure}

To the lowest non-trivial orders in the strength $U$ of interactions, two extra terms need to be added to the action of the field theory of the NLSM of the non-interacting case.
The first is a 
``mass''
term
that breaks the continuous
$\rSU(R)\times \rSU(R)$ symmetry of the non-interacting NLSM to a combination of the discrete permutation symmetry $S_R \times S_R$
amongst the replicas and the charge conservation within each replica. This mass of
strength $M$, whose
bare value $\propto U^2/({\rm measurement \ rate})^2$, gaps out the non-diagonal Goldstone modes of the SU($R$) matrix field ${\hat X}$.
At a small measurement rate, this generates the volume-law phase with the expected~\cite{Jian20,Bao20,VasseurRTN2019}
spontaneous symmetry breaking of
$S_R \times S_R$ to the diagonal $S_R$ subgroup,
characterized by a corresponding
expectation value of
the matrix of square moduli
$|X_{j\bar{k}}|^2$ of the matrix elements of ${\hat X}$.
[For details, see the paragraph above Eq.~\eqref{eq:ChargeSharpDOF} and Sec.\ \ref{sec:CFVLphase}.]
The remnant $(R-1)$ diagonal phase modes of ${\hat X}$ end up giving rise to the ``charge-sharpening" physics {\it within} the volume-law phase. For large measurement rate,
the $S_R\times S_R$ symmetry is {\it not} spontaneously broken,
corresponding to a vanishing expectation value of ${\hat X}$, and smoothly extending this property of the non-interacting system to the interacting case. This is the area-law phase.
The phase diagram summarizing this discussion is depicted in Fig.~\ref{fig:PhaseDiagram_AIII}.
The entanglement transition between volume- and area-law
phases (associated with the spontaneous symmetry breaking of permutation symmetry $S_R\times S_R$) appears as a separatrix in the RG flow of the interacting field
theory (for a brief summary, see below).

The second term that appears 
due to the interactions
involves squares of Noether currents of the NLSM for the non-interacting dynamics.
Its bare coupling constant is proportional to the interaction strength. 
We reveal the key physics of the Noether-current interaction through the calculation of the RG equations. First, the mass term $M$ discussed above is generated by the Noether-current interaction, even if initially absent. 
Second, the Noether-current interaction flows towards a finite value in the strong-coupling limit. 
This interaction is enabled by the replica-resolved U(1) symmetries of the dynamics. 
We cannot controllably extend our weak-coupling RG to the putative MIPT, but a finite 
replica-resolved U(1) Noether-current interaction coupling could influence the nature of the MIPT itself.

The complementary methods to obtain the field theory description for the interacting dynamics 
are summarized as follows.
We obtain the field theory in the presence of interactions first by using a simple and exact ``gauging trick,'' within the Keldysh formulation (Sec.~\ref{sec:Results}), and discuss details of the physics resulting from the RG equations (Sec.~\ref{sec:RG}), whose derivation is delegated to App.~\ref{app:RG}. Regarding the Keldysh methodology, we moreover obtain the presence of the two interaction terms 
discussed above, alternatively to and independent from the ``gauging'' procedure, also from a microscopic 
calculation. 
The latter employs a saddle point and fourth-order trace-log expansion,
utilizing standard diagrammatic analysis for the Keldysh fermions.
We also verify the U(1) Ward identity in this formulation (Sec.~\ref{sec:KDeriv}). In Appendix~\ref{app:Brownian}, we derive a similar microscopic Keldysh theory for a circuit with Brownian interactions~\footnote{These can naturally be thought of as arising from a space-time constant interaction by coarse-graining, given the other sources of space-time randomness
in the circuit,
i.e. the measurement outcomes, and they thus are expected to yield equivalent results.}
using the $G$-$\Sigma$ formalism.

\subsection{Dynamical phases and physical interpretation}

\begin{figure}[t]
    \centering
    \includegraphics[width=0.4\textwidth]{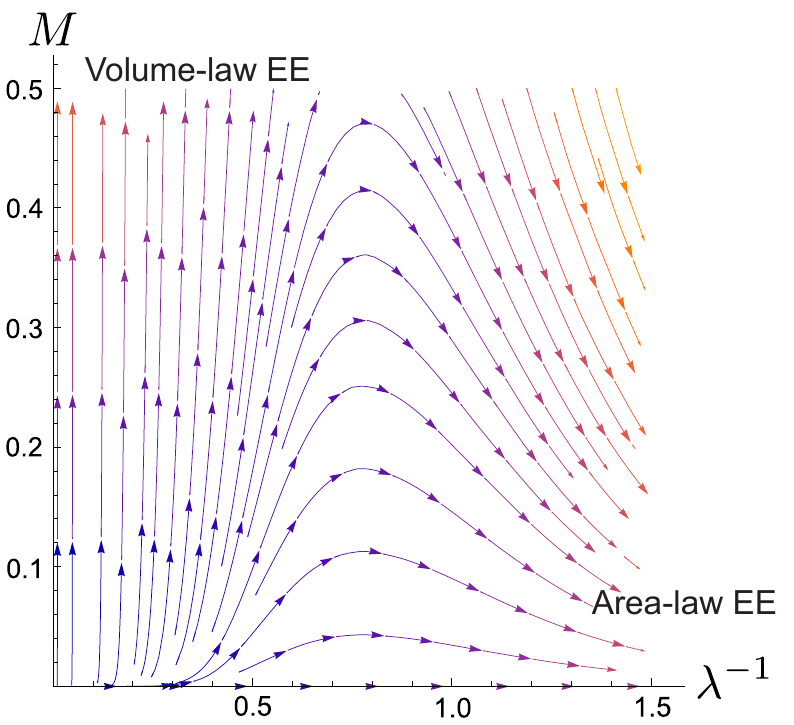}
    \caption{One-loop renormalization group flow for the effective field theory.
    The parameter $\lambda^{-1}$ is the inverse stiffness [Eq.~(\ref{eq:PCM})], a dimensionless measure of the measurement rate. The parameter $M \equiv \lambda m^2$ [Eqs.~(\ref{eq:SI}) and (\ref{eq:SIY})] is an interaction-generated ``mass'' that gaps out the off-diagonal Goldstone modes of the monitored non-interacting fermion theory. The large-mass limit leaves only the diagonal $[\textrm{U}(1)]^{(R-1)}$ degrees of freedom that describe the charge-fuzzy and charge-sharpened regimes of the volume-law phase, see Fig.~\ref{fig:PhaseDiagram_AIII} and Eq.~(\ref{eq:ChargeSharpDOF}).
    At the same time, a stronger measurement rate renormalizes the mass to smaller values; beyond a certain threshold, this drives a flow to the area-law phase with $\lambda^{-1} \rightarrow \infty$.
    The flow plotted here obtains from the one-loop RG equations (\ref{eq:1loop}) with $\bar{\Gamma}_U$ set equal to zero 
    for simplicity.
    The weak-coupling RG flow shown here must be truncated once either $\lambda^{-1}$ or $M$ become order one;
    in the latter case, the $[\rU(1)]^{R-1}$ phase fluctuations and associated charge-fuzzy and sharpened phases (and ensuing Kosterlitz-Thouless-transition) take over.
    See Sec.~\ref{sec:RG} for a summary of the RG flow
    that includes possible implications of $\bar{\Gamma}_U \neq 0$ unique to the MIPT for U(1) conserving systems,
    and Sec.~\ref{sec:PhaseandFlow} for the detailed discussion and the relationship of the flow to the dynamical phases exhibited in Fig.~\ref{fig:PhaseDiagram_AIII}.}
    \label{fig:RGFlow_AIII}
\end{figure}

The phase diagram in the regime of small interaction strength and small measurement rate
can 
be analyzed using a controlled 1-loop RG calculation. 
Important conclusions from the RG flow 
are exemplified via the flow diagram depicted in Fig.~\ref{fig:RGFlow_AIII}.
At a small measurement rate, sufficiently large interactions generate a strongly RG-relevant ``mass'' $M$ that gaps out the Goldstone modes of the
NLSM (for the non-interacting dynamics) that
are responsible for the flow towards the area-law phase. At the same time, at larger measurement rates
(as compared to the interaction strength),
the fluctuations of these same modes
act to suppress $M$, turning the growth of $M$ around. This produces a separatrix in the flow diagram, indicating the presence at strong coupling of the MIPT fixed point.

The fermion perspective provides a unique insight into the physics of the mass term. The microscopic derivation of the latter arises from Feynman diagrams that encode the analog of ``dephasing'' processes in weak localization, as shown in Fig.~\ref{fig:masses}. Dephasing is the \emph{dynamical} infrared process responsible for cutting off single-particle quantum interference effects in dirty, equilibrium interacting fermion systems at nonzero temperature \cite{AAK82,Liao17}. Similarly, interaction-mediated quantum chaos scrambles the ``localizing'' effects of measurements on monitored fermions. A remarkable aspect of our field theory is that \emph{both} ultraviolet (localizing) and infrared (dephasing) processes are simply encoded at the \emph{level of the effective action}. 
These effects usually require separate treatments. For example, real processes play the key role in the scenario of fermion many-body localization \cite{BAA}, while virtual processes control zero-temperature metal-insulator transitions \cite{Finkelstein83,BK94,Burmistrov19}.

Another important element of the present paper is the independent derivation of the same interacting field theory as that obtained using the Keldysh formalism, without using Keldysh in the first place but by employing instead the technically rather different (quantum information-theory-based) methodology of the
Choi-Jamiolkowski isomorphism in the doubled (``bra-ket'') Hilbert space of the density matrix (Sec.~\ref{sec:doublehilbert}, and App.~\ref{app:pathintegral}.)
This derivation provides a formulation of the same interacting field theory as 
an {\it equilibrium}
``Statistical Mechanics Model'' {\it in spacetime} of the type employed in the original work in
Refs.~\cite{Agrawal22,Barratt_U1_FT} on the ``charge-sharpening transition,'' as well as in many other similar or related quantum circuits including, e.g., those in Refs.~\cite{Jian20, Bao20,VasseurRTN2019,LiVasseurFisherLudwig,Zabalo2021,KumarKemalChakrabortyLudwigGopalakrishnanPixleyVasseur,LiChenLudwigFisher2021,MajidyAgrawalGopalakrishnanPotterVasseurHalpern2023,NahumWiese}.
It is interesting that this rather different formulation exists in parallel to that 
resulting from the use of
the Keldysh technique. 
Physically, this alternative formulation is possible because heating effects arising from the measurements done on the system render the Fermi distribution function of the Keldysh formalism trivial.
A non-trivial Fermi distribution function is, on the other hand, an essential and indisposable element of the Keldysh formalism necessary to describe the physics of other types of non-equilibrium quantum systems \cite{Kamenev23}. Recall that a non-trivial Fermi distribution function represents the ``anomaly'' referred to in  Sec.~\ref{sec:Introduction} (3rd and 4th paragraph)
which, as we show, is thus absent in the quantum circuit formulations of the monitored dynamics 
as an  equilibrium
``Statistical Mechanics Model''  in spacetime, and the corresponding doubled Hilbert space formulation of the density matrix.

Having provided a derivation of the
same interacting field theory using both Keldysh and doubled Hilbert space methodologies, we provide in Sec.~\ref{sec:PhysCons} a very general symmetry-based analysis of the implications of the interaction for the various phases, the phase transitions, and the resulting phase diagram
(see Fig.~\ref{fig:PhaseDiagram_AIII}). Here are some of the key aspects we would like to highlight (and reiterate) for this brief survey of results. The volume-law entanglement scaling arises from the spontaneous breaking of the $S_R \times S_R$ replica permutation symmetry characterized by the non-trivial expectation value of $|X_{j\bar{k}}|^2$. The $S_R \times S_R$ 
symmetry is fully restored after an entanglement transition into the area-law phases where the field $\hat{X}$ is fully disordered. In the presence of the volume-law entanglement, the diagonal modes of $\hat{X}$, which are equivalent to $(R-1)$ compact bosons in 1+1D, can either develop a quasi-long-range order or become disordered, giving rise to the charge-fuzzy and charge-sharpened phases, respectively. The charge-sharpening transition between the charge-fuzzy and charge-sharpened phases is of the Kosterlitz-Thouless type, consistent with the results in
Ref.~\cite{Barratt_U1_FT} obtained for the monitored qudit systems. 

Another interesting observation is that, based on general symmetry considerations, we can rule out the existence of a charge-fuzzy area-law phase, implying that the charge-sharpening transition should only occur in the presence of volume-law entanglement scaling. Refs.~\cite{Agrawal22,Barratt_U1_FT} made a similar observation for qudit systems and provided an argument for it using large-qudit-dimensions. Here, our analysis is based on general symmetry consideration in the unifying field theory.

\subsection{Outline}

The rest of the paper is organized as follows.
In Sec.~\ref{sec:Results}, we provide an overview of the main ideas underlying our approach to and of our key results on the effective field theory describing the 1+1D charge-conserving monitored dynamics of interacting fermions.
Measurement-induced heating is discussed in Sec.~\ref{sec:MIH}.
In Sec.~\ref{sec:RG} we summarize our main RG results. The basic definition of the quantum dynamics, its observables, and its formulation using replicas is given in Sec.~\ref{sec:circuit_micro}.
Its description in the language of Keldysh field theory is provided in
Sec.~\ref{sec:Kel_Description}. Sec.~\ref{sec:doublehilbert} demonstrates, using the path integral, that this description is equivalent to the time evolution of the system's replicated density matrix mapped to the doubled Hilbert space. In Sec.~\ref{sec:KDeriv}, we outline the derivation of the effective field theory from a microscopic Keldysh formulation, and verify the Ward identity for the physical density response in Sec.~\ref{sec:Ward}.

A unifying 
Landau-Ginzburg
perspective of the various phases and their transitions all emerging from our interacting field theory is given in
Sec.~\ref{sec:PhysCons} in terms of its fundamental symmetries and their possible breaking. After a brief review of the non-interacting case in Sec.~\ref{sec:ReviewNonInteracting}, the various possible phases are described and characterized in the language of the fundamental field of our field theory in Sec.~\ref{sec:phasesintmodel}. The implications of this field theory for the structure of the phase diagram are explained in Sec.~\ref{sec:PhaseandFlow}, together with the relevant information obtained from the weak coupling RG flow. 

Various technical details are delegated to Appendices. 
The saddle-point solution and effective field theory for non-interacting fermions at arbitrary filling is summarized in App.~\ref{app:ahf}.
A kinetic equation used to demonstrate heating is derived in App.~\ref{app:KE}.
App.~\ref{app:RG} further addresses 
the RG equations and the methodologies used to obtain them.
App.~\ref{app:pathintegral} provides the details for the coherent state path integral used in Sec.~\ref{sec:doublehilbert}  for the equivalence of double-Hilbert space and Keldysh formalisms. In
App.~\ref{app:Brownian} we derive an alternative but equivalent microscopic Keldysh theory for the dynamics of the quantum dynamics with Brownian interactions using the $G-\Sigma$ formalism.
In App.~\ref{app:WardU(1)} we derive the Hubbard-Stratonovich decoupling used in Eq.~(\ref{eq:barS}), and show how 
a particular parameter choice is required
in order for the theory to give the diffusive response for the average density.

\subsection{Effective field theory, symmetry, and phase diagram\label{sec:Results}}

In this section, we summarize our key results on the effective field theory describing the
1+1D
charge-conserving monitored dynamics of interacting fermions. This field theory is formulated to capture the
time evolution of $R$ replicas of the physical system, averaged over all quantum trajectories (labeled by different measurement outcomes).
Average physical observables, including,
the average entanglement entropy and general-moment charge correlations, can be extracted from the replica limit $R\rightarrow 1$ \cite{Jian20,Bao20,Barratt_U1_FT}. More importantly, this effective theory provides a powerful tool to study the
dynamical
phase diagram of the
system.
In this section, we will present the effective field theory
for
interacting
fermions
as the result of a ``gauging trick" applied to the field theory
for
non-interacting monitored dynamics. The latter was previously obtained in Refs.~\cite{Mirlin23,Mirlin2024_Above1D,Chahine23,FavaU1}. The full microscopic description of the interacting monitored dynamical system and
alternative
derivations of the effective field theory
from microscopics
are provided in Secs.~\ref{sec:circuit2continuum }, \ref{sec:KDeriv}, and Appendix~\ref{app:Brownian}. 
In this 
and the following subsections, 
we will also highlight the key symmetry aspects of our effective field theory, 
the role of measurement-induced heating,
the results of a weak-coupling renormalization group (RG)
study, and their consequences on the phase diagram of the interacting monitored dynamics. The in-depth discussion on these aspects will be provided in Sec.\ \ref{sec:PhysCons}.

The effective theory for 1+1D charge-conserving monitored dynamics of non-interacting fermions is described by the path integral \cite{Mirlin23,FavaU1}
\begin{align}\label{eq:Z}
    Z_0
    =
    \int\calD \hat{X} \, \calD \theta \, \calD \phi
    \,
    e^{
        -
        S_{\textrm{SU($R$)}}
        -
        S_{\textrm{U(1)}}
    },
\end{align}
where in 1+1-dimensions
\begin{align}
\label{eq:PCM}
\!\!
    S_{\textrm{SU($R$)}}
    =&\,
    \lambda
    \int
    \frac{dt dx}{8}
    \,
    \trr_R
    \left[
    \parr_t \hx^\dagger \parr_t \hx
    +
    v^2
    \parr_x \hx^\dagger \parr_x \hx
    \right],
\\
\label{eq:U(1)}
    S_{\textrm{U(1)}}
    =&\,
    \int
    \frac{dt dx}{8 R}
    \left[
    \begin{aligned}
    &\,
        (\parr_t \theta)^2 + v^2 (\parr_x \theta)^2
        -
        4
        \gamma
        \,
        \phi
        \,
        \parr_t
        \theta
    \\
        -
    &\,
        (\parr_t \phi)^2 - v^2 (\parr_x \phi)^2
    \end{aligned}
    \right]\!.
\end{align}
The field $\hat{X} \rightarrow X_{j \bar{k}}$ is an $R \times R$ matrix representing an SU($R$) group element; it satisfies the constraint $\hat{X}^\dagger \hat{X} = \hat{1}_R$.
Here, our notation is such that $\hat{X}$ represents the entire $R \times R$ matrix whose matrix elements are given by $X_{j\bar{k}}$. The action in Eq.~(\ref{eq:PCM}) is a principal chiral nonlinear sigma model (PCM) in class AIII. \cite{Mirlin23,Mirlin2024_Above1D,Chahine23,FavaU1}
(see also Ref. \cite{CMJian2022} for the identification of the symmetry class AIII using a different quantum-circuit-based method). It is important to note that even though this PCM captures the dynamics in real time $t$, the spacetime signature is effectively Euclidean (without doing any Wick rotation).
The fields $\theta$ and $\phi$ are free bosons
with diffusive dynamics that characterize the \emph{trajectory-averaged} particle density.
The parameters $\gamma$ (dimensions of energy) and $\lambda^{-1}$ (dimensionless) are both proportional to the measurement rate; $v$, the emergent ``speed of light,'' is proportional to 
the root-mean-square band velocity.

In the Keldysh description, fermion fields reside along the time ($T$)-ordered or anti-time ($\bar{T}$)-ordered branches of the Keldysh contour \cite{Kamenev23}. We denote $T$- and $\bar{T}$-ordered replicated fermion fields as $\psi_{+,j}$  and $\psi_{-,\bar{k}}$, respectively, where $j,\bar{k} \in \{1,2,\ldots,R\}$ are the replica indices. The derivation of the effective theory implies the correspondence (see Sec.~\ref{sec:KDeriv})
\begin{align}\label{eq:XFieldIdentify}
    e^{i \theta/R} \, X_{j \bar{k}}
    \Leftrightarrow
    \psi_{+,j} \, \bar{\psi}_{-,\bar{k}},
    \quad
    e^{-i \theta/R} \, X^\dagger_{\bar{j} k}
    \Leftrightarrow
    \psi_{-,\bar{j}} \, \bar{\psi}_{+,k}.
\end{align}

The entanglement dynamics of 1+1D non-interacting, monitored fermions are described by the PCM in Eq.~(\ref{eq:PCM}).
This theory has a large SU($R$) $\times$ SU($R$) symmetry, corresponding to separate continuous replica rotations of the $T$ and $\bar{T}$-ordered fermions.
This symmetry acts on the field $\hat{X}$ as left- and right-$\rSU(R)$ multiplications.
In the usual Keldysh formulation for non-equilibrium quantum dynamics, fields are interlinked by cross-contour correlation functions that encode the distribution function $f(\varepsilon)$ \cite{Kamenev23}. A key simplification
of the Keldysh formulation for fermion systems with quantum dynamics subject to measurements, such as the one under consideration,
is that the fermion distribution function
$f(\varepsilon)$ assumes an energy $\varepsilon$-independent value (equal to 1/2 at half-filling), formally equivalent to infinite temperature. This simplification occurs due to measurement-driven ``heating'' in a generic quantum trajectory \cite{Mirlin23}, and
leads to the effective decoupling of the two contours, i.e.\ the almost total erasure of the initial condition. 
(Only the total particle number is remembered.)
We demonstrate the heating effect for monitored, interacting fermions below in Sec.~\ref{sec:MIH}.
Eq.~(\ref{eq:PCM}) applies to non-interacting monitored fermions at exactly half-filling. SU($R$) $\times$ SU($R$) symmetry (i.e, the decoupling of the two contour branches) is preserved away from half-filling as well; the only new feature is a modulation of the stiffness $\lambda$ by the dynamical density fluctuations in the U(1) sector [Eq.~(\ref{eq:U(1)})], as anticipated in Ref.~\cite{Ha2024}. This is demonstrated in Appendix \ref{app:ahf}.

The (replicated) monitored dynamics of interacting fermions is 
expected to exhibit
lower
$S_R$ $\times$ $S_R$ permutation symmetry among the replicas, where $S_R$ is the permutation group on $R$ objects \cite{Jian20,Bao20}, as well as U(1) charge conservation for each replica. In this paper, we show that this pattern of symmetry indeed arises and is made possible by the heating effect that persists in the interacting case. Note that the replica-resolved U(1)'s do not commute with the $S_R$ $\times$ $S_R$ replica permutation symmetry. The precise structure of the symmetry group of the interacting theory will be articulated in Sec. \ref{sec:PhysCons}.

Interactions for the fermions can be incorporated using a ``gauging trick.''
To implement density-density interactions,
we ``gauge" the
\emph{replica-resolved} contour-axial charge U(1) symmetries via the auxiliary bosonic 
fields 
$\varphi_{+,j} $ and $\varphi_{-,\bar{k}}$:
\begin{align}\label{eq:IntGauging}
\begin{aligned}
    \parr_t X_{j \bar{k}}
    \rightarrow&\,
    \parr_t X_{j \bar{k}}
    +
    \ii \left(\varphi_{+,j} - \varphi_{-,\bar{k}}\right) X_{j \bar{k}},
    \\
    \parr_t X^\dagger_{\bar{j} k}
    \rightarrow&\,
    \parr_t X^\dagger_{\bar{j} k}
    +
    \ii \left(\varphi_{-,\bar{j}} - \varphi_{+,k}\right) X^\dagger_{\bar{j} k}.
\end{aligned}
\end{align}
This is a standard minimal coupling prescription for the PCM, which is exact and non-perturbative.
In Sec.~\ref{sec:KDeriv}, we detail an alternative perturbative derivation of the effective field theory for the interacting dynamics that incorporates terms up to fourth order in a trace-log expansion, but leads to identical conclusions.

Assuming short-ranged density-density interactions on the Keldysh contour,
the action for the auxiliary bosonic fields $\varphi_{+,j}$ and $\varphi_{-,\bar{j}}$
on the two Keldysh branches
is \cite{Kamenev23}
\begin{align}\label{eq:Sphi}
    S_\varphi
    =
    -\frac{1}{2 \ii U}
    \int
    dt dx
    \left(
    \sum_{j = 1}^R
    \varphi_{+,j}^2
    -
    \sum_{\bar{j} = 1}^R
    \varphi_{-,\bar{j}}^2
    \right),
\end{align}
where $U > 0$ is a repulsive local interaction strength.
The auxiliary bosonic fields can be integrated out exactly via the Gaussian integration
\begin{align}
    Z
    \equiv
    \int
    \calD \varphi
    \,
    e^{-S_\varphi}
    \,
    Z_0[\varphi],
\end{align}
where $Z_0$ is defined in Eq.~(\ref{eq:Z}), subject to the gauging transformation in Eq.~(\ref{eq:IntGauging}).
The integration over the auxiliary fields $\varphi_{+,j}$ and $\varphi_{-,\bar{j}}$ induces a shift of the PCM action
$S_{\textrm{SU($R$)}} \rightarrow S_{\textrm{SU($R$)}} + S_I$. Expanding $S_I$ to quadratic order in $U$ yields
\begin{align}\label{eq:SI}
    S_I
    =
    2 \ii U
    \sum_{j = 1}^R
    \int
    d t d x
    \,
    n_{\cl,j}
    n_{\q,j}
    -
    \frac{\lambda m^2}{16}
    \int d t d x
    \sum_{j,\bar{k} = 1}^R
    \left|X_{j \bar{k}}\right|^4\!\!,
\end{align}
where $m^2 = \lambda U^2/ 2 a^2$, and $a$ is a short-distance cutoff.
The classical and quantum \emph{replica-resolved} density operators
in Eq.~(\ref{eq:SI})
are related to the Noether currents of the PCM,
\begin{align}\label{eq:varrhoclqDef}
    n_{\cl,\q,j}
    \equiv&\,
    \frac{\ii \lambda}{8}
    \left[ \left(\hat{J}_t\right)_{jj} \pm \left(\hat{\overline{J}}_t\right)_{\bar{j}\bar{j}}\right].
\end{align}
The Noether currents that implement replica rotations on $T$-ordered and $\bar{T}$-ordered fermions are
\begin{align}\label{eq:Noether}
\begin{aligned}
    \hat{J}_\mu
    =&\,
    (\parr_\mu \hat{X})\hat{X}^\dagger
    \rightarrow
    \left[(\parr_\mu \hat{X})\hat{X}^\dagger\right]_{jk},
\\
    \hat{\overline{J}}_\mu
    =&\,
    \hat{X}^\dagger(\parr_\mu \hat{X})
    \rightarrow
    \left[\hat{X}^\dagger(\parr_\mu \hat{X})\right]_{\bar{j}\bar{k}}.
\end{aligned}
\end{align}

The full interacting field theory can be expressed by the path integral
\begin{align}\label{eq:ZFull}
    Z
    =
    \int\calD \hat{X} \, \calD \theta \, \calD \phi
    \,
    e^{
        -
        S_{\textrm{U(1)}}
        -
        S_{\textrm{SU($R$)}}
        -
        S_I
    }.
\end{align}
The same form for the theory arises in general dimensions.
The action consists of the average density sector
$S_{\textrm{U(1)}}$
[Eq.~(\ref{eq:U(1)})]
and the entanglement sector $S_{\textrm{SU($R$)}} + S_I$ [Eqs.~(\ref{eq:PCM}) and (\ref{eq:SI})]. We stress that the entanglement sector governs the dynamics of entanglement, as well as other observables that are non-linear functions of the quantum-trajectory-resolved density matrix.
The $S_I$ term partially reduces the original symmetry $\rSU(R)\times \rSU(R)$ of $S_{\textrm{SU($R$)}}$. For each copy of $\rSU(R)$,
what remains is
generated by a $S_R$ permutation symmetry and $[\rU(1)]^{R-1}$ symmetries,
as expected for an interacting circuit. The $[\rU(1)]^{R-1}$ symmetries manifest through the diagonal elements of the SU($R$) field
[see Eq.~(\ref{eq:ChargeSharpDOF}) below],
and are the key to the charge-sharpening transition \cite{Agrawal22} in the volume-law phase,
while
the $S_R \times S_R$ symmetry is responsible for the
transition between volume-law and area-law entanglement scaling \cite{Jian20,Bao20}.

The diffusive global U(1) sector governed by Eq.~(\ref{eq:U(1)}) is not 
structurally modified by the interactions, which can indirectly renormalize the bare diffusion constant [defined in Eq.~(\ref{Ward1+1})]. This renormalization can occur, e.g., through interaction corrections to the root-mean-square band velocity $v$.
This is consistent with the idea that quantities linear in the average density matrix  (such as the average particle density) are oblivious to phase transitions that can occur in the entanglement
or generally within \emph{moments} of physical observables, including those of the charge density.

The Noether current-current perturbation in Eq.~(\ref{eq:SI}) is marginal
in the RG sense
at tree level,
while the quartic interaction is relevant.
If we parameterize
$\hat{X} = \exp(\ii \hy/\sqrt{\lambda})$, with $\hy$ a traceless Hermitian matrix, then the interaction
part of the action
has the expansion
\begin{align}\label{eq:SIY}
    S_I
    \sim&\,
    \int dt d x
    \left\{
    \begin{aligned}
    &\,
        \frac{m^2}{8}
        \sum_{j\neq k = 1}^R
        Y_{j k} \, Y_{k j}
    \\
        +
    &\,
        \frac{\sqrt{\lambda} \, U}{16}
        \sum_{j = 1}^R
        \left(\parr_t Y_{j j}\right)
        \left[(\parr_t \hat{Y}),\hat{Y}\right]_{jj}
    \end{aligned}
    \right\}
\nonumber\\
    &\,
    +
    \ord{Y^4}.
\end{align}
Thus the interactions introduce a ``mass'' for off-diagonal fluctuations of the principal chiral field; the $n_{\cl,j} n_{\q,j}$ term in Eq.~(\ref{eq:SI}) becomes a
cubic
interaction in terms of the field coordinate. In Eq.~(\ref{eq:SIY}),
$[\hat{A},\hat{B}] = \hat{A} \, \hat{B} - \hat{B} \, \hat{A}$ is the matrix commutator.

The appearance of the mass and the cubic couplings due to the interactions drastically changes the phase diagram of the 1+1D monitored dynamics of fermions with charge-conservation.
Without interactions, the PCM [Eq.~(\ref{eq:PCM})] exhibits only a slow logarithmic RG flow towards the area-law phase \cite{Mirlin23,FavaU1}, see Eq.~(\ref{eq:betalambda}) below. Instead, the mass term represents entanglement generation that suppresses (or ``dephases,'' see Sec.~\ref{sec:KDeriv}) the quantum Zeno effect responsible for the area-law phase. For a slow measurement rate, a very small (but nonzero)
interaction-generated
mass is sufficient to
drive a rapid RG flow into the volume-law phases, as pictured in Fig.~\ref{fig:PhaseDiagram_AIII}.

From the symmetry perspective, the volume-law entanglement scaling arises from the spontaneous breaking of the $S_R \times S_R$ symmetry to its diagonal subgroup
characterized by the development of an expectation value
of the form
$\langle |X_{j\bar{k}}|^2 \rangle = B \, \delta_{j\bar{k}} + (1-B) R^{-1}$ with $B>0$
\footnote{Note that this form of the expectation value
$\langle |X_{j\bar{k}}|^2 \rangle= B \, \delta_{j\bar{k}} + (1-B) R^{-1}$ is due to constraint $\hat{X}^\dag \hat{X}=\hat{1}_R$. $S_R \times S_R$ is spontaneously broken to its diagonal subgroup for any $B>0$. More details of the order parameter of the spontaneous breaking of $S_R \times S_R$ is provided in Sec. \ref{sec:CFVLphase}.
}.
With the volume-law entanglement present, one can focus on the energies and inverse-length scales below the mass scale $m$ where the ${\rm SU}(R)$ field $\hat{X}$ can be reduced to its diagonal fluctuations:
\begin{align}\label{eq:ChargeSharpDOF}
    \hat{X} \rightarrow \diag\{e^{(\ii/\sqrt{\lambda}) Y_{jj}(t,x)}\}, \quad j \in \{1,2,\ldots,R\},
\end{align}
where $Y_{jj}$ are compact boson fields subject to $\sum_j Y_{jj} =0$ and the identification $Y_{jj}/\sqrt{\lambda} \sim  Y_{jj}/\sqrt{\lambda} + 2\pi\mathbb{Z}$.
The charge-sharpening transition (first introduced in Refs.~\cite{Agrawal22,Barratt_U1_FT}) occurs due to vortex proliferation in our theory with the residual $[\rU(1)]^{R-1}$ symmetry that shifts the boson fields $ Y_{jj}$. Depending on whether the vortices proliferated or not, the system's dynamics are charge-sharpened or charge-fuzzy in its quantum-trajectory-resolved global charge fluctuation. Both the charge-fuzzy and the charge-sharpened volume-law phases are absent in the non-interacting limit. As the measurement rate increases, the fluctuations of $\hat{X}$ will eventually restore the $S_R \times S_R$ replica permutation symmetry, driving the system into the area-law phase
(where $\langle |X_{j\bar{k}}|^2 \rangle = \frac{1}{R}$).
The detailed discussion of these phases and transitions are given in Sec.~\ref{sec:PhysCons}.

\subsection{Measurement-induced heating\label{sec:MIH}}

Although the derivation
of the effective theory using the gauging trick [Eq.~(\ref{eq:IntGauging})]
above is extremely simple, we emphasize that this trick works here only because of the measurement-induced heating effect, which trivializes the fermion distribution function. 
The heating effect can be directly seen at a semiclassical level. In Appendix~\ref{app:KE}, we derive the following kinetic equation for $F(\omega;t) \equiv 1 - 2 f(\omega,t)$, where  $f(\omega,t)$ is the Fermi occupation factor:
\begin{align}\label{eq:KE}
    \frac{d}{d t} F(\omega;t) 
    =&\,
    \mathfrak{St}[F],
\end{align}
\begin{align}\label{eq:St}
    \mathfrak{St}[F]
    =&\,
    \intl{\Omega}
    \frac{\varrho(\Omega)}{v}
    \left\{
    \begin{aligned}
        &\,
        F_B(\Omega;t)
        \left[
            F(\Omega + \omega;t) - F(\omega;t)
        \right]
    \\
    &\,    
        -
         \left[
            1 - F(\Omega + \omega;t) \, F(\omega;t)
        \right]
    \end{aligned}
    \right\}
\nonumber\\
&\,
    -
    2
    \gamma
    \,
    F(\omega;t), 
\end{align}
with $\intl{\Omega} \equiv \int_{-\infty}^\infty d \Omega / 2 \pi$.
Here $\varrho(\Omega) \propto U^2 \, \Omega /v \gamma$ denotes the local spectral function for 
the auxiliary interaction bosons $\varphi$ 
[Eqs.~(\ref{eq:IntGauging}) and (\ref{eq:Sphi})], evaluated via the self-consistent Eliashberg equations,
see Appendix~\ref{app:KE} for details. 
The boson distribution function $F_B$ is not independent, but determined via
\begin{align}\label{eq:FB}
    \Omega \, F_B(\Omega;t)
    =
    -
    \pi
    \int_{-\infty}^\infty
    \frac{d \omega}{2 \pi}
    \left[
        F(\omega - \Omega;t) \, F(\omega;t) - 1
    \right].
\end{align}
The first term in Eq.~(\ref{eq:St}) is the inelastic collision integral due to fermion-fermion scattering. Although this term can re-distribute energy, it cannot raise or lower the total energy density. 
This can be immediately seen by noting that a thermal distribution at any temperature $T$ $F(\omega) = \tanh(\omega/2T)$ is a zero mode of this term 
\footnote{
For an equilibrium $F(\omega) = \tanh(\omega/2T)$, $F_B(\omega) =  \coth(\omega/2T)$ and the first term in Eq.~(\ref{eq:St}) vanishes due to the identity
$
\coth(x - y)\left[\tanh(x) - \tanh(y)\right] = 1 - \tanh(x) \, \tanh(y)
$.
}
(detailed balance \cite{Kamenev23}). The second term in Eq.~(\ref{eq:St})  arises from the measurements. The only steady-state solution to the kinetic equation (\ref{eq:KE}) has $F = 0$ at half filling, the same as infinite temperature $T \rightarrow \infty$. 
(A constant nonzero $F$ arises away from half filling, see Appendix~\ref{app:ahf}.)
The ``infinite temperature'' induced by measurements is only a formal analogy, however, as the trivialization of $F$ applies to both volume- and area-law entangled phases.
Note that Eq.~(\ref{eq:KE}) holds separately for each replica, because the semiclassical kinetic equation obtains from the replica-diagonal saddle point in the Keldysh technique (see Appendix~\ref{app:KE}).

In an interacting, disordered fermion system at finite temperature that can also be described by a
(more complicated)
type of nonlinear sigma model
(see e.g.\ Refs.~\cite{Finkelstein83,BK94,Liao17,Burmistrov19}), the form of the interactions incorporates the Fermi function in a crucial way that cannot be reproduced by a minimal coupling scheme applied to the non-interacting effective theory.
Starting from the microscopic Keldysh description, we have also performed an explicit fourth-order trace-log expansion around the measurement-induced saddle-point and used it to independently verify the form of 
the interacting theory for monitored fermions
Eq.~(\ref{eq:SIY}), see Sec.~\ref{sec:KDeriv} for details.

We note that the trivialization of the distribution function also manifests in the global U(1) sector (identically for both non-interacting and interacting fermions). We find that the only nonvanishing correlation function of the average density is the Keldysh one, which satisfies the U(1) Ward identity and encodes the diffusive average density fluctuations with the diffusion constant $D = v^2/2\gamma$. By contrast, external electric fields or quenched disorder do not couple to the trajectory-averaged steady state, leading to vanishing linear response functions. These points are demonstrated explicitly in Sec.~\ref{sec:Ward}.

\subsection{Renormalization group and mass generation \label{sec:RG}}

The entanglement sector of the theory in Eq.~(\ref{eq:ZFull}) is an anisotropic, massive deformation of the PCM in Eq.~(\ref{eq:PCM}). In Appendix~\ref{app:RG}, we detail the one-loop renormalization group calculation for the beta functions of this model. Here we summarize the main results and the key physics.
We state the results of the one-loop RG calculation in terms of the stiffness $\lambda$, the mass parameter $M \equiv \lambda m^2$, and the effective interaction strength defined via
\begin{align}\label{eq:bGUDef}
    \bar{\Gamma}_U \equiv \frac{\lambda \, U}{32}.
\end{align}
The couplings $\lambda \, \bar{\Gamma}_U$ and $M$ are precisely the coefficients of the current-current and quartic terms in Eq.~\eqref{eq:SI}. We choose to work with $\bar{\Gamma}_U$ (instead of $\lambda \, \bar{\Gamma}_U$) for the simplicity of the beta function.
Note that in the massless case, $\bar{\Gamma}_U$ measures the \emph{relative anisotropy} of the sigma-model target manifold.
Since $\lambda$ is proportional to the inverse of the measurement rate, $\bar{\Gamma}_U$ is a measure of the interaction energy $U$ relative to the measurement rate.

For SU($R$), the one-loop beta functions for $\lambda$ and $M$ are
\begin{subequations}\label{eq:1loop}
\begin{align}
    \frac{d \lambda}{d \ln L}
    =&\,
    -
    \frac{R}{\pi}
    +
    \frac{c \, \bar{\Gamma}_U^2}{\pi},
\label{eq:betalambda}
\\
    \frac{d M}{d \ln L}
    =&\,
    \left[
        2
        -
        \Delta_M(R)
        -
        \frac{c \, \bar{\Gamma}_U^2}{\pi \, \lambda}
    \right]
    M
    +
    \Lambda^2
    \,
    \frac{c \, \bar{\Gamma}_U^2}{\pi}.
\label{eq:betaM}
\end{align}
\end{subequations}
In Eq.~(\ref{eq:1loop}) $c = 128$, $\Lambda$ is a hard
frequency-momentum cutoff, and
the scaling dimension of the mass operator
\footnote{One-loop anomalous dimension function.}
is
\begin{align}\label{eq:DeltaM}
    \Delta_M(R)
    =&\,
    \frac{4(R+1)}{\pi \, \lambda}.
\end{align}
The entanglement physics is captured by the replica limit $R \rightarrow 1$.
In this limit, we find that the beta function for the interaction coupling $\bar{\Gamma}_U$
is given by
\begin{align}
    \frac{d \bar{\Gamma}_U}{d \ln L}
    =&\,
    \frac{1}{\pi \, \lambda}
    \left[
        \frac{5}{6}
        -
        2 c \, \bar{\Gamma}_U^2
        \right]
    \bar{\Gamma}_U.
\label{eq:betaGamma}
\end{align}

Eq.~(\ref{eq:betalambda}) implies that localizing effects induced by the measurements compete with the entangling interactions; the former (latter) suppresses (enhances) the stiffness $\lambda$.
We can view the interaction enhancement of the stiffness as analogous to an \emph{anti}-localizing Altshuler-Aronov quantum conductance correction in equilibrium, disordered and interacting 
fermion systems
\cite{AA85,Liao17}. 

Eq.~(\ref{eq:betaM}) describes two key effects. First, measurements suppress the scaling dimension of the mass parameter, because $\Delta_M(1) > 0$ [Eq.~(\ref{eq:DeltaM})]. For large stiffness $\lambda$ and small interactions $\bar{\Gamma}_U \ll 1$, $M$ remains a strongly relevant perturbation to the principal chiral model.
The second key effect concerns the last term in Eq.~(\ref{eq:betaM}), which implies that $M > 0$ is \emph{generated} by the current-current interactions, even if initially absent.
The flow diagram in the $\bar{\Gamma}_U = 0$ plane is shown in Fig.~\ref{fig:RGFlow_AIII}.
The competition of the logarithmically slow reduction of $\lambda$ and the rapid RG flow of $M$ implies
a separatrix in the flow, exponentially close to the $M = 0$ non-interacting limit for a slow measurement rate, evident from Fig.~\ref{fig:RGFlow_AIII}. This separatrix is the area-to-volume law transition, which presumably terminates at a strong-coupling MIPT RG critical point.

The $M$ parameter suppresses replica-off-diagonal fluctuations of the sigma-model field $\hx$,
enabling the volume-law phase and manifesting the diagonal degrees of freedom responsible for charge sharpening \cite{Agrawal22}.
In this work, we have identified three different ways of viewing the generation of the mass term:
(1) the formal gauging argument presented in Sec.~\ref{sec:Results},
(2) a perturbative fourth-order trace-log expansion presented in Sec.~\ref{sec:KDeriv}, wherein the mass term arises due to the interaction-mediated inelastic self-energy that cuts off (``dephases'') PCM-mediated interference processes (see Fig.~\ref{fig:masses}),
and
(3) generation of $M$
by the current-current interactions via the renormalization group.

Eq.~(\ref{eq:betaGamma}) suggests that the Noether-current interaction strength $\bar{\Gamma}_U$ flows towards a nonzero constant, indicating a finite target-manifold anisotropy at the MIPT.
While the RG flow is not under full control in the strong-coupling regime, this result has potentially
important implications for the nature of the MIPT itself. The diagonal Noether-current operators are directly tied to the relative U(1) symmetries of the monitored fermions studied here, see
Sec.~\ref{sec:Results}, above. 
A value $\bar{\Gamma}_U \neq 0$ could influence the nature of the entanglement transition.

Target-manifold
anisotropy is also manifested in an additional stiffness term that only involves the diagonal components of the Noether currents. This term does not qualitatively change the behavior of the RG flow so we have ignored it here. The full results for the 
RG flow 
are
presented in Appendix~\ref{app:RG}. 

A remarkable feature of the effective statistical mechanics model in Eq.~(\ref{eq:ZFull}) is that it encodes both ultraviolet effects (such as ``localization'') and infrared effects (such as mass generation) for the interacting fermion system, via the renormalization group. By contrast, in equilibrium, disordered and interacting fermion systems,
real inelastic processes responsible for the dephasing of UV interference effects must be treated separately from the RG, as dynamically generated infrared regularization processes \cite{AAK82,Liao17}.

The physical interpretation of our theory is expounded in Sec.~\ref{sec:PhysCons}.
In Sec.~\ref{sec:PhaseandFlow}, the implications of the weak-coupling RG for the three phases of the monitored, interacting fermions are explained. We define the effective order parameters in terms of the matrix field $X_{j \bar{k}}$ for the charge-fuzzy and charge-sharpened volume-law and charge-sharp area-law phases
in Sec.~\ref{sec:phasesintmodel}. The phenomenology of these dynamical phases is explained in this section.
In Sec.~\ref{sec:phasesintmodel}, we also argue why a charge-fuzzy area-law phase does not exist as it would be
incompatible with the patterns of symmetry breaking that can be articulated using
$X_{j \bar{k}}$.

\section{Interacting monitored dynamics with ${\rm U}(1)$ charge conservation: from microscopics to continuum}
\label{sec:circuit2continuum }

\subsection{Microscopic model, observables, and the replica formulation}
\label{sec:circuit_micro}
In this work, we are interested in the monitored dynamics of the interacting fermions with ${\rm U}(1)$ charge conservation. In the following, we introduce the microscopic model for the monitored dynamical system and the observables that probe the nature of the dynamical phase of systems.

We consider the monitored dynamics of spinless fermions driven by unitary evolution with ${\rm U}(1)$ charge conservation and local fermion density measurements. The unitary part of the evolution is generated by a ${\rm U}(1)$-symmetric interacting Hamiltonian on a $d$-dimensional lattice:
\begin{align}
    H = \sum_{\vex{r},\vex{r}'} \left( t_{\vex{r}\vex{r}'} \psi^\dag_{\vex{r}} \psi_{\vex{r}'} +\rm{h.c.}\right)+ \frac{1}{2} \sum_{\vex{r},\vex{r}'} U_{\vec{r}\vec{r'}} n_{\vex{r}} n_{\vex{r}'},
    \label{eq:latticeHam}
\end{align}
which includes a short-ranged hopping term $t_{\vex{r}\vex{r}'}$ and a short-ranged interaction $U_{\vex{r}\vex{r}'}$. Here, $\psi_{\vex{r}}$ is the fermion operator on the lattice site located at $\vec{r}$ and $n_{\vex{r}} \equiv \psi_{\vex{r}}^\dag \psi_{\vex{r}}$ is the particle number operator at $\vec{r}$. The specific shape of the lattice and the details of $ t_{\vex{r}\vex{r}'}$ are unimportant for the universal properties of the dynamical phases we will investigate. We will keep the spatial dimension $d$ general in the discussion of the microscopic models and the general formulation of the continuum path integral.

The measurements that ``monitor" the system are the weak measurements \cite{NielsenChuang2010} of the particle number operator $n_{\vex{r}}$ at every site. At a given site $\vec{r}$, every one of such a measurement yields two possible outcomes $\upsilon_{\vec{r}}=0,1$ resulting in two different quantum trajectories in which the system's density matrix $\rho$ evolves as $\rho \rightarrow  K_{\upsilon_{\vec{r}}} \rho  K_{\upsilon_{\vec{r}}}^\dag $. Here, $K_{\upsilon_{\vec{r}}}$ is the
(trajectory-dependent)
Kraus operator that describes the wavefunction collapse caused by the measurement:
\begin{align}\label{eq:K_upsilon}
    K_{\upsilon_{\vec{r}}} = \frac{1}{\sqrt{2\cosh 2\kappa}} e^{4 \kappa (\upsilon_{\vec{r}}-\frac{1}{2}) (n_{\vec{r}}-\frac{1}{2})},
\end{align}
with $\kappa>0$ the measurement strength, a tuning parameter for the system's
dynamics. The Born-rule probability of the trajectory with measurement outcome $\upsilon_{\vec{r}}$ is given by
\begin{align}
    p_{\upsilon_{\vec{r}}} = \frac{\Tr\left(K_{\upsilon_{\vec{r}}} \rho  K_{\upsilon_{\vec{r}}}^\dag\right)}{\Tr\rho}.
\end{align}
This weak measurement of local particle number preserves the ${\rm U}(1)$ symmetry of the system (because the Kraus operator $K_{\upsilon_{\vec{r}}}$ commutes with total particle number operator $\sum_\vec{r} n_\vec{r}$). Here, the Kraus operators in Eq. \eqref{eq:K_upsilon} satisfy the condition for a positive operator-valued measure $\sum_{\upsilon_{\vec{r}}=0,1} K^\dag_{\upsilon_{\vec{r}}}K_{\upsilon_{\vec{r}}} = \mathds{1}$, which ensure that the Born-rule probabilities are always normalized. 

In the limit of $\kappa \rightarrow +\infty$, the weak measurement reduces to the standard projective measurement of the operator $n_{\vec{r}}$. The Kraus operator $K_{\upsilon_{\vec{r}}}$ reduces to the projector onto the states with $n_{\vec{r}}=\upsilon_{\vec{r}}=0,1$.

The interacting monitored dynamics we will study are given by the iteration of the unitary evolution $e^{-\ii H \Delta t}$ (for a time step $\Delta t$) and the weak measurement of $n_\vec{r}$ on every site. Let's denote the measurement outcomes at time $t=m \Delta t$ as $\upsilon_{\vec{r},t=m\Delta t}$. The system's density matrix evolves as $\rho_0 \rightarrow \rho=  {\cal V}_{\{\upsilon_{\vec{r},t}\}} \rho_0 {\cal V}_{\{\upsilon_{\vec{r},t}\}}^\dag$.  ${\cal V}_{\{\upsilon_{\vec{r},t}\}}$ is the total evolution operator ${\cal V}_{\{\upsilon_{\vec{r},t}\}} $ associated with quantum trajectory labeled by the measurement outcomes $\{\upsilon_{\vec{r},t}\}$:
\begin{align}
    {\cal V}_{\{\upsilon_{\vec{r},t}\}} = \prod_m \left( \hat{K}_{\{\upsilon_{\vec{r},t=m T}\} }\cdot e^{-\ii H \Delta t} \right).
\end{align}
Here, we've used $\rho_0$ to denote the (normalized) initial density matrix of the system and defined
\begin{align}\label{eq:K_many_upsilon}
    \hat{K}_{\{\upsilon_{\vec{r},t=m \Delta t}\} } = \prod_{\vec{r}} K_{\upsilon_{\vec{r},t=m\Delta t}}.
\end{align}
The total Born-rule probability for the quantum trajectory $\{\upsilon_{\vec{r},t}\}$ is then given by
\begin{align}\label{eq:Kraus_evol}
    p_{\{\upsilon_{\vec{r},t}\}} = \Tr\left( {\cal V}_{\{\upsilon_{\vec{r},t}\}} \rho_0 {\cal V}_{\{\upsilon_{\vec{r},t}\}}^\dag \right).
\end{align}

The nature of the dynamical phase of the system can be probed by the behavior of observables, such as entanglement entropy and higher-moment charge fluctuation, averaged over all quantum trajectories in the long-time limit. Such average quantities, many of them non-linear functions of the density matrix before averaging, can be systematically treated using a replica formulation as we explain below.

Take the entanglement entropy as an example to illustrate the replica formulation. For a subsystem $A$, its (normalized) reduced density matrix $\rho_{A}$ in the long-time limit in a given quantum trajectory $\{\upsilon_{\vec{r},t}\}$ is given by $\rho_A = \Tr_{\bar{A}}\left( {\cal V}_{\{\upsilon_{\vec{r},t}\}} \rho_0 {\cal V}_{\{\upsilon_{\vec{r},t}\}}^\dag \right) / \Tr\left( {\cal V}_{\{\upsilon_{\vec{r},t}\}} \rho_0 {\cal V}_{\{\upsilon_{\vec{r},t}\}}^\dag \right)$, where $\Tr_{\bar{A}}$ represents the trace over the degrees of the freedom in $A$'s compliment $\bar{A}$. The average (von Neumann) entanglement entropy $\bar{S}_A$ is defined by
\begin{align} \label{eq:EE_SA}
    \bar{S}_A \equiv \mathbb{E} \Big( p_{\{\upsilon_{\vec{r},t}\}} S_A \Big)= \mathbb{E} \Big( p_{\{\upsilon_{\vec{r},t}\}} \Tr (-\rho_A \log \rho_A) \Big),
\end{align}
where $\mathbb{E} $ represents the averaging over quantum trajectories. Note that each quantum trajectory ${\{\upsilon_{\vec{r},t}\}}$ is weighted by its Born-rule probability $p_{\{\upsilon_{\vec{r},t}\}}$ in this average.

A general technique to study such an average quantity is to apply the replica trick. Notice that $\log \rho_A = \lim_{n\rightarrow 0}\frac{1}{n}\left\{ \left[\Tr_{\bar{A}}\left( {\cal V} \rho_0 {\cal V}^\dag \right) \right]^n - \left[\Tr \left( {\cal V} \rho_0 {\cal V}^\dag \right)  \right]^n \mathds{1}_A
\right\}$. Here, the subscript of the evolution operator ${\cal V}$ is suppressed, and $\mathds{1}_A$ represents the identity operator on the subsystem $A$. Additionally, we see that $p_{\{\upsilon_{\vec{r},t}\}} \rho_A = \Tr_{\bar{A}}\left( {\cal V} \rho_0 {\cal V}^\dag\right)$. These observations relate the average entropy $\bar{S}_A$ to the {\it un-normalized} average $R$-replica density matrix
\begin{align}
    \rho^{(R)}
    = \mathbb{E} \left[ \left( {\cal V}_{\{\upsilon_{\vec{r},t}\}} \rho_0 {\cal V}_{\{\upsilon_{\vec{r},t}\}}^{\dag} \right)^{\otimes R} \right]
    \label{eq:ave_evo}
\end{align}
in the replica limit $R\rightarrow 1$ (with $R=n+1$ and $n\rightarrow 0$). More generally, one can consider the average $k$th Renyi entropy 
$\mathbb{E} \Big[ p_{\{\upsilon_{\vec{r},t}\}} \frac{1}{1-k}\Tr (\log \rho_A^k) \Big]$ 
or the average $k$th moment of the {\it normalized} density matrix (or other related quantities). The replica limit is $R = nk+1$ with $n\rightarrow 0$, which still results in $R\rightarrow 1$, as shown in Refs. \cite{Jian20,Bao20}. Physically, different Renyi entropies (or different moments) of the system at a large enough time $t$ should be viewed as different ``boundary probes" (at the temporal boundary $t$) of the same dynamical phase living in the bulk of spacetime. Therefore, it is natural that the Renyi index or the order of the moments does not change the replica limit $R\rightarrow 1$, which is intrinsic to the dynamical phase.  
In the long-time limit, the specific choice of the initial state $\rho_0$ is insignificant for the universal behavior of the average entanglement entropy $\bar{S}_A$ (as well as other average observables). Calculating $\bar{S}_A$ amounts to choosing a certain boundary condition at the final time slice for spacetime dynamics of  $\rho^{(R)}$\cite{VasseurRTN2019,ZhouRUC2019,Jian20,Bao20}. Therefore, understanding the behavior of $\rho^{(R)}$ is the key to understanding the monitored dynamics in the $(d+1)$-dimensional spacetime.

It is worth emphasizing that studying $\rho^{(R)}$ (in the replica limit $R\rightarrow 1$) is not just to understand the dynamics of entanglement. Average observables that probe the
charge-sharpening
dynamics, for example, the average second-moment density-density correlator
\begin{align} \label{eq:charge_correl}
    \mathbb{E}\left( p_{\{\upsilon_{\vec{r},t}\}} \left[ \big\langle n_\vec{r} \big \rangle_{\!\! \upsilon} \, \big\langle n_{\vec{r}'} \big\rangle_{\!\! \upsilon} -  \big\langle n_\vec{r} n_{\vec{r}'}  \big \rangle_{\!\! \upsilon} \right] \right),
\end{align}
can also be captured by $\rho^{(R)}$ in the same replica limit $R\rightarrow 1$ \cite{Barratt_U1_FT}. Here, $\big\langle \cdot \big \rangle_{\!\! \upsilon}$ represents the quantum expectation value (at the final time slice) in a given quantum trajectory. This correlator is called a second-moment correlator because the $\big\langle n_\vec{r} \big \rangle_{\!\! \upsilon} \, \big\langle n_{\vec{r}'} \big\rangle_{\!\! \upsilon}$ part essentially depends on two copies of the system's quantum-trajectory-resolved density matrix.

For a generic monitored dynamical system of qudits, the corresponding average evolution of $\rho^{(R)}$ can be mapped to a statistical model that captures the dynamical phases and measurement-induced phase transition of the monitored dynamics \cite{Jian20,Bao20}. In the subsequent sections, we will develop a field-theoretic description of $\rho^{(R)}$ as a general tool to understand the phase diagram of the monitored dynamics of interacting fermions with ${\rm U}(1)$ charge conservation.

\subsection{Continuum replica Keldysh field theory and symmetries }
\label{sec:Kel_Description}
In the following, we will present the description of the continuum version of $\rho^{(R)}$ in the language of Keldysh field theory. This Keldysh field theory
describing the quantum dynamics of the system subjected to measurements is nothing other than the path integral representation of the continuum version of $\rho^{(R)}$ in the ($R$-replica) doubled Hilbert space. The underlying physical reason for this is, as will be explained in Sec. \ref{sec:KDeriv}, the trivialization of the fermion distribution function caused by measurement-induced heating effects.
In this subsection, we will focus on presenting the Keldysh field theory. The connection between the Keldysh field theory and the evolution in the doubled Hilbert space will be presented in the next subsection.

As shown in Eq.~\eqref{eq:ave_evo}, $\rho^{(R)}$ involves the averaging of $R$ replicas of ${\cal V}$'s and $R$ replicas of ${\cal V}^\dag$'s. The path integral representation of ${\rho}^{(R)}$ should include the complex Grassmann fields $\psi_{+,j}(t,\vec{r})$ and $\bar{\psi}_{+,j}(t,\vec{r})$ (with $j=1,..,R$) for the $R$ replicas of ${\cal V}$'s, and $\psi_{-,\bar{j}}(t,\vec{r})$ and $\bar{\psi}_{-,\bar{j}}(t,\vec{r})$ (with $\bar{j}=1,..,R$) for the $R$ replicas of ${\cal V}^\dag$'s. In the language of the Keldysh field theory, the subscripts $+/-$ correspond to the $T$-ordered and $\bar{T}$-ordered branches of the Keldysh contour. The averaging over all quantum trajectories will be implemented as the path integral of the field $\upsilon(t,\vec{r})$ that represents the measurement outcomes associated with each quantum trajectory, i.e., the continuum counterpart of $\upsilon_{\vec{r}, t}$ in Sec. \ref{sec:circuit_micro}.
Note that all the replicas share the same measurement outcomes. Hence, the field $\upsilon(t,\vec{r})$ does not carry any replica index. The Keldysh path integral of the average evolution operator takes the following form:
\begin{align}
    {\rho}^{(R)}  = \int {\cal D}\upsilon \int {\cal D}\bar{\psi}_\pm  {\cal D}\psi_\pm ~ P[\upsilon] \cdot e^{\ii S^{(R)}_K}.
    \label{eq:path_integral_rhoR}
\end{align}
The action $S^{(R)}_K$ and weight factor $P[\upsilon]$ for the quantum trajectories are constructed as follows.

The action $S^{(R)}_K$ takes the form
\begin{align}
    S^{(R)}_K = S^{\rm u}_+ - S^{\rm u}_- + S_+^{\rm m} + S_-^{\rm m},
    \label{eq:SK}
\end{align}
where the subscript $+/-$ labels the contributions from ${\cal V}^{\otimes R}$/${\cal V}^{\dag \otimes R}$. The superscripts ``${\rm u}$" and ``${\rm m}$" distinguish the contributions from the unitary evolution and those from the measurements. The unitary parts of the action is given by
\begin{align}\label{}
  S^\text{u}_{+} = \int \rd t \rd^d\vec{r} \left[\sum_{j=1}^R \Big\{ \bar{\psi}_{+,j} \ii\partial_t \psi_{+,j}-\calH[\bar{\psi}_{+,j},\psi_{+,j}]\Big\} \right], \nonumber \\
  S^\text{u}_{-} = \int \rd t \rd^d\vec{r} \left[\sum_{\bar{j}=1}^R \Big\{ \bar{\psi}_{-,\bar{j}} \ii\partial_t \psi_{-,\bar{j}}-\calH[\bar{\psi}_{-,\bar{j}},\psi_{-,\bar{j}}]\Big\} \right]
\end{align}
with the Hamiltonian density defined as
\begin{equation}\label{}
  {\mathcal H}[\bar{\psi},\psi]= \bar{\psi}\left(-\frac{\nabla^2}{2m}-\mu\right){\psi} +\frac{U}{2} n^2
  \,.
\end{equation}
where $n=\bar{\psi} \psi$ is the local particle density. Here, we have replaced the lattice Hamiltonian in Eq.~\eqref{eq:latticeHam} with its continuum counterpart. We have made a technical choice of the kinetic term $\bar{\psi}\left(-\frac{\nabla^2}{2m}-\mu\right){\psi}$ and the local interaction $\frac{U}{2} n^2$. However, we stress that our final results on the universal behavior of the monitored dynamics are independent of the specific form of the kinetic term and interaction. Also, we will choose to work with a time-independent interaction $U$ for most of our analysis. As shown in App. \ref{app:Brownian}, a time-dependent random interaction also leads to the same conclusions.

As shown in Eq.~\eqref{eq:SK}, $S^\text{u}_{+}$ and $S^\text{u}_{-}$ contributes to $S^{(R)}_K$ with opposite signs, just like the standard Keldysh path integral without replicas. The opposite signs reflect the opposite ``arrows of time" on the two branches of the Keldysh contour. $S^\text{u}_{+}$ arises from ${\cal V}^{\otimes R}$ and lives on the $T$-ordered branch, while $S^\text{u}_{-}$ comes from ${\cal V}^{\dag\otimes R}$ and lives on the $
\bar{T}$-ordered branch.

For the measurement-induced terms $S_\pm^{\rm m}$, we can write
\begin{align}
    &S^\text{m}_{+}+S^\text{m}_{-}
    \nonumber \\
    &= -\ii\int\rd t\rd^d\vec{r} ~ \upsilon(t,\vec{r}) \left( \sum_{j=1}^R n_{+,j}(t,\vec{r})   + \sum_{\bar{j}=1}^R n_{-,\bar{j}}(t,\vec{r}) \right),
    \label{eq:S_measure}
\end{align}
which captures the continuum limit of the weak measurement of charge density
[Eq.~\eqref{eq:K_upsilon}]
on every site. On the lattice, the quantum trajectories are labeled by $\upsilon_{\vec{r},t}$, which could take values $0$ and $1$. In the continuum, upon coarse-graining, we can replace the combination $\kappa(\upsilon_{\vec{r},t}-\frac{1}{2})$ for the lattice system, which takes values $\pm \frac{\kappa}{2}$, by a Gaussian variable $\upsilon(t,\vec{r})$ with zero mean, i.e. $\overline{\upsilon(t,\vec{r})} = 0$, and a variance $\overline{\upsilon(t,\vec{r})\upsilon(t',\vec{r}')} = g^2 \delta(t-t')\delta(\vec{r}-\vec{r}')$. The physical meaning of the parameter $g$ is as follows. In the quantum circuit description with discretized time steps (see Sec. \ref{sec:circuit_micro}), every fermion mode undergoes a weak measurement of strength $\kappa$ per time interval $\Delta t$. In the continuum, the measurement strength $\kappa$ and the frequency $(\Delta t)^{-1}$ are coarse-grained into this one parameter $g^2$ that describes the effective measurement rate of the local charge density. The Gaussian distribution of $\upsilon(t, \vec{r})$ is imposed by the Gaussian weight
\begin{align}\label{eq:noiseavg}
    P[\upsilon]
    =
    \exp\left[- \frac{1}{2 g^2} \int dt d^d\vec{r} \, \upsilon^2(t,\vec{r})\right]
\end{align}
in Eq.~\eqref{eq:path_integral_rhoR}. Extra constant shifts to the terms in the parenthesis of Eq.~\eqref{eq:S_measure} have been dropped without affecting our final results. Imposing the Gaussian distribution of $\upsilon(t, \vec{r})$ is for technical convenience and is not expected to be crucial for the universal behavior of the monitored dynamics.

In contrast to the unitary parts $S^\text{u}_{\pm}$ of the action, the measurement-induced parts $S^\text{m}_{\pm}$ share the same sign for the $\pm$ time branches due to the fact that the Kraus operator Eq.~\eqref{eq:K_upsilon} associated with the measurements is Hermitian, i.e. $K_{\upsilon_\vec{r}}=K_{\upsilon_\vec{r}}^\dag $. Therefore, the measurement-induced action from ${\cal V}^{\otimes R}$ are the same as those from ${\cal V}^{\dag\otimes R}$.

Now, we are ready to discuss the symmetry of this path integral [Eq.~\eqref{eq:path_integral_rhoR}]. Let's first start with the non-interacting limit, i.e. $U=0$. The action $S_K^{(R)}$ is entirely built from bilinears of the fields $\psi_{+,j}$, $\bar{\psi}_{+,j}$, $\psi_{-,\bar{j}}$, and $\bar{\psi}_{-,\bar{j}}$. The action is invariant under a continuous $\rU(R)_+ \times \rU(R)_-$ symmetry:
\begin{align}
    \psi_{+,j}\rightarrow \sum_{j'} ({\cal W}_+)_{jj'}\psi_{+,j'}, ~~~~ \bar{\psi}_{+,j} \rightarrow \sum_{j'} ({\cal W}_+^*)_{jj'}\bar{\psi}_{+,j'},\nonumber \\
   \psi_{-,\bar{j}}\rightarrow \sum_{\bar{j}'} ({\cal W}_-)_{\bar{j}\bar{j}'}\psi_{-,\bar{j}'}, ~~~~ \bar{\psi}_{-,\bar{j}} \rightarrow \sum_{\bar{j}'} ({\cal W}_-^*)_{\bar{j}\bar{j}'}\bar{\psi}_{-,\bar{j}'},
\end{align}
where ${\cal W}_\pm \in \rU(R)_\pm$ are $R\times R$ unitary matrices that continuously mix different replicas on the $T$-ordered ($+$) and $\bar{T}$-ordered ($-$) time branches, respectively. In addition, there is an extra anti-unitary symmetry
that keeps $e^{\ii S_K^{(R)}}$ invariant:
\begin{align}
\begin{gathered}
        \psi_{+,j} \to \bar{\psi}_{-,\bar{j}= j},~~ \bar{\psi}_{+,j} \to -\psi_{-,\bar{j}= j}, \\
         \psi_{-,\bar{j}} \to \bar{\psi}_{ +,j=\bar{j}},~~
        \bar{\psi}_{-,\bar{j}} \to -\psi_{+,j=\bar{j}},
\end{gathered}
 ~~~~\ii \rightarrow - \ii.
    \label{eq:antiU_Sym}
\end{align}
Here, $\ii \to -\ii$ means complex conjugating anything other than the fermion operators. This symmetry arises from the fact that the evolution on the $T$-ordered time branch ($+$) and the $\bar{T}$-ordered time branch ($-$) are closely related. The former is generated by ${\cal V}$, and the latter is generated by ${\cal V}^\dag$.

When we turn on interactions, the $\rU(R)_\pm $ symmetry is reduced to the semi-direct product $[\rU(1)_\pm]^{ R} \rtimes S_{R,\pm}$. Here, $[\rU(1)_\pm]^{ R}$ represents the $R$ copies of the $\rU(1)$ symmetries generated by the diagonal unitary matrices in $\rU(R)_\pm $. Physically, it corresponds to the $\rU(1)$ charge conversation in each replica of ${\cal V}$ and ${\cal V
}^\dag$. $S_{R,\pm}$ is the permutation symmetry amongst the $R$ replicas of ${\cal V}$/${\cal V}^\dag$. There is no longer any symmetry that can continuously rotate the different replicas into each other when the interaction is present. In addition to $[\rU(1)_\pm]^{ R}$, the anti-unitary symmetry Eq.~\eqref{eq:antiU_Sym} is still a symmetry of the interacting field theory.

We emphasize that the symmetries independently acting on the $T$-ordered and the $\bar{T}$-ordered Keldysh branches are the symmetries in the bulk of the $(d+1)$-dimensional spacetime. Even though the $R$ replicas of
$\rho_0$ at the initial time $t=0$ do not enjoy the independent symmetries on the two branches, the resulting symmetry-breaking effect merely occurs on the spacetime boundary at $t=0$. It is inconsequential for the universal dynamical behavior of the system in the long-time limit. Physically, the microscopic details of $\rho_0$ are eliminated by the repeated measurements in the monitored dynamics and, hence, are unimportant for the long-time universal behavior. In the language of the conventional Keldysh formalism, the heating effects from the local measurements trivialize the fermion distribution function, rendering the microscopic details of the initial state $\rho_0$ not essential (see Sec. \ref{sec:KDeriv} for more discussions). Therefore, we should focus on the symmetries of the replica Keldysh field theory in the bulk of the $(d+1)$-dimensional spacetime (without worrying about the effects from the initial state $\rho_0$).

In the next subsection, we will show that the same field theory Eq.~\eqref{eq:path_integral_rhoR} can be re-derived as a path integral in a doubled Hilbert space. This re-derivation follows the same logic that led to the established statistical mechanics model description \cite{Jian20,Bao20} for the monitored dynamics of qudit systems.

\subsection{Relation to the coherent-state path integral in the doubled Hilbert space description}\label{sec:doublehilbert}

The path integral Eq.~\eqref{eq:path_integral_rhoR} was introduced using the language of the Keldysh field theory in Sec. \ref{sec:Kel_Description}. In the following, we will show that it is equivalent to the standard fermionic coherent-state path integral that captures the time evolution of $R$ replicas of the system's density matrix mapped to the $R$-replica doubled Hilbert space. This path integral in the ($R$-replica) doubled Hilbert space is the path integral representation of the statistical mechanics model of the monitored dynamics of interacting fermions, analogous to the established statistical mechanics model description\cite{Jian20,Bao20} for the monitored dynamics of qudit systems. The details of the initial state ${\rho}_0$ are inconsequential in both types of monitored dynamics. It is the universal behavior of the dynamics in the spacetime bulk that reflects the underlying phases and transitions. To illustrate the equivalence between the Keldysh field theory and the doubled Hilbert space formulation, we will focus on the case with one single replica. Generalizing to a generic replica number $R$ is straightforward.

Using the Choi–Jamio{\l}kowski isomorphism, we can map the density matrix ${\rho}$ (and its evolution) into the doubled Hilbert space via
\begin{align}
    | \rho \drangle \equiv ({\rho} \otimes \mathds{1}) \, |\Phi_0\drangle,
    \label{eq:CJ_Iso}
\end{align}
where $| \rho \drangle$ is known as the Choi representation of the density matrix $\rho$ in the doubled Hilbert space. The operator $(\rho \otimes \mathds{1})$ is essentially the density matrix $\rho$ acting only on the first copy of the Hilbert space. $|\Phi_0\drangle$ is a maximally entangled state between the two copies of the Hilbert spaces, which can be chosen as follows. For every fermionic mode $\psi_\vec{r}$ in the original physical Hilbert space, there is a corresponding mode $\chi_\vec{r}$ in the other auxiliary copy of this Hilbert space. In the state $|\Phi_0\drangle$, every pair of $\psi_\vec{r}$ and $\chi_\vec{r}$ are maximally entangled in their occupation basis:
\begin{align}
   |\Phi_0\drangle =\bigotimes_\vec{r} \left(\ket{0}_{\psi_\vec{r}} \ket{1}_{\chi_\vec{r}}  +\ket{1}_{\psi_\vec{r}} \ket{0}_{\chi_\vec{r}} \right).
\end{align}
Equivalently, the state $|\Phi_0\drangle$ is defined by identities
\begin{align}
    (\psi_\vec{r}^\dag + \chi_\vec{r}^\dag)|\Phi_0\drangle = (\psi_\vec{r} - \chi_\vec{r})|\Phi_0\drangle = 0.
    \label{eq:EPR_defining_relation}
\end{align}

Under the Choi–Jamio{\l}kowski isomorphism Eq.~\eqref{eq:CJ_Iso}, the evolution $\rho_0\rightarrow \rho = {\cal V}\rho_0{\cal V}^\dag$ is mapped to the evolution in the doubled Hilbert space
\begin{align}
 |\rho_0 \drangle \rightarrow |\rho \drangle = ({\cal V} \otimes {\cal V}^{\dag \mathsf{T}}) |\rho_0 \drangle
\end{align}
The transpose ${}^\mathsf{T}$
of any operator $O$ is defined via the relation $(O \otimes \mathds{1}) |\Phi\drangle = (\mathds{1} \otimes O^\mathsf{T}) |\Phi\drangle$.

Now, we examine the evolution operator ${\cal V} \otimes {\cal V}^{\dag \mathsf{T}} $ in greater detail. We will use the lattice version of this evolution to demonstrate the key steps to express it as a coherent-state path integral. As we shall see, this coherent-state path integral has an action of the same form as in the Keldysh field theory Eq.~\eqref{eq:SK}.

The unitary part ${\cal V} \otimes {\cal V}^{\dag \mathsf{T}} $ takes the form
\begin{align}\label{eq:H_otimes_HT}
    e^{-\ii H \Delta t } \otimes e^{ \ii H^\mathsf{T} \Delta t }.
\end{align}
When $H$ is put into the general normal-ordered form
\begin{align}
    H =  \sum_{\vec{r}_1,  \vec{r}_2} t_{\vec{r}_1\vec{r}_2}\psi_{\vec{r}_1}^\dagger \psi_{\vec{r}_2} +\sum_{\vec{r}_1,\vec{r}_2,\vec{r}_3,\vec{r}_4} V_{{\vec{r}_1}{\vec{r}_2}{\vec{r}_3}{\vec{r}_4}} \psi_{\vec{r}_1}^\dagger \psi_{\vec{r}_2}^\dagger \psi_{\vec{r}_3} \psi_{\vec{r}_4}\, ,
\end{align}
its transpose takes the form
\begin{align}
    H^\mathsf{T} =   \sum_{\vec{r}_1,  \vec{r}_2} t_{\vec{r}_1\vec{r}_2} \chi_{\vec{r}_2} \chi_{\vec{r}_1}^\dagger +\sum_{\vec{r}_1, \vec{r}_2, \vec{r}_3, \vec{r}_4} V_{{\vec{r}_1}{\vec{r}_2}{\vec{r}_3}{\vec{r}_4}} \chi_{\vec{r}_4} \chi_{\vec{r}_3} \chi_{\vec{r}_2}^\dag \chi_{\vec{r}_1}^\dag\, ,
\end{align}
which has a parallel structure as $H$ but with all the fermion operators $\chi_{\vec{r}}$ anti-normal ordered.

For the measurement-induced terms (at a given time step) in ${\cal V} \otimes {\cal V}^{\dag \mathsf{T}}$ take the form
\begin{align} \label{eq:K_otimes_KT}
    &\prod_{\vec{r}} K_{\upsilon_{\vec{r}}} \otimes K^{\dag\mathsf{T}}_{\upsilon_{\vec{r}}}
    \nonumber \\
    &~~~~ = \exp\left\{\kappa\sum_\vec{r} \left(\upsilon_\vec{r}-\frac{1}{2}\right)(\psi_\vec{r}^\dag \psi_\vec{r} + \chi_\vec{r} \chi_\vec{r}^\dag - 1 )\right\}.
\end{align}
Notice that the contributions from the $\psi$ and $\chi$ fermions are parallel except that the $\psi$ operators are normal-ordered and $\chi$'s are anti-normal-ordered.

One naive approach to express ${\cal V} \otimes {\cal V}^{\dag \mathsf{T}} $ using a coherent-state path integral is first to bring the terms built from the $\chi$ fermion operators to normal order and then introduce the coherent states for the $\psi$ and $\chi$ fermion operators following the same conventional recipe. However, the normal ordering of the $\chi$ fermion operators can generally lead to extra terms, resulting in an action that does not fit the form shown in Eq.~\eqref{eq:SK}. Interestingly, for the non-interacting system, this subtlety from normal ordering does not cause any extra complication because normal ordering the quadratic operators in $\chi$ only leads to additional unimportant constants. When we apply the standard recipe of coherent states to both the $\psi$ and $\chi$ fermion modes on equal footing, we will obtain a path integral whose action has a different sign structure than Eq.~\eqref{eq:SK}: The two unitary components $S^{\rm u}_\pm$ will contribute to the action with the same sign, while the measurement-induced terms $S^{\rm m}_\pm$ will carry opposite signs. This sign structure is consistent with the previous studies \cite{MFava2023,FavaU1} of the field theory of non-interacting monitored fermions using the doubled Hilbert space formulation. For our current work, we want to establish the equivalence between the {\it interacting}
Keldysh field theory Eq.~\eqref{eq:SK} and the doubled-Hilbert-space formulation (and to avoid the complications from normal ordering $\chi$). Therefore, we need to use an alternative definition for the coherent states to construct the path integral.

To correctly reproduce the form of the
{\it interacting}
Keldysh action Eq.~\eqref{eq:SK} (in particular, the relative signs between the different terms therein), we need to use different coherent states for $\psi$ and $\chi$ fermion operators. For the $\psi_\vec{r}$ fermion mode, we use the coherent state $|\psi_+(\vec{r})\rangle$ and $\langle\bar{\psi}_+(\vec{r})|$ that obeys $\psi_\vec{r}|\psi_+(\vec{r})\rangle = \psi_+(\vec{r})|\psi_+(\vec{r})\rangle$ and $\langle\bar{\psi}_+(\vec{r})| \psi_\vec{r}^\dag = \langle\bar{\psi}_+(\vec{r})| \bar{\psi}_+(\vec{r})$. $\psi_+(\vec{r})$ and $\bar{\psi}_+(\vec{r})$ are the complex Grassmann fields on the $+$ time fold of the Keldysh path integral. For the fermion operator $\chi_\vec{r}$ , we define the coherent states $|\bar{\psi}_-(\vec{r})\rangle$ and $\langle\psi_-(\vec{r})|$ according to $\chi_\vec{r}^\dag|\bar{\psi}_-(\vec{r})\rangle = \bar{\psi}_-(\vec{r})|\bar{\psi}_-(\vec{r})\rangle$ and $\langle \psi_-(\vec{r})|\chi_\vec{r} = -\langle\psi_-(\vec{r})| \psi_-(\vec{r}) $.
$\psi_-(\vec{r})$ and $\bar{\psi}_-(\vec{r})$ are the complex Grassmann field on the $\bar{T}$ time branch $(-)$
of the Keldysh path integral. The so-defined coherent states allow us to work directly with the anti-normal-ordered terms for the $\chi$ fermion operators. The resulting action will have the form shown in Eq.~\eqref{eq:SK}. Specifically, the unitary parts of ${\cal V} \otimes {\cal V}^{\dag \mathsf{T}} $, consisting of terms like $e^{-\ii H \Delta t } \otimes e^{ \ii H^\mathsf{T} \Delta t }$, result
in the $S^{\rm u}_+ - S^{\rm u}_-$ part of the action [Eq.~\eqref{eq:SK}].
In particular, the forms of $S^{\rm u}_\pm$ are identical. The measurement-induced terms of ${\cal V} \otimes {\cal V}^{\dag \mathsf{T}} $, consisting of operators like $\prod_{\vec{r}} K_{\upsilon_{\vec{r}}} \otimes K^{\dag\mathsf{T}}_{\upsilon_{\vec{r}}} $, lead
to the $S^{\rm m}_+ + S^{\rm m}_-$ part of the action [Eq.~\eqref{eq:SK}]. The forms of $S^{\rm m}_\pm$ are identical. Furthermore, the fermionic fields in the path-integral satisfy the standard anti-periodic boundary condition. More details of the above discussion are presented in Appendix.~\ref{app:pathintegral}. Generalizing these coherent states to a generic replica number $R$ is straightforward.

\section{Derivation from Keldysh microscopics \label{sec:KDeriv}}

In this section, we describe an alternative derivation of the theory presented in Sec.~\ref{sec:Results}. Instead of the gauging procedure used to obtain
Eq.~(\ref{eq:SI}) from the effective non-interacting fermion
theory in Eq.~(\ref{eq:PCM}),
we begin with the microscopic Keldysh formulation of monitored, interacting fermions and proceed via a saddle-point analysis and trace-log expansion.  In this section, we
focus on the case where the interaction $U$ is a constant term. In App.~\ref{app:Brownian}, we consider another
formulation where the interactions are generated by random unitaries and are modeled by a Brownian interaction, yielding qualitatively similar results.

The formal $R$-replica Keldysh path integral has been set up in Sec.~\ref{sec:circuit2continuum }. To set the stage
for the saddle-point analysis and trace-log expansion,  we consider first the formulation for one replica and
a single, fixed quantum trajectory $\upsilon(t,\vex{r})$.
Later we replicate the theory in order to perform trajectory averaging.
The starting point is a Keldysh path integral \cite{Liao17,Kamenev23}
\begin{align}
    Z
    =
    \int
    \calD\bar{\psi} \, \calD \psi \,  e^{\ii S},
\end{align}
with the action
\begin{align}
\label{eq:SDef0}
    S
    =&\,
    S_\psi
    -
    \frac{U}{2}
    \intl{t,\vex{r}}
    \left[
        \left(\bar{\psi}_{+} \psi_{+}\right)^2
        -
        \left(\bar{\psi}_{-} \psi_{-}\right)^2
    \right]
\nonumber\\
&\,
    +
    \ii
    \intl{t,\vex{r}}
    \upsilon(t,\vex{r})
    \,
    n_{\cl}.
\end{align}
Here, $\intl{t,\vex{r}} \equiv \int dt \, d^d\vex{r}$
and
$\psi_{\pm}(t,\vex{r})$ are $T$ ($\bar{T}$)-contour-ordered fermions in $d$+$1$ dimensions.
In Eq.~(\ref{eq:SDef0}),
$U > 0$ is the strength of repulsive density-density interactions,
and $\upsilon(t,\vex{r})$ captures the measurements of the so-called
``classical'' (contour-summed) density operator
\begin{align}
    n_{\cl} \equiv \bar{\psi}_{+} \psi_{+} + \bar{\psi}_{-} \psi_{-}.
\end{align}
The non-interacting, unitary part of the theory is encoded in
\begin{align}
    S_\psi
    =
    \intl{t,\vex{r}}
    \bar{\psi}
    \,
    \hat{G}_0^{-1}
    \,
    \psi,
\end{align}
where the Keldysh Green's function is a matrix in $T$/$\bar{T}$ space \cite{Kamenev23}
\begin{multline}
\!\!\!\!\!
    \ii \, \hat{G}_0(t,\vex{r};t',\vex{r'})
    \\
    \rightarrow
    \begin{bmatrix}
    \left\langle T \, \psi_{+}(t,\vex{r}) \, \bar{\psi}_+(t',\vex{r'})\right\rangle
    &
    - \left\langle \bar{\psi}_{-}(t',\vex{r'}) \, {\psi}_+(t,\vex{r})\right\rangle
    \\[4pt]
    \left\langle \psi_{-}(t,\vex{r}) \, \bar{\psi}_+(t',\vex{r'})\right\rangle
    &
    \left\langle \bar{T} \, \psi_{-}(t,\vex{r}) \, \bar{\psi}_-(t',\vex{r'})\right\rangle
    \end{bmatrix}_{T,\bar{T}}\!\!.
\!\!
\end{multline}

It is convenient to decouple the interactions using Keldysh bosonic Hubbard-Stratonovich fields
$a_{\pm}$, leading to
\begin{align}\label{eq:Zadecoup}
    Z
    =
    \int
    \calD{a} \,
    \calD\bar{\psi} \, \calD \psi \,  e^{\ii S},
\end{align}
where
\begin{align}
    S
    =
    S_\psi
    +
    S_a
    -
    \intl{t,\vex{r}}
    \left[
        n_{\cl}
        \,
        \tilde{a}_\q
        +
        n_{\q}
        \,
        \tilde{a}_\cl
        -
        \ii
        \,
        \upsilon(t,\vex{r})
        \,
        n_{\cl}
        \right],
    \label{SDef}
\end{align}
and
\begin{align}\label{Sa}
    S_a
    =
    \frac{2}{U}
    \intl{t,\vex{r}}
    a_\cl
    \,
    a_\q,
\end{align}
with $a_{\cl,\q} \equiv (1/2)(a_+ \pm a_-)$.
The quantum component of the fermion density is
$n_{\q} = \bar{\psi}_{+} \psi_{+} - \bar{\psi}_{-} \psi_{-}$.

Following standard practice,
we switch to the Larkin-Ovchinnikov (LO) basis \cite{Kamenev23}.
The LO basis is obtained from a rotation in Keldysh space.
The fermions residing on the $T$ and the $\bar{T}$ time branches
are combined into a two-component Keldysh spinor $\psi \equiv \{\psi_{+},\psi_{-}\}$.
Then the transformation
\begin{align}\label{eq:LO}
\begin{gathered}
    \psi \rightarrow \tauh^3 \, \hat{U}_{\mathsf{LO}} \, \psi,
\quad
    \bar{\psi} \rightarrow \bar{\psi} \, \hat{U}_{\mathsf{LO}}^\dagger,
\\
    \hat{U}_{\mathsf{LO}} \equiv \frac{1}{\sqrt{2}}\left(\hat{1} + \ii \, \tauh^2\right)
\end{gathered}
\end{align}
brings the free-fermion Green's function into the standard form \cite{Kamenev23}
\begin{align}\label{eq:G0LO}
    \hat{G}_0
    \rightarrow
    \begin{bmatrix}
    G_0^R & G_0^K \\
    0 & G_0^A
    \end{bmatrix}_{\mathsf{LO}},
\end{align}
where $G_0^{R,A,K}$ denotes the retarded, advanced, and Keldysh Green's functions, respectively.
In Eq.~(\ref{eq:LO}), the Pauli matrices $\tauh^{i}$ act on the Keldysh spinor components. $\hat{1}$ is a $2\times 2$ identity matrix.

In the LO basis, the classical and quantum fermion density operators take the forms $n_{\cl,\q} \equiv \{\bar{\psi}\tauh^1\psi,\bar{\psi}\psi\}$. Physical external electric potentials couple to $n_{\q}$,
while $n_{\cl}$ measures the particle density in physical response functions \cite{Kamenev23}.
In Eq.~(\ref{SDef}),
$\tilde{a}_{s} \equiv a_s + V_s$, where $V_s$ denotes an additional source field that we have incorporated. These can be used to
generate response functions or incorporate external electric fields.

The ``noise'' $\upsilon(t,\vex{r})$ in Eq.~(\ref{SDef}) encodes weak measurements of the conserved particle density,
as explained above in Sec.~\ref{sec:Kel_Description} [see Eq.~(\ref{eq:S_measure})].
Because it couples to $n_{\cl}(t,\vex{r})$, $\upsilon$ enters the action identically to the source utilized to compute full counting statistics of shot noise in a mesoscopic quantum device \cite{Kamenev23,Klich09}. We replicate all fields $\psi \rightarrow \psi_{j}$, $\tilde{a}_s \rightarrow \tilde{a}_{s,j}$, with $j \in \{1,2,\ldots,R\}$.
The actions $S_a$ and $S_\varphi$ in Eqs.~(\ref{Sa}) and (\ref{eq:Sphi}), respectively, are the same with the identification
$a_{\cl,\q,j} = (\ii/2)(\varphi_{+,j} \pm \varphi_{-,\bar{j}})$.
In the absence of interactions or source terms, the action in Eq.~(\ref{SDef}) is invariant under U($R$) $\times$ U($R)$ transformations that preserve the form of
$n_{\cl} = \bar{\psi} \tauh^1 \psi$.
This implies that the effective field theory for non-interacting
fermions has the target manifold associated with class AIII \cite{Mirlin23,FavaU1} (see also Ref.~\cite{CMJian2022} for the identification of the symmetry class). Interactions break the symmetry down to
$S_R$ $\times$ $S_R$, where $S_R$ is the permutation group, as well as the $[\rU(1)]^{R}$ symmetry associated with the conservation of charge within each replica.

The replica trick enables us to average over quantum trajectories described by different measurement outcomes. This is equivalent to averaging over $\upsilon(t,\vex{r})$, which we take to be Gaussian correlated with a white noise variance $g^2$.
The latter is the effective measurement rate in the monitored dynamics [see Eq.~(\ref{eq:noiseavg}) in Sec.~\ref{sec:Kel_Description}].
In the end, we must take the replica $R \to 1$ limit in order to correctly weigh different trajectories according to the Born rule \cite{Jian20,Bao20}.
Up to irrelevant operators, the same structure is obtained when averaging over projective measurements and outcomes \cite{Mirlin23}.
We decouple the resulting four-fermion bilinear with a generalized Hubbard-Stratonovich
transformation \cite{Chahine23} and integrate out the fermions to get the effective field theory
\begin{align}
    \bar{Z}
    =
   \int
   \mathcal{D}\hqq \,
   \mathcal{D}a \,
   e^{\ii S_a - \bar{S}},
\end{align}
where
\begin{align}\label{eq:barS}
\!\!\!
    \bar{S}
    =&
    \frac{\lambda }{8}
    \intl{t,\vex{r}}
        \left\{
        \trr\left[\left(\hqq \, \tauh^1\right)^2\right]
        -
        f_\alpha
        \left[\trr\left(\hqq \, \tauh^1\right)\right]^2
        \right\}
\nonumber\\
    &\,
    -
    \Tr\log
    \!
\left[
    \hat{G}_0^{-1}
    +
    \hqq
    -
    f_\alpha
    \,
    \tauh^1
    \,
    \tr(\hqq \, \tauh^1)
    -
    \tilde{a}_\cl
    -
    \tilde{a}_\q
    \,
    \tauh^1
\right]\!.
\!\!\!
\end{align}
Here $\hqq \rightarrow Q^{\tau,\tau'}_{i j}(t,\vex{r})$ is a spacetime-local matrix-valued field that carries (LO)
Keldysh ($\tau,\tau'$)
and
replica ($i,j$) indices.
Averaging over $\upsilon$ introduces a spacetime-local term of the form
$(\sum_i\bar{\psi}_i\hat{\tau}^1\psi_i)^2
=
-
\trr
( \sum_{i,j} \psi_i \bar{\psi}_j \hat{\tau}^1 \psi_j \bar{\psi}_i \hat{\tau}^1)
$.
This four-fermion term can be decoupled in two different ways \cite{Mirlin23,Chahine23},
as shown in the first line of Eq.~\eqref{eq:barS}. We therefore retain both terms and their weights are parameterized by $\alpha$ and $1-\alpha$, respectively.
The coupling strengths
$\lambda \equiv 4/g^2 \alpha$ and $f_\alpha \equiv (1-\alpha)/\alpha$
depend on the
arbitrary Hubbard-Stratonovich parameter $\alpha$.
A particular value of this parameter turns out to be required in order
for the effective field theory
to exhibit diffusive average-density dynamics, which
satisfy the U(1) Ward identity.
The Ward identity is summarized in the next section, while the Hubbard-Stratonovich decoupling that produces Eq.~(\ref{eq:barS}) and the necessity of choosing a particular $\alpha$ are explicated
in Appendix~\ref{app:WardU(1)}.
In the rest of this section we focus on the case of 1+1D, where we must take
$\alpha = 2R/(1 + 2 R)$; here $2 R$ counts the number of left- and right-moving fermions.
In Eq.~(\ref{eq:barS}),
$\trr$ ($\Tr$) denotes a trace over replica, Keldysh, left-right mover indices (and spacetime).

We extremize $\bar{S}$ to obtain the saddle point. We note that due to the choice $\alpha=2R/(1+2R)$ and $f_\alpha=1/(2R)$, $\tr(\hat{Q} \, \hat{\tau}^1)$ becomes a free parameter
of the theory, which describes the (replica summed) charge density. For simplicity, we will assume the system is at half-filling and demand that $\tr(\hat{Q} \, \hat{\tau}^1)=0$ at the saddle. Although both continuous U$(R)$-symmetric and $S_R$-symmetric saddles are possible, the latter
turns out to give
a theory with additional unphysical massless modes when interactions are included. \footnote{The $S_R$-symmetric saddle is
$
    \hqq_{\spp} = \ii \, \gamma \, \tauh^3 \, \hat{\mathcal{S}}_R \otimes \hat{1}_2,
$
where $(\hat{\mathcal{S}}_R)_{i j} = (2/R) - \delta_{i j}$.
Parameterizing fluctuations around this saddle gives an identical effective theory for non-interacting fermions [Eq.~(\ref{eq:PCM})]; however, the resulting interacting theory possesses additional, unphysical gapless modes.
}
We therefore calculate for the continuous
U$(R)$-symmetric
saddle,
\begin{align}\label{eq:qSP}
    \hqq_{\spp} = \ii \, \gamma \, \tauh^3 \, \hat{1}_R\otimes\hat{1}_2,
\end{align}
where $\gamma \equiv (2/\lambda) (\Lambda/2\pi)$ is the decay rate for the fermions induced by the measurements and $\Lambda$ is a hard momentum cutoff. Here, $\hat{1}_R$ is the identity matrix in the replica indices and $\hat{1}_2$ is the identity matrix in the basis of left- and right-movers.
This saddle preserves the causal structure of the fermion Green's functions, but corresponds formally to ``infinite temperature:'' the Keldysh fermion Green's function $G^K$ vanishes at half-filling [cf.\ Eq.~(\ref{eq:G0LO})], consistent with a vanishing distribution function $F(\omega) = 0$.
[In equilibrium at temperature $T$, $F(\omega) = \tanh(\omega/2 T) = 1 - 2 \, f(\omega)$, where
$f(\omega)$ is the Fermi-Dirac function.]
A similar form of the saddle-point in Eq.~(\ref{eq:qSP}) is found away from half-filling, with a nonzero but frequency-independent $F$ determined by the total particle density. We work out the half-filled case here;
deviations from half-filling are considered in Appendix \ref{app:ahf}.
Measurement-induced ``heating'' for a generic quantum trajectory produces a steady state with a featureless average level occupation number, 
irrespective of the entanglement properties (area vs.\ volume law)---see Sec.~\ref{sec:MIH}.

In the LO basis, the U($R$) $\times$ U($R)$ fluctuations around the saddle point can be expressed using the decomposition:
\begin{align}\label{QDecomp}
    \hqq(t,x)
    &=
    \frac{\ii \gamma}{2}
    \left[
        \ii \tauh^2
        \left(\hxc^\dagger - \hxc\right)
        +
        \tauh^3
        \left(\hxc + \hxc^\dagger\right)
    \right],
\end{align}
where $\hxc$ is a replica-space U($R$) matrix field with saddle-point $\hxc_\spp = \hat{1}_R$.
In Eq.~(\ref{QDecomp}), we restrict to fluctuations diagonal in the left- and right-mover basis, and drop the $\hat{1}_2$ factor to lighten the notation.
Translating this equation from the LO basis to the one for $T$- and $\bar{T}$-ordered fermion fields, Eq.~(\ref{QDecomp}) becomes
\begin{align}
    \hqq
    \rightarrow
    \gamma
    \begin{bmatrix}
    0 & \ii \, \hxc \\
    -\ii \, \hxc^\dagger & 0
    \end{bmatrix}_{T,\bar{T}},
\end{align}
leading to the association in Eq.~(\ref{eq:XFieldIdentify}),
with $\hxc \equiv e^{i \theta/R} \, \hx$.

We expand Eq.~(\ref{QDecomp}) in Gaussian fluctuations around the saddle point,
\begin{align}\label{YDef}
    \hqq - \hqq_{\spp}
    \equiv
    \ii
    \frac{\gamma}{\sqrt{\lambda}}
    \,
    \tauh^2
    \,
    \left(\hy + \frac{\theta}{R}\right)
    -
    \frac{\gamma}{\sqrt{\lambda}}
    \,
    \tauh^1
    \,
    \frac{\phi}{R}.
\end{align}
Here $\hy$ ($\theta$) generates SU($R$) [U(1)] transformations on the AIII manifold,
while
$\phi$ is a nominally massive mode that was excluded from the standard Goldstone parameterization in Eq.~(\ref{QDecomp}).
In fact, it is necessary to retain this mode, because it is \emph{made massless} by
the
parameter choice for $\alpha$ described above
(see Appendix~\ref{app:WardU(1)} for an explicit demonstration of this).
Because $n_\cl = \bar{\psi} \tauh^1 \psi$ in the LO basis, we expect that
$\phi$ plays a role in the effective field-theory encoding of the replica-averaged fermion density. This is indeed the case [Eq.~(\ref{eq:rhocl})], 
see Appendix~\ref{app:ahf} for the articulation of the precise connection.

Performing a fourth-order trace-log expansion and integrating out the $a_{s}$ fields, we obtain the effective field theory
\begin{align}\label{eq:SExpansion}
    \bar{S}
    =
    S_{\textrm{SU($R$)}}^\pup{2}
    +
    S_{\textrm{U(1)}}
    +
    S_I
    +
    S_V,
\end{align}
where in 1+1D
the components $S_{\textrm{U(1)}}$ and $S_I$ are respectively given by
Eqs.~(\ref{eq:U(1)}) and (\ref{eq:SIY}) in Sec.~\ref{sec:Results}. 
$S_{\textrm{SU($R$)}}^\pup{2}$ is the Gaussian fluctuation approximation to the PCM action in Eq.~(\ref{eq:PCM}),
\begin{align}
\label{eq:Y_2nd}
    S_{\textrm{SU($R$)}}^\pup{2}
    =
    \frac{1}{8}
    \intl{t,x}
    \trr_R
    \left[
    (\parr_t \hy)^2
    +
    v^2
    (\parr_x \hy)^2
    \right].
\end{align}
The Euclidean Laplacian appears
here via the loop integral 
\begin{align}
    \int \frac{d \Omega \, d q}{(2 \pi)^2}
    &\,
    \trr_\sigma\left[\hat{G}^R_\spp(\Omega,q) \, \hat{G}^A_\spp(\Omega-\omega,q-k)\right]
\nonumber\\
    =&\,
    \left(\frac{\Lambda}{2 \pi}\right)
    \left[
    \frac{1}{2 \gamma - \ii(\omega - v \, k)}
    +
    \frac{1}{2 \gamma - \ii(\omega + v \, k)}
    \right]
\nonumber\\
    \simeq&\,
    \frac{\lambda}{2}
    \left[
        1
        +
        \frac{\ii \, \omega}{2 \gamma}
        -
        \frac{v^2 \, k^2 + \omega^2}{4 \gamma^2}
    \right],
\end{align}
where the trace runs over right- and left-movers, and $\hat{G}^{R,A}_{\spp}$ denotes the saddle-point Green's function.
The action $S_V$ 
in Eq.~(\ref{eq:SExpansion})
encodes the source field that couples only to the average density in the U(1) sector; it is discussed in the next subsection, see Eq.~(\ref{eq:SV}).

To the working order in the expansion, the \emph{replica-resolved} classical and quantum fermion density operators in Eq.~(\ref{eq:varrhoclqDef}) take the forms
\begin{align}
    n_{\cl,j}
    \sim
    -
    \frac{\sqrt{\lambda}}{4}
    \parr_t Y_{jj},
\quad
    n_{\q,j}
    \sim
    \frac{\ii}{8}
    \left[
    (\parr_t \hy),
    \hy
    \right]_{jj},
\end{align}
where
$[\hat{A},\hat{B}] = \hat{A} \, \hat{B} - \hat{B} \, \hat{A}$ denotes the matrix commutator.
The product of these gives the
cubic
interaction in Eq.~(\ref{eq:SIY}).

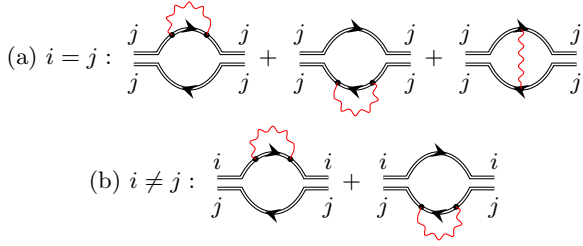
\begin{figure}[t!]
    \centering
    (a) $i=j:$
    \begin{tikzpicture}[scale=0.6,baseline=(y0.base)]
    \coordinate (y0) at (0,-3pt);
        \draw[double, mid arrow] (-35pt,3.5pt) node[above] {$j$}--(-20pt,3.5pt).. controls (-10pt,24pt) and (10pt,24pt) .. (20pt,3.5pt) coordinate[pos=0.25] (M1) coordinate[pos=0.75] (M2)
        --(35pt,3.5pt) node[above] {$j$} ;
        \draw[fill=black] (M1) circle (1.5pt);
        \draw[fill=black] (M2) circle (1.5pt);
        \draw[boson,red] (M1)..controls (-20pt,38pt) and (20pt,38pt)..(M2);
        \draw[double, mid arrow] (35pt,-3.5pt) node[below] {$j$}--(20pt,-3.5pt).. controls (10pt,-24pt) and (-10pt,-24pt) .. (-20pt,-3.5pt)--(-35pt,-3.5pt) node[below]{$j$};
    \end{tikzpicture}+
    \begin{tikzpicture}[scale=0.6,baseline=(y0.base)]
        \coordinate (y0) at (0,-3pt);
        \draw[double, mid arrow] (-35pt,3.5pt) node[above] {$j$}--(-20pt,3.5pt).. controls (-10pt,24pt) and (10pt,24pt) .. (20pt,3.5pt)
        --(35pt,3.5pt) node[above] {$j$};
        \draw[double, mid arrow] (35pt,-3.5pt) node[below] {$j$}--(20pt,-3.5pt).. controls (10pt,-24pt) and (-10pt,-24pt) .. (-20pt,-3.5pt) coordinate[pos=0.25] (M1) coordinate[pos=0.75] (M2) --(-35pt,-3.5pt) node[below] {$j$};
        \draw[fill=black] (M1) circle (1.5pt);
        \draw[fill=black] (M2) circle (1.5pt);
        \draw[boson,red] (M1)..controls (20pt,-38pt) and (-20pt,-38pt)..(M2);
    \end{tikzpicture}+
    \begin{tikzpicture}[baseline=(y0.base),scale=0.6]
        \coordinate (y0) at (0,-3pt);
        \draw[double, mid arrow] (-35pt,3.5pt) node[above] {$j$}--(-20pt,3.5pt).. controls (-10pt,24pt) and (10pt,24pt) .. (20pt,3.5pt) coordinate[pos=0.5](M1)
        --(35pt,3.5pt) node[above] {$j$};

        \draw[double, mid arrow] (35pt,-3.5pt) node[below] {$j$}--(20pt,-3.5pt).. controls (10pt,-24pt) and (-10pt,-24pt) .. (-20pt,-3.5pt)  coordinate[pos=0.5] (M2) --(-35pt,-3.5pt) node[below] {$j$};
        \draw[fill=black] (M1) circle (1.5pt);
        \draw[fill=black] (M2) circle (1.5pt);
        \draw[boson,red] (M1)--(M2);
    \end{tikzpicture} \\
    (b) $i\neq j:$
    \begin{tikzpicture}[scale=0.6,baseline=(y0.base)]
    \coordinate (y0) at (0,-3pt);
        \draw[double, mid arrow] (-35pt,3.5pt) node[above] {$i$}--(-20pt,3.5pt).. controls (-10pt,24pt) and (10pt,24pt) .. (20pt,3.5pt) coordinate[pos=0.25] (M1) coordinate[pos=0.75] (M2)
        --(35pt,3.5pt) node[above] {$i$} ;
        \draw[fill=black] (M1) circle (1.5pt);
        \draw[fill=black] (M2) circle (1.5pt);
        \draw[boson,red] (M1)..controls (-20pt,38pt) and (20pt,38pt)..(M2);
        \draw[double, mid arrow] (35pt,-3.5pt) node[below] {$j$}--(20pt,-3.5pt).. controls (10pt,-24pt) and (-10pt,-24pt) .. (-20pt,-3.5pt)--(-35pt,-3.5pt) node[below]{$j$};
    \end{tikzpicture}+
    \begin{tikzpicture}[scale=0.6,baseline=(y0.base)]
        \coordinate (y0) at (0,-3pt);
        \draw[double, mid arrow] (-35pt,3.5pt) node[above] {$i$}--(-20pt,3.5pt).. controls (-10pt,24pt) and (10pt,24pt) .. (20pt,3.5pt)
        --(35pt,3.5pt) node[above] {$i$};
        \draw[double, mid arrow] (35pt,-3.5pt) node[below] {$j$}--(20pt,-3.5pt).. controls (10pt,-24pt) and (-10pt,-24pt) .. (-20pt,-3.5pt) coordinate[pos=0.25] (M1) coordinate[pos=0.75] (M2) --(-35pt,-3.5pt) node[below] {$j$};
        \draw[fill=black] (M1) circle (1.5pt);
        \draw[fill=black] (M2) circle (1.5pt);
        \draw[boson,red] (M1)..controls (20pt,-38pt) and (-20pt,-38pt)..(M2);
    \end{tikzpicture}

    \caption{Diagrams from the fourth-order trace-log expansion responsible for the mass terms $\propto m^2$ that appear in the replica off-diagonal modes of the matrix field $Y_{j k}$   in Eq.~(\ref{eq:SIY}), due to interactions. The doubled lines with arrows indicate the measurement-averaged saddle-point fermion Green's function, while the red wavy lines are instantaneous Keldysh interaction propagators.
  These diagrams describe the dressing by interactions of the entanglement matrix-field fluctuation
  $\left\langle \hy(\omega,\vex{k}) \, \hy(-\omega,-\vex{k})\right\rangle$ [Eq.~(\ref{YDef})].
  The density-boson Keldysh propagator is generated by incorporating a constant term analogous to $V_{\q}^2$ in the action Eq.~(\ref{eq:SV}),
  which arises from the expansion and ensures the Ward identity [Eq.~(\ref{Ward1+1})]. Replica-diagonal fermion self-energies are \emph{canceled} by vertex corrections (a), while no such cancelation occurs for off-diagonal fluctuations (b).
  }
	\label{fig:masses}
\end{figure}

The other key element induced by the interactions are the mass terms in Eq.~(\ref{eq:SIY}),
which suppress the fluctuations of the off-diagonal modes $Y_{j \neq k}$. The masses arise through a combination of interaction-mediated self-energies and vertex corrections in the trace-log expansion, see Fig.~\ref{fig:masses}.
These are due to induced Keldysh fluctuations of the density, which produce self-energies similar to that of dephasing in weak localization theory \cite{AAK82,Liao17}.

The entanglement SU($R$) sector of the theory obtained in Eq.~(\ref{eq:SExpansion}) takes the form of Goldstone modes $Y_{i j}(t,x)$, gapless in the absence of interactions, made partially massive by turning on the interactions. 
As is standard in the derivation of non-linear sigma model field theories for disordered, interacting fermions in equilibrium \cite{BK94,Liao17}, we ``lift'' the theory to the full nonlinear target manifold for monitored, non-interacting fermions to obtain Eqs.~(\ref{eq:PCM}) and (\ref{eq:SI}). This promotion is completely constrained up to irrelevant operators and parameter renormalization by the $G \times G$ structure of the replicated, microscopic Keldysh action 
\emph{in every fixed quantum trajectory}
[with $G = \textrm{SU}(R)$ and $G = S_R$ for non-interacting and interacting fermions, respectively].

\subsection{U(1) response and Ward identity \label{sec:Ward}}

In the U(1) sector of the theory Eq.~(\ref{eq:U(1)}),
the field $\theta$ is the boson mode that transforms under the Keldysh contour-axial U(1), while $\phi$ is a dual partner mode.
The physics of the average density dynamics can be uncovered via the 
explicit
trace-log calculation
of the source-field action,
\begin{align}\label{eq:SV}
    S_V
    =
    \int dt dx \, \left(\frac{R}{2} V_\q^2 + n_\cl \, V_\q\right).
\end{align}
The same result can be obtained by simply gauging the contour-axial U(1) in Eq.~(\ref{eq:U(1)}),
\begin{align}
    \parr_t \theta \rightarrow \parr_t \theta + 2 R \, V_\q,
\end{align}
which sends $S_{\textrm{U(1)}} \rightarrow S_{\textrm{U(1)}} + S_V$.
In Eq.~(\ref{eq:SV}), $V_\q$ is the quantum component of the potential in the Keldysh formalism, which couples to the average classical (observable) density operator (measured relative to half-filling)
\begin{align}\label{eq:rhocl}
    n_\cl = \frac{1}{2}\left(\parr_t \theta - 2 \gamma \phi\right).
\end{align}

Although our microscopic action in Eq.~(\ref{SDef}) includes
coupling to both quantum and classical sources $V_{\q,\cl}$
[via $\tilde{a}_{\cl,\q} = a_{\cl,\q} + V_{\cl,\q}$],
the classical source disappears from the final 
U(1) sector action given by Eqs.~(\ref{eq:U(1)}) and (\ref{eq:SV}).
The trivialization of the steady-state distribution function induced
by measurements leads to a vanishing coupling to external electric
fields or quenched potential disorder, both of which would
appear in $V_\cl$. Retarded and advanced average-density response functions
vanish.
These results obtain because the measurements ``scramble'' the fermions
across the Bloch band, so that there is no well-defined notion of a carrier 
group velocity. At the same time, however, density still fluctuates stochastically in space and time,
with a diffusion constant determined by the measurement rate [Eq.~(\ref{Ward1+1}), below].
Away from half-filling, density fluctuations directly modulate the stiffness of the entanglement sector, see Appendix~\ref{app:ahf}.

The only non-vanishing average-density correlation function is the Keldysh one.
Using Eqs.~(\ref{eq:SV}) and (\ref{eq:rhocl}),
\begin{align}
    \!\!\!\!
    \!\!
    \ii \, \Pi^K(t,\vex{r})
    =&\,
    -
    \left.
    \frac{\delta^2 \bar{Z}}{\delta V_{\q}(t,\vex{r}) \, \delta V_{\q}(0,0)}
    \right|_{V = 0}
\nonumber\\
    =&\,
    R \, \delta(t) \, \delta^{(d)}(\vex{r})
    -
    \left\langle n_{\cl}(t,\vex{r}) \, n_{\cl}(0,0)\right\rangle.
\end{align}
In 1+1D, this gives
\begin{align}\label{Ward1+1}
    \ii \, \Pi^K(\omega,k)
    =&\,
    R
    \frac{
    v^2 k^2(4 \gamma^2 + v^2 k^2 + \omega^2)
    }{
    (v^2 k^2 + \omega^2)^2 + (2 \gamma \omega)^2
    }
\nonumber\\
    \simeq&\,
    R \, \gamma
    \frac{2 D \, k^2}{(D \, k^2)^2 + \omega^2},
    \quad
    D = \frac{v^2}{2 \gamma},
\end{align}
valid for $v |k|,|\omega| \ll \gamma$.
Thus the Keldysh response is diffusive,
and satisfies the Ward identity
$\Pi^K(\omega,k\rightarrow 0) = 0$
(i.e., the total charge does not fluctuate).
The diffusion constant is determined by the
effective measurement rate $\gamma$, which also
plays the role of an effective ``temperature''
for the noisy density fluctuations encoded by
$\ii \, \Pi^K(t,x)$. 
Here, the diffusion constant $D$ appears to only depend upon the measurement rate, but is independent of the interaction strength. This is an artifact of the perturbative expansion in the interaction strength. In Appendix~\ref{app:Brownian}, we demonstrate that the diffusion constant is renormalized by interactions 
after summing over higher order diagrams using Eliashberg-type equations.

\section{Physical implications of the field theory}
\label{sec:PhysCons}

In this section, we investigate the phase diagram of the 1+1D charge-conserving monitored dynamics of interacting fermions. We treat the unifying field theory Eq.~\eqref{eq:ZFull} as a Laudau-Ginzburg-type field theory, which allows us to discuss the global structure of the phase diagram, the universal characters of each phase, and the nature of the phase transitions based on both symmetry-centeric analysis and RG studies. In particular, we discuss how interactions enable new phases and transitions that do not exist in the non-interacting limit.

\subsection{Brief review of the field theory in the non-interacting limit}
\label{sec:ReviewNonInteracting}

Before we dive into the full theory for interacting monitored dynamics, it is helpful to first review the basic properties of the field theory [Eq.~\eqref{eq:Z}] that describes the monitored dynamics in the non-interacting limit \cite{Mirlin23,Mirlin2024_Above1D,Chahine23,FavaU1}. This field theory contains two sectors: $S_{\rU(1)}$ [Eq.~\eqref{eq:U(1)}] and $S_{\rSU (R)}$ [Eq.~\eqref{eq:PCM}]. The former is responsible for the dynamics of the average charge density, while the latter governs the dynamics of average observables that are non-linear in the system's density matrix, for example, the average entanglement entropy [Eq.~\eqref{eq:EE_SA}] and second-moment charge correlation [Eq.~\eqref{eq:charge_correl}]. As discussed in Sec.~\ref{sec:Kel_Description}, this field theory
in
Eq.~\eqref{eq:Z} enjoys a $\rU(R)_+ \times \rU(R)_-$ symmetry.

The $S_{\rU(1)}$ sector with the fields $\theta$ and $\phi$ controls the dynamics of the average charge density $n_\cl$ of the system. The field  $\theta$ shifts under the $\rU(1)$ subgroup of  $\rU(R)_\pm$ that generates a uniform phase rotation for each replica. For any finite measurement rate, the coupling $\gamma$ in Eq.~\eqref{eq:U(1)}
is finite and, consequently, the average charge density $n_\cl$ always follows diffusive dynamics, as discussed underneath Eq.~\eqref{eq:rhocl}.

The $S_{\rSU (R)}$ sector is a class-AIII PCM field theory that governs the dynamics of average observables that are non-linear functions of the system's density matrix. The $R\times R$ matrix field $\hat{X}$ takes value in the target space $\rSU(R)$, which is dictated
by the Altland-Zirnbauer (AZ) symmetry class AIII of the dynamics. The AZ symmetry class of the charge-conserving monitored dynamics of non-interacting fermions was first identified in Ref.~\cite{CMJian2022}. The target space $\rSU(R)$ was also identified using other field-theoretic methods in Refs.~\cite{Mirlin23,FavaU1}. Under the $\rSU(R)_\pm$ subgroup of $\rU(R)_\pm$, the field $\hat{X}$
transforms
as $\hat{X}\rightarrow W_+ \hat{X} W_-^\dag$ with $W_\pm \in \rSU(R)_\pm$. A remarkable feature of this 1+1D PCM field theory is that even though it is formulated in real time $t$, it follows an Euclidean signature (without Wick rotation). As is pointed out in Refs.~\cite{Mirlin23,FavaU1}, this PCM always flows under the renormalization group (RG) to strong coupling $\lambda^{-1}\rightarrow +\infty$ (including in the replica limit $R\rightarrow 1$). Physically, this RG analysis implies that any saddle point with a finite expectation value of $\hat{X}$ (that breaks the $\rSU(R)_\pm$ symmetry) is unstable against Goldstone-mode fluctuations. In other words, this PCM only exhibits one phase with $\langle \hat{X} \rangle = 0$. Therefore, the 1+1D monitored dynamics of non-interacting fermions can only exhibit a dynamical phase with area-law entanglement entropy scaling in the long-time limit. In the subsequent subsections, we will discuss how new dynamical phases and transitions arise when interactions are introduced.

\subsection{Phases of charge-conserving monitored dynamics of interacting fermions in 1+1D \label{sec:phasesintmodel}}

Now, we discuss the possible dynamical phases of charge-conserving interacting fermions using the full interacting field theory
in
Eq.~\eqref{eq:ZFull}. As we will see, the interaction allows new dynamical phases and transitions
in 1+1D that are absent in the non-interacting limit. We will present the field theory description of these new phases and study their transitions. We will also provide a schematic phase diagram parameterized by the measurement rate and interactions.

Just like the non-interacting limit, the $S_{\rU(1)}$ sector [Eq.~\eqref{eq:U(1)}] still controls the dynamics of the average charge density $n_{\cl}$. The symmetry action on the fields $\theta$ and $\phi$ remains the same. More importantly, the dynamics of $n_{\cl}$ is still diffusive.

The most important effects of interactions reside in the entanglement sector $S_{\rSU(R)}+S_I $ that governs observables that are nonlinear in the quantum-trajectory-resolved density matrix of the system. First of all, the $\rSU(R)_+ \times \rSU(R)_-$ symmetry of $S_{\rSU(R)}$ is reduced by $S_I$ to $\calG_+ \times \calG_-$. Here,  $\calG_\pm$ are the subgroups of $\rSU(R)_\pm$ generated by the $R\times R$ matrices $W_\pm$ of the form:
\begin{align} \label{eq:Gpm}
    (W_+)_{jk} = e^{\ii \eta_{+,j}} \delta_{\tau_+(j),k}, \nonumber \\
    (W_-)_{\bar{j}\bar{k}} = e^{\ii \eta_{-,\bar{j}}} \delta_{\tau_-(\bar{j}),\bar{k}},
\end{align}
where $\tau_\pm$ are elements of the permutation groups $S_{R,\pm}$ (on a set of $R$ objects), $\eta_{+,j}, \eta_{-,\bar{j}} \in \mathbb{R}$, and the matrices $W_\pm$ are subject to the constraint $\det W_\pm =1$. Let's describe the key elements of $\calG_\pm$ to illustrate their mathematical structures. Take $\calG_+$ as an example. When we set $\tau_+$ to be the trivial permutation, the factors $e^{\ii \eta_{+,j}}$ parameterize the diagonal (or Cartan) subgroup of $\rSU(R)_+$, which is isomorphic to $[\rU(1)_+]^{(R-1)}$, i.e. all the relative $\rU(1)$ phase rotation between the $R$ replicas of the $+$ time branch. These U(1) symmetries are the consequence of the independent charge conversation in each replica. The element $\tau_+\in S_{R,+}$ is responsible for the permutation amongst the $R$ replicas of the $+$ time branch. $\calG_-$ acting on the $-$ time branch shares the similar mathematical structure as $\calG_+$. For physical intuition,
we can identify $\calG_\pm$ as the semi-direct product
\begin{align}\label{eq:calG_pm}
    \calG_\pm \sim [\rU(1)_\pm]^{(R-1)} \rtimes S_{R,\pm}.
\end{align}
There are mathematical subtleties with this identification due to the constraint $\det W_\pm =1$. The more rigorous statement is that $\calG_\pm = \{[\rU(1)_\pm]^{R} \rtimes S_{R,\pm}\}/\rU(1)_\pm$, where $[\rU(1)_\pm]^{R}$ is the replica-resolved $\rU(1)$ symmetry for the $R$ replicas on the $\pm$ time branch and the $\rU(1)_\pm$ that is quotiented out
is the global $\rU(1)$ subgroup of $\rU(R)_\pm$. The semi-direct product captures the fact that the U(1) 
conserved
charges in different replicas are permutated by $S_{R,\pm}$. The difference between Eq.~\eqref{eq:calG_pm} and the more rigorous version does not affect the subsequent analysis of the phases and transitions. Therefore, for notational simplicity, we will use this Eq.~\eqref{eq:calG_pm} when the mathematical subtleties are unimportant. We emphasize that the semi-direct product structure in $\calG_\pm$ is crucial. As we elaborate later, one of its fundamental consequences is a constraint between the entanglement scaling and the charge fluctuations, which dictates the order of the entanglement phase transition and the charge-sharpening transition.

Under the reduced symmetry group, the matrix field $\hat{X}$ still transforms as
\begin{align}
\hat{X}\rightarrow W_+ \hat{X} W_-^\dag,
\end{align}
where $W_\pm \in \calG_\pm$ are parameterized by Eq.~\eqref{eq:Gpm}.

With the symmetry $\calG_+ \times \calG_-$ of the interacting theory
$S_{\rSU(R)}+S_I $ clarified, we are ready to discuss the phases of 1+1D fermions undergoing interacting monitored dynamics.

\subsubsection{Charge-fuzzy volume-law phase} \label{sec:CFVLphase}
With the symmetry $\calG_+ \times \calG_-$ of the interacting entanglement sector $S_{\rSU(R)}+S_I$, a new phase with $\langle \hat{X} \rangle \neq 0$ becomes stable. In the following, we will discuss why this phase is stable and, more importantly, that this phase exhibits a volume-law scaling of average entanglement entropy and ``charge fuzziness" in the quantum-trajectory-resolved charge fluctuation.

The $\sum_{j,\bar{k} = 1}^R\left|X_{j \bar{k}}\right|^4$ term in $S_I$ [Eq.~\eqref{eq:SI}] introduces anisotropy to the PCM's target manifold $\rSU(R)$ and favors the saddle points with
\begin{align}
  \langle X_{j\bar{k}} \rangle \propto e^{\ii \zeta_j} \delta_{\sigma(j),\bar{k}},
    \label{eq:EEVol_CFuzzy_Saddle}
\end{align}
where $\sigma \in S_R$ is a permutation group element and $e^{\ii \zeta_j} $ is a $\rU(1)$ phase for each $j=1,2,...,R$. The proportionality constant in Eq.~\eqref{eq:EEVol_CFuzzy_Saddle} is {\it independent } of the matrix indices $j$ and $\bar{k}$.
All saddle points above are equivalent since they are related to each other by the symmetry $\calG_+\times \calG_-$ (up to an unimportant global multiplicative constant).  Note that the first term in $S_I$ [Eq.~\eqref{eq:SI}] does not directly affect the saddle point at the mean-field level because of its dependence on $\partial_\mu X$. However, it renormalizes the strength of the $\sum_{j,\bar{k} = 1}^R\left|X_{j \bar{k}}\right|^4$ term under RG as shown in Eq. \eqref{eq:betaM}.

Using the following saddle point
\begin{align}
    \langle X_{j\bar{k}} \rangle  \propto \delta_{j\bar{k}} ,
    \label{eq:EEVol_CFuzzy_Saddle_Id}
\end{align}
as a representative, it is straightforward to see that the phase represented by
Eq.~\eqref{eq:EEVol_CFuzzy_Saddle} spontaneously breaks the $\calG_+ \times \calG_-$ down to the diagonal subgroup, denoted as $\calG_{\rm diag} \sim [\rU(1)_{\rm diag}]^{(R-1)} \rtimes S_{R,{\rm diag}} $. The order parameter for this spontaneous symmetry breaking is obviously $\langle \hat{X} \rangle$. 

It is also helpful to introduce a ``secondary order parameter" that only probes the spontaneous breaking of
$S_{R,+}\times S_{R,-}$ to $S_{R, {\rm diag}}$, which is responsible for the volume-law entanglement entropy scaling, as we show below. Notice that $V_{j\bar{k}} \equiv |X_{j\bar{k}}|^2$ transform non-trivially under $S_{R,+}\times S_{R,-}$ but trivially under $[\rU(1)_+]^{(R-1)} \times [\rU(1)_-]^{(R-1)} $.
The saddle point Eq.~\eqref{eq:EEVol_CFuzzy_Saddle} implies
\begin{align} \label{eq:EE_OrderParameter_V}
  \langle V_{j\bar{k}} \rangle = \langle |X_{j\bar{k}}|^2 \rangle = B \, \delta_{\sigma(j),\bar{k}} + (1-B)/R,
\end{align}
with $B>0$. This expression satisfies $\sum_j\langle |X_{j\bar{k}}|^2 \rangle = \sum_{\bar{k}}\langle |X_{j\bar{k}}|^2 \rangle = 1$, which are constraints due to $\hat{X}^\dag\hat{X} = \hat{1}_R$. Notice that, in Eq.~\eqref{eq:EE_OrderParameter_V}, the first term non-trivially depends on the matrix indices $j$ and $\bar{k}$, while the second term does not. In the following, we will use $\langle V_{j\bar{k}} \rangle_\mathsf{p}$ to denote the part of $\langle V_{j\bar{k}} \rangle$ that non-trivially depends on the matrix indices. The spontaneous breaking of $S_{R,+}\times S_{R,-}$ to $S_{R, {\rm diag}}$ is reflected by the order parameter
\begin{align} \label{eq:Vp_Def}
     \langle V_{j\bar{k}} \rangle_\mathsf{p} \propto \delta_{\sigma(j),\bar{k}},
\end{align}
for $\sigma \in S_R$. Even though $\langle V_{j\bar{k}} \rangle_\mathsf{p}$ is treated as a secondary order parameter in the charge-fuzzy volume-law phase represented by the saddle point Eq.~\eqref{eq:EEVol_CFuzzy_Saddle}, it can, in principle, be non-trivial, signaling the
volume-law entanglement scaling, 
even when $\langle \hat{X} \rangle = 0$ due to $[\rU(1)]^{R-1}$ fluctuations. The discussion of such a scenario will be reserved for Sec.~\ref{sec:CSVLphase}.

Now, we investigate the stability of the phase Eq.~\eqref{eq:EEVol_CFuzzy_Saddle_Id} against fluctuations. As discussed in Sec. \ref{sec:Results}, the fluctuation can be parameterized by $\hat{X} = \exp{\left(\ii \hat{Y}/\sqrt{\lambda}\right)}$ using a traceless Hermitian matrix $\hat{Y}$. The action that governs the fluctuation of $\hat{Y}$ is given by the combination of Eqs.~\eqref{eq:Y_2nd} and \eqref{eq:SIY}, which are obtained from expanding the action $S_{\rSU(R)}+S_I$. Notice that the mass term in Eq.~\eqref{eq:SIY} suppresses the fluctuation of off-diagonal modes in $\hat{Y}$, i.e. $Y_{jk}$ with $j\neq k$. Hence, the off-diagonal modes in $\hat{Y}$ do not affect the stability of the saddle point Eq.~\eqref{eq:EEVol_CFuzzy_Saddle_Id}, which is responsible for the volume-law entanglement scaling, as we show below. The $S_{R,+} \times S_{R,-}$ subgroup of $\calG_+ \times \calG_-$ remains broken (to its diagonal $S_{R,{\rm diag}}$ subgroup).

The more important fluctuations are the diagonal components of $\hat{Y}$, which are not suppressed by any mass term. Hence, we only need to focus on the fluctuation of the form 
\begin{align}
    \hat{X}= {\rm diag}(
e^{\ii Y_{11}/\sqrt{\lambda}},e^{\ii Y_{22}/\sqrt{\lambda}},...,e^{\ii Y_{RR}/\sqrt{\lambda}})
\end{align}
with the constraint $\sum_j Y_{jj} = 0$ and the compactification $Y_{jj}/\sqrt{\lambda} \sim Y_{jj}/\sqrt{\lambda} + 2\pi \mathbb{Z}$. Such massless fluctuations of $\hat{X}$ are equivalent to $R-1$ compact boson fields in 1+1D. 

The saddle point Eq.~\eqref{eq:EEVol_CFuzzy_Saddle_Id} breaks the $[\rU(1)_+]^{(R-1)} \times [\rU(1)_-]^{(R-1)} $ subgroup of $\calG_+ \times \calG_-$ down to its diagonal subgroup $[\rU(1)_{\rm diag}]^{(R-1)}$. In 1+1D, under the fluctuation of the $R-1$ compact bosons, namely the modes
$Y_{jj}$,
the breaking of $[\rU(1)_+]^{(R-1)} \times [\rU(1)_-]^{(R-1)} $ around the saddle point Eq.~\eqref{eq:EEVol_CFuzzy_Saddle_Id} should be understood as a {\it quasi-long-range order}. At this point, we can conclude the phase represented by the saddle point Eq.~\eqref{eq:EEVol_CFuzzy_Saddle} is stable under fluctuations. This phase does not exist in the non-interacting limit. As we will see below, the spontaneous breaking of the $S_{R,+} \times S_{R,-}$ subgroup in this phase implies the volume-law entanglement scaling, while the quasi-long-range order associated with $[\rU(1)_+]^{(R-1)} \times [\rU(1)_-]^{(R-1)} $ implies the fuzziness of quantum-trajectory-resolved global charge fluctuation.
\footnote{The notion of charge-fuzzy phase is first introduced in
Ref.~\cite{Agrawal22} in a different setting with dynamical qubit systems.
}

\begin{figure*}
    \centering
    \includegraphics[width=1\linewidth]{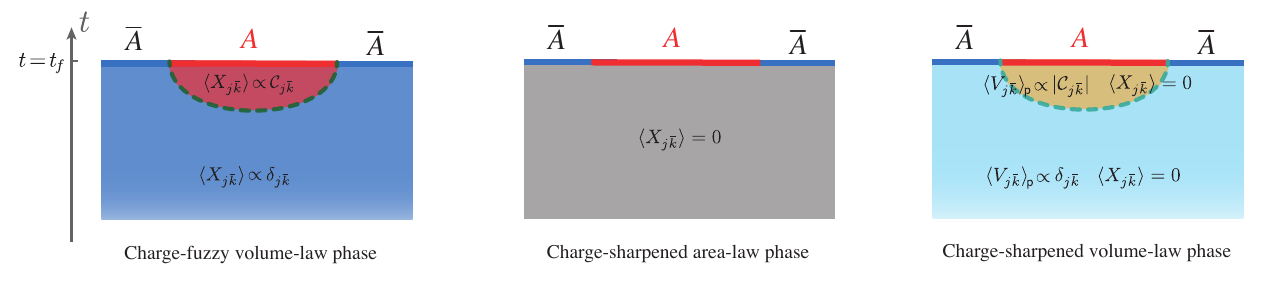}
    \caption{Domain wall configurations of $\langle\hat{X}\rangle$ and $\langle\hat{V}\rangle_{\mathsf{p}}$ that dictates the scaling of the average entanglement entropy $\bar{S}_A$ at a late time $t=t_f$ in the three different phases. In the charge-fuzzy volume-law phase, there is a domain wall (depicted as the dashed line) separating the two domains with $\langle X_{jk}\rangle \propto \delta_{j\bar{k}}$ and $\langle X_{jk}\rangle \propto {\cal C}_{j\bar{k}}$. The domain terminates at the $\partial A$ at $t=t_f$. In the charge-sharpened volume-law phase, there is also a domain wall separating the two domains with $\langle V_{jk}\rangle_\mathsf{p} \propto \delta_{j\bar{k}}$ and $\langle V_{jk}\rangle_\mathsf{p} \propto |{\cal C}_{j\bar{k}}|$ (see the discussion around Eq.~\eqref{eq:Vp_Def} for the definition of $\langle \hat{V} \rangle_\mathsf{p}$). There is no domain wall in the charge-sharpened area-law phase.
    }
    \label{fig:DomainWall}
\end{figure*}

The volume-law entanglement scaling in the phase represented by the saddles
in
Eq.~\eqref{eq:EEVol_CFuzzy_Saddle} can be understood as follows. Consider the average entanglement entropy $\bar{S}_A$ of a subsystem interval $A$ [see Eq.~\eqref{eq:EE_SA}] at time $t = t_f$ (assuming the $t=t_f$ is already in the long-time limit when the entanglement scaling has saturated). The way to calculate $\bar{S}_A$ using the interacting field theory is to study the sector $S_{\rSU(R)}+S_I $ on a half-infinite spacetime with a boundary at $t=t_f$. $\bar{S}_A$ can be generally mapped to domain-wall free energy, which can be expressed as the $\log$ of the ratio between the two field-theory partition functions, i.e. $\bar{S}_A \propto - \log (Z_{\rm dw}/Z_{0})$, with different boundary conditions at $t=t_f$ (in the replica limit $R\rightarrow 1$)\cite{Jian20,Bao20,MFava2023}. Here, $Z_{0}$ is the partition function under the boundary condition $X_{j\bar{k}} (t=t_f, x) \propto \delta_{j\bar{k}}$. $Z_{0}$ essentially calculates the normalization factor for the $R$-replica average density matrix $\rho^{(R)}$ [Eq.~\eqref{eq:path_integral_rhoR}]. The partition function $Z_{\rm dw}$ is obtained under the boundary condition \cite{MFava2023,Jian20,Bao20}:
\begin{align} \label{eq:DomainWall_bc}
   X_{j\bar{k}}( t= t_f, x) \propto \begin{cases}
        \delta_{j\bar{k}}, & x \in \bar{A}, \\
        {\cal C}_{j\bar{k}}, & x \in A,
    \end{cases}
\end{align}
where
\begin{align}
    {\cal C}   = \left( \begin{array}{ccccccc}
        0 & 0 & 0 & &  0 & 1\\
        -1 & 0 & 0 & \cdots & 0 & 0 \\
        0 & -1 & 0 &  & 0 & 0\\
         & \vdots &  &  \ddots  \\
        0 & 0 & 0 &  & 0 & 0 \\
        0 & 0 & 0 &  & -1 & 0
    \end{array}
    \right).
\end{align}
${\cal C} $ relates the different replicas in a cyclic manner. The $-1$ entries ensure that $\det {\cal C} = 1$. In the phase represented by the saddle points in
Eq.~\eqref{eq:EEVol_CFuzzy_Saddle}, the boundary condition Eq.~\eqref{eq:DomainWall_bc} forces the configuration of $\langle \hat{X} \rangle$ in the 1+1D spacetime to have two domains:  $\langle X_{j\bar{k}} \rangle \propto \delta_{j\bar{k}}$ and $\langle X_{j\bar{k}} \rangle \propto {\cal C}_{j\bar{k}}$ (see Fig.~\ref{fig:DomainWall} left panel). Note that both domains belong to the class of saddle points given in Eq.~\eqref{eq:EEVol_CFuzzy_Saddle}. However, they belong to different connected components of the saddle-point manifold (due to the broken discrete $S_{R,+}\times S_{R,-}$ symmetry), which implies there must be a domain wall separating them. The domain wall free energy $- \log (Z_{\rm dw}/Z_{0})$ is expected to be proportional to the size of the domain wall, which should scale as the size $|A|$ of the interval $A$. This expectation should hold for any $R$, including the limit $R\rightarrow 1$. Therefore, the phase represented by the saddle points Eq.~\eqref{eq:EEVol_CFuzzy_Saddle} with finite $\langle \hat{X}\rangle$ has a volume-law entanglement entropy scaling, i.e., $\bar{S}_A \sim |A|$. Such a stable volume-law phase has been observed in numerical simulations of 1D monitored interacting monitored chains \cite{YoheiAshida2020,Lumia2024} with finite interactions. As will be discussed in Sec.~\ref{sec:PhaseandFlow}, the renormalization group analysis based on the field theory Eq.~\eqref{eq:ZFull} suggests that the volume-law entanglement scaling can be stable even in the weakly interacting regime.  

In principle, one can estimate $\bar{S}_A$ by estimating the domain-wall free energy $- \log (Z_{\rm dw}/Z_{0})$ as follows. We begin with the equation of motion of $\hat{X}$ subject to the boundary condition
Eq.~\eqref{eq:DomainWall_bc}. We make a further approximation by considering only the action $S_{\rSU(R)}$ [Eq.~\eqref{eq:PCM}] and the anisotropy term $\sum_{j,\bar{k} = 1}^R\left|X_{j \bar{k}}\right|^4$ in $S_I$
[Eq.~\eqref{eq:SI}], whose variation over $\hat{X}$ leads to the following equation of motion
\begin{align}\label{eq:dw_EOM}
    -\frac{\lambda}{8} (\partial_t^2 +v^2 \partial_x^2)X_{j\bar{k}} + \nu X_{j\bar{k}} +  \frac{-\lambda m^2}{8} |X_{j\bar{k}}|^2  X_{j\bar{k}} = 0,
\end{align}
where $\nu(t,x)$ is the Lagrange multiplier that enforces the constraint $\hat{X}^\dag \hat{X} = \hat{1}_R$. The action associated with the solutions of Eq.~\eqref{eq:dw_EOM} (when exponentiated) provides an estimate of $Z_{\rm dw}$ and $Z_0$ when we use the corresponding boundary condition at $t=t_f$. Admittedly, it is challenging to take the $R\rightarrow 1$ in this estimate. At least, the estimate of $-\log (Z_{\rm dw}/Z_{0})$ at finite $R$ can be treated as a proxy to $\bar{S}_A$. As we argue above, $- \log (Z_{\rm dw}/Z_{0})$ at any integer $R>1$ is expected to scale as $\sim |A|$, implying the volume-law scaling of average entanglement entropy, i.e. $\bar{S}_A \sim |A|$. The proportionality constant is identified with the domain-wall tension in the limit $R\rightarrow 1$. It is a non-universal quantity that depends on the microscopic details.

Next, we discuss the charge-fuzziness in this phase. The physical picture of the charge-fuzzy phase (first introduced in Ref.~\cite{Agrawal22} in a monitored qudit system) is as follows. When the system's initial state is a coherent superposition of states with different eigenvalues under the total charge operator $\hat{Q}_{\rm tot} = \int dx\,n(x)$, the charge-fuzzy dynamics is such that
the coherence between different charge sectors does not vanish and $\hat{Q}_{\rm tot}$ can still fluctuate within each quantum trajectory even in the long-time limit, despite the repeated measurement of the local charge density in the dynamics.  In other words, the information of the global charge $\hat{Q}_{\rm tot}$ remains scrambled under the charge-fuzzy dynamics in the sense that it cannot be reconstructed from the outcomes of the local charge density measurements occurring in the dynamics.

An observable that is sensitive to the charge fuzziness is given by
\begin{align} \label{eq:FQ}
    F_Q (\eta) \equiv \mathbb{E} \left( p_{\{v(x, t)\}} \langle e^{\ii \eta \hat{Q}_{\rm tot}}\rangle_{\! \upsilon} \langle e^{-\ii \eta \hat{Q}_{\rm tot}}\rangle_{\! \upsilon} \right).
\end{align}
Recall that $\mathbb{E}$ represents the average over quantum trajectories $\upsilon(t,x)$, $\langle \cdot \rangle_{\! \upsilon}$ represents the quantum operator expectation value in a given quantum trajectory (to be distinguished from the average value of a field in the field theory), and $p_{\{\upsilon(x, t)\}}$ is the Born-rule probability of the quantum trajectory $\upsilon(x, t)$. When the system is undergoing charge-fuzzy dynamics, the quantum state in each quantum trajectory is not an eigenstate of $\hat{Q}_{\rm tot}$ (assuming the initial state is in a coherent superposition of states with different $\hat{Q}_{\rm tot}$ eigenvalues). Therefore, $F_Q (\eta)$ should deviate from 1 and depend on $\eta$ in the charge-fuzzy phase.

It is helpful to contrast the behavior of $F_Q (\eta)$ in the charge-fuzzy phase to that in the ``opposite" phase, known as the charge-sharpened phase. When the dynamics is charge-sharpened, the local charge density measurements have a high enough rate so that the information on $\hat{Q}_{\rm tot}$ can be reconstructed. In other words, regardless of the initial state, the quantum state in each quantum trajectory becomes an eigenstate of $\hat{Q}_{\rm tot}$ (i.e., a state without any fluctuations of $\hat{Q}_{\rm tot}$) in the long-time limit. Hence, we expect $F_Q (\eta) = 1 $ in the long-time limit for any value $\eta$ in the charge-sharpened phase.

The behavior of $F_Q (\eta)$ in the charge-fuzzy and charge-sharpened phases can be understood from a symmetry perspective. The axial subgroup of the $[\rU(1)_+]^{(R-1)} \times [\rU(1)_-]^{(R-1)} $ symmetry action in the replica field theory can generate a shift $\eta \rightarrow \eta + \delta \eta$ of the argument of $F_Q (\eta)$ (while the diagonal subgroup keeps $\eta$ invariant).  This statement is based on the interpretation that the two factors $\langle e^{\ii \eta \hat{Q}_{\rm tot}}\rangle_{\! \upsilon}$ and $ \langle e^{-\ii \eta \hat{Q}_{\rm tot}}\rangle_{\! \upsilon}$ in $F_Q (\eta)$ can be viewed as coming from two different replicas of the dynamical system. In the charge-fuzzy phase, the non-trivial dependence of $F_Q (\eta)$ on $\eta$ signifies the spontaneous breaking of symmetry $[\rU(1)_+]^{(R-1)} \times [\rU(1)_-]^{(R-1)} $ to the diagonal subgroup $[\rU(1)_{\rm diag}]^{(R-1)}$ at the saddle point
Eq.~\eqref{eq:EEVol_CFuzzy_Saddle}. In contrast, the charge-sharpened phase respects the full $[\rU(1)_+]^{(R-1)} \times [\rU(1)_-]^{(R-1)} $ symmetry. Consequently, $F_Q (\eta)$ should be independent of $\eta$. At this point, we can conclude that the saddle point Eq.~\eqref{eq:EEVol_CFuzzy_Saddle} represents a phase with charge-fuzzy dynamics.

Having established that the saddle point Eq.~\eqref{eq:EEVol_CFuzzy_Saddle} corresponds to a dynamical phase with volume-law entanglement scaling and charge-fuzzy dynamics, it is interesting to discuss another characteristic feature of this phase coming from the massless fluctuation around the saddle point. As discussed above, the massless fluctuations around the saddle Eq.~\eqref{eq:EEVol_CFuzzy_Saddle_Id} are the diagonal elements $e^{\frac{\ii}{\sqrt{\lambda}}Y_{jj}}$ of $\hat{X}$.
They are equivalent to $R-1$ compact bosons in 1+1D in the quasi-long-range order phase. The compact boson fields $Y_{jj}$ are dual to the fields $\tilde{Y}_{jj}$ which are related to the charge density in the $j$th replica via $n_j \sim \partial_x \tilde{Y}_{jj}$.  Consequently, the volume-law charge-fuzzy phase exhibits the following scaling behavior in the average second-momentum density-density correlator in the long-time limit:
\begin{align} \label{eq:sec_monent_scaling}
    \mathbb{E}\Big[ p_{\{v(x, t)\}}  \Big(\langle n(x_1) \rangle_{\! \upsilon} & \langle n(x_2) \rangle_{\! \upsilon} -   \langle n(x_1) n(x_2) \rangle_{\! \upsilon}\Big) \Big]
\nonumber\\
    \propto&\, \langle \partial_x \tilde{Y}_{jj}(x_1) \partial_x \tilde{Y}_{jj}(x_2) \rangle
\nonumber\\
    \sim&\,
    |x_1-x_2|^{-2}.
\end{align}
Here, all the correlation functions (and quantum expectation values) are calculated at equal time, and the time variables $t$ are suppressed.

The scaling behavior Eq.~\eqref{eq:sec_monent_scaling} is consistent with the signature of the charge-fuzzy phase found in the previous work Ref.~\cite{Barratt_U1_FT} in a monitored qudit system. In fact, Ref.~\cite{Barratt_U1_FT} obtained a very similar field theory description of the 1+1D charge-fuzzy phase in terms of $R-1$ compact bosons. This field theory was obtained with the volume-law entanglement scaling as an assumption that draws intuition from the large qudit dimension limit. In our setting of charge-conserving monitored dynamics of interacting fermions, we see that both volume-law entanglement scaling and the charge fuzziness are immediate consequences of the saddle point Eq.~\eqref{eq:EEVol_CFuzzy_Saddle} (and its associated fluctuations). Later, we will see that our interacting field theory can provide a systematic view of the relation between the entanglement scaling and the charge fuzziness in different phases and at different transitions.

\subsubsection{Charge-sharpened area-law phase}

Having discussed the volume-law charge-fuzzy phase represented by the saddle point
Eq.~\eqref{eq:EEVol_CFuzzy_Saddle}, we switch our attention to the phase where the field $\hat{X}$ is fully ``disordered", i.e. $\langle \hat{X} \rangle = 0$. This phase fully preserves the entire $\calG_+ \times \calG_-$ symmetry of the interacting field theory. It smoothly connects to the only phase that exists in the non-interacting limit. The non-interacting limit of this phase has already been studied in
Refs.~\cite{Mirlin23,FavaU1}. For completeness, we briefly discuss the area-law entanglement scaling and the sharpness of the global charge fluctuation in this disordered phase in the following.

Let's first discuss the area-law entanglement scaling. As discussed in
Sec.~\ref{sec:CFVLphase}, the calculation of the average entanglement entropy $\bar{S}_A$ in the long-time limit amounts to calculating the domain wall free energy $- \log (Z_{\rm dw}/Z_{0})\big|_{R\rightarrow 1}$. For $Z_{\rm dw}$, even though the field theory still follows the boundary condition Eq.~\eqref{eq:DomainWall_bc} at the final time $t_f$, the expectation value $\langle \hat{X} \rangle$ rapidly decays to 0 as we move away from the boundary (see Fig.~\ref{fig:DomainWall} middle panel). Hence, there is no actual domain wall that extends into the bulk of the 1+1D spacetime. Consequently, $- \log (Z_{\rm dw}/Z_{0})\big|_{R\rightarrow 1}$ is expected to scale as $\sim |\partial A|$, namely the size of the boundary of the interval $A$. In other words, the average entanglement entropy follows an area-law scaling, i.e. $\bar{S}_A \sim |\partial A|$.

Regarding the quantum-trajectory-resolved global charge fluctuation, the unbroken $[\rU(1)_+]^{(R-1)} \times [\rU(1)_-]^{(R-1)} $ symmetry ensures that the quantity $F_Q(\eta)$ [Eq.~\eqref{eq:FQ}] is independent of $\eta$ (whose value shifts
under this symmetry). Hence, we can conclude that the dynamical phase with $\hat{X}$ full disordered is charge-sharpened.

\subsubsection{Charge-sharpened volume-law phase} \label{sec:CSVLphase}

The interacting field theory also admits a third possible phase with volume-law entanglement scaling and charge-sharpened dynamics. Based on the previous discussions in Sec. \ref{sec:CFVLphase}, we know that the volume-law entanglement scaling arises from the spontaneous breaking of the $S_{R,+} \times S_{R,-}$ symmetry to the diagonal subgroup $S_{R,{\rm diag}}$. A charge-sharpened phase requires the full $[\rU(1)_+]^{(R-1)} \times [\rU(1)_-]^{(R-1)} $ symmetry to remain unbroken. These symmetry patterns exactly match the saddle point with
\begin{align} \label{eq:EEVol_CSharp_Saddle}
    \langle X_{j\bar{k}} \rangle = 0,~~~ \langle V_{j\bar{k}} \rangle_\mathsf{p} \propto \delta_{\sigma(j),\bar{k}},
\end{align}
where $\sigma \in S_R$ is an element of the permutation group $S_R$. Recall that $\langle V_{j\bar{k}} \rangle_\mathsf{p}$ is the part of $\langle V_{j\bar{k}} \rangle \equiv \langle |X_{j\bar{k}}|^2 \rangle$ that depends non-trivially on the matrix indices $j$ and $\bar{k}$ (as introduced in Sec. \ref{sec:CFVLphase}). $\langle V_{j\bar{k}} \rangle_\mathsf{p}$ is the order parameter that detects the sponatanous breaking of  $S_{R,+} \times S_{R,-}$ and it transforms trivially under $[\rU(1)_+]^{(R-1)} \times [\rU(1)_-]^{(R-1)} $. This saddle point Eq.~\eqref{eq:EEVol_CSharp_Saddle} does not exist in the non-interacting limit.

When we consider the average entanglement entropy $\bar{S}_A$, related to $ - \log (Z_{\rm dw}/Z_{0})\big|_{R\rightarrow 1} $, in this phase, $Z_{\rm dw}$ is still governed by the domain wall-configuration separating two spacetime domains with $\langle V_{j\bar{k}} \rangle_\mathsf{p} \propto \delta_{j,\bar{k}}$ and $\langle V_{j\bar{k}} \rangle_\mathsf{p} \propto |{\cal C}_{j,\bar{k}}|$ (see the right panel of
Fig.~\ref{fig:DomainWall}). Hence, $\bar{S}_A$ should follow the same scaling as the domain-wall free energy and, consequently, exbibit a volume law, i.e.\ $\bar{S}_A \sim |A|$.

Regarding the quantum-trajectory-resolved global charge fluctuation, the saddle point Eq.~\eqref{eq:EEVol_CSharp_Saddle} fully respects
the $[\rU(1)_+]^{(R-1)} \times [\rU(1)_-]^{(R-1)} $ symmetry (which acts trivially on $\langle \hat{V} \rangle_\mathsf{p}$). Hence, the saddle Eq.~\eqref{eq:EEVol_CSharp_Saddle} represents a charge-sharpened phase.

\subsubsection{Incompatibility between area-law entanglement scaling and charge-fuzzy dynamics} \label{sec:Incompatibility}
At this point, one may ask if there can be a charge-fuzzy area-law phase in charge-conserving monitored dynamics of interacting fermions. In the following, we will argue that such a phase is likely impossible from the field theory perspective.

A hypothetical charge-fuzzy area-law phase would require the spontaneous breaking of the $[\rU(1)_+]^{(R-1)} \times [\rU(1)_-]^{(R-1)} $ symmetry to its diagonal subgroup $[\rU(1)_{\rm diag}]^{(R-1)}$ while preserving the replica permutation symmetry $S_{R,+} \times S_{R,-}$. However, using the matrix field $\hat{X}$ as the fundamental degrees of freedom in the field theory, one cannot write down any composite field that transforms non-trivially under $[\rU(1)_+]^{(R-1)} \times [\rU(1)_-]^{(R-1)} $ but behaves as a singlet under $S_{R,+} \times S_{R,-}$. Hence, a charge-fuzzy area-law phase, if it exists, can not be represented by any simple saddle point with non-trivial expectation values of fields built from $\hat{X}$.  Moreover, the replica permutation symmetry $S_{R,\pm}$ permutes the different copies of the $\rU(1)$ subgroups of $[\rU(1)_\pm]^{(R-1)}$. One cannot identify an unambiguous diagonal subgroup $[\rU(1)_{\rm diag}]^{(R-1)}$ of $[\rU(1)_+]^{(R-1)}\times  [\rU(1)_-]^{(R-1)} $ that is compatible with unbroken $S_{R,+} \times S_{R, -}$ symmetry.  This is fundamentally due to the semi-direct product structure of the two components $[\rU(1)_\pm]^{(R-1)}$ and $S_{R,\pm}$ in $\mathcal{G}_\pm$. Therefore, the symmetry-breaking pattern required by the hypothetical charge-fuzzy area-law phase might not be possible. In other words, a charge-fuzzy area-law phase might not exist. 

The absence of a charge-fuzzy area-law phase leads to the immediate consequence that a charge-sharpening transition between a charge-fuzzy phase and a charge-sharpened phase should occur {\it only} in the presence of volume-law entanglement scaling. Previously, Ref. \cite{Barratt_U1_FT} formulated the field theory of the charge-sharpening transition of a monitored qudit system under the assumption that the entanglement entropy follows a volume law. This assumption was based on theoretical reasoning obtained from systems of large qudit dimensions and numerical studies in Ref. \cite{Agrawal22}.
Here, we provide a natural explanation of the volume-law entanglement scaling at the charge-sharpening transition, which is fully based on the
symmetry analysis of our unifying field theory.

\subsection{Phase diagram and phase transitions of 1+1D charge-conserving interacting monitored dynamics \label{sec:PhaseandFlow}}

Now, we discuss the schematic phase diagram (see Fig.~\ref{fig:PhaseDiagram_AIII}) and the possible phase transitions of the 1+1D charge-conserving monitored dynamics of interacting fermions.

In the schematic phase diagram Fig.~\ref{fig:PhaseDiagram_AIII}, the horizontal axis is the strength of the local charge density measurement, which is proportional to $\lambda^{-1}$ in the replica field theory [Eq.~\eqref{eq:ZFull}]. The vertical axis characterizes the overall strength of interaction, which includes the effects of both interaction terms in $S_I$ [Eq.~\eqref{eq:SI}]. The origin of the phase diagram (marked by the orange dot in Fig.~\ref{fig:PhaseDiagram_AIII}) corresponds to the limit with vanishing interaction and infinitesimal measurement rate.

At the origin of the phase diagram, the monitored dynamics of the system is governed by the class-AIII PCM model Eq.~\eqref{eq:PCM} in the limit $\lambda^{-1} \rightarrow 0$. Recall that, without interactions, the coupling constant $\lambda^{-1}$, which is proportional to the measurement rate, always runs to strong coupling under the RG
\cite{Mirlin23,FavaU1}. Therefore, on the entire horizontal axis, the (non-interacting) charge-conserving monitored dynamics is in the charge-sharpened area-law phase with $\langle \hat{X} \rangle = \langle \hat{V} \rangle_\mathsf{p} =0$.

At a finite interaction strength, as discussed above, there are generically three phases present: charge-fuzzy volume-law phase,  charge-sharpened volume-law phase, and charge-sharpened area-law phase. Consider a generic horizontal cut of the phase diagram parameterized by the measurement rate at a finite interaction strength. When the measurement rate is low, the system's dynamics is in the charge-fuzzy volume-law phase represented by the saddle point with a finite $\langle \hat{X}\rangle$
[Eq.~\eqref{eq:EEVol_CFuzzy_Saddle}]. In this phase, the $\calG_+ \times \calG_-$ symmetry of the interacting field theory is spontaneously broken to its diagonal subgroup $\calG_{\rm diag}$.

As the measurement rate increases, the massless fluctuation of the diagonal modes $X_{jj} = e^{\frac{\ii}{\sqrt{\lambda}} Y_{jj}}$, $j=1,...,R$ can first restore the $[\rU(1)_+]^{(R-1)}\times [\rU(1)_-]^{(R-1)} $ symmetry of the system (while leaving the $S_{R,+} \times S_{R,-}$ subgroup still broken). Consequently, we enter the charge-sharpened volume-law phase captured by the saddle point $\langle \hat{X} \rangle = 0$ and $\langle \hat{V} \rangle_\mathsf{p} \neq 0$ (see Sec.~\ref{sec:CSVLphase}). This charge-sharpening transition between the charge-fuzzy and charge-sharpened is a Kosterlitz-Thouless transition associated with the compact bosons $Y_{jj}$ (subject to the constraint $\sum_j Y_{jj} = 0$). This result is consistent with the charge-sharpening transition found in the monitored qudit system \cite{Agrawal22, Barratt_U1_FT}. The average entanglement entropy maintains its volume-law scaling (due to the broken $S_{R,+} \times S_{R,-}$ symmetry) across the charge-sharpening transition. The broken $S_{R,+} \times S_{R,-}$ symmetry ensures that the off-diagonal fluctuations $X_{j\bar{k}}$ with $j\neq k$ are massive and do not affect the universality of the charge-sharpening transition driven by diagonal fluctuations $X_{jj} = e^{\frac{\ii}{\sqrt{\lambda}} Y_{jj}}$.

When we further increase the measurement rate, it is natural to expect the system to go through an entanglement transition into a charge-sharpened area-law phase where the full $\calG_+ \times \calG_-$ symmetry of the interacting field theory is completely restored. Such an entanglement transition has been found in numerical simulations of monitored interacting fermion chains with finite interactions. \cite{YoheiAshida2020,Lumia2024}. Based on our field theory analysis, with increasing measurement rate, the entanglement transition (between volume- and area-law entanglement scaling) should always happen after the charge-sharpening transition because of the incompatibility between the area-law entanglement scaling and the fuzziness of global charge fluctuation (as discussed in Sec. \ref{sec:Incompatibility}).

The discussion above pertains to a horizontal cut of the phase diagram at a generic interaction strength. When the interaction is weak, there might be two possible scenarios, as depicted in the two dashed boxes in Fig.~\ref{fig:PhaseDiagram_AIII}. In scenario 1, any horizontal cut with a finite interaction strength can access the three dynamical phases. The charge-sharpening transition and the entanglement transition are always separated along the horizontal cut. In scenario 2, we speculate that possibly, at weak enough interaction strength, there could be a direct transition between the charge-fuzzy volume-law phase and the charge-sharpened area-law phase. This transition can, in principle, be either
first-order or continuous. Possibly, due to the presence of randomness, a first-order transition might not occur~\cite{Greenblatt2009RoundingQuantum}. In this scenario, the charge-sharpened volume-law phase only exists when the interaction strength exceeds a finite threshold.
There would then also be a multicritical point where the charge-sharpening transition and the entanglement transition merge. Note both scenarios include the charge-fuzzy volume-law phase in the low measurement rate limit and the charge-sharpened area-law phase in the high measurement rate limit, which is naturally expected. Their difference lies in whether or not
there is always an intermediate charge-sharpened volume-law phase as we move along a horizontal cut of the phase diagram.

The phase diagram near the origin (where the interaction is weak and the measurement rate is low) can be examined through the lens of the weak-coupling RG presented in Sec. \ref{sec:RG}. Recall that a low measurement 
rate results in a small coupling constant $\lambda^{-1}$ in $S_{\rSU(R)}$ [Eq.~\eqref{eq:PCM}]. When the interaction strength is weak enough, the mass parameter $M= \lambda m^2$ and the strength $\bar{\Gamma}_U$
[Eq.~\eqref{eq:bGUDef}] reside within the weak-coupling region where the RG analysis in Sec.~\ref{sec:RG} is applicable.

In the non-interacting limit with $M= 0$ and $\bar{\Gamma}_U = 0$, the coupling $
\lambda^{-1}$ will always flow to strong coupling, and the non-interacting monitored fermion system always exhibits the area-law phase \cite{Mirlin23,FavaU1}. For interacting monitored dynamics, based on the beta functions in Eq.~\eqref{eq:1loop}, for any given small UV value of $(\lambda^{-1})_\mathsf{uv}
$, there is always a finite window for (the UV values of) $M_\mathsf{uv}$ and $(\bar{\Gamma}_{U})_\mathsf{uv}$ such that $\lambda^{-1}$ flows to strong coupling first, i.e. $\lambda^{-1} \sim O(1)$, while $M$ and $\bar{\Gamma}_U$ remain small
(see Fig.~\ref{fig:RGFlow_AIII}). In this window, the interacting monitored dynamics is expected to be captured by a fully disordered $\hat{X}$ field and exhibits the area-law entanglement entropy scaling. In other words, the beta functions [Eqs.~\eqref{eq:1loop}] imply that the area-law phase can be stable within a finite range of interactions. For example, for $\bar{\Gamma}_U = 0$, the area-law phase is stable
when $ M_\mathsf{uv} \ll a^{-2} e^{-2\pi \lambda_\mathsf{uv} }$. This estimate is made by setting $R=1$ and dropping the $\Delta_M(R)$ term in Eq.~\eqref{eq:betaM}, which underestimates the upper limit of $M_\mathsf{uv}$ in this window.

On the other hand, if $ M_\mathsf{uv}$ is much larger than $a^{-2} e^{-2\pi \lambda_\mathsf{uv}}$ (where $a$ is the lattice scale), $M$ will flow to a large value, i.e. $M a^2 \sim O(1)$, and stop the RG flow
 before the coupling $\lambda^{-1}$ becomes large.
(see Fig.~\ref{fig:RGFlow_AIII}). In this case, the off-diagonal modes of $\hat{X}$ (or equivalently, the off-diagonal models of $\hat{Y}$) become massive. If the initial value of $\lambda^{-1}$ is small, the renormalized $\lambda^{-1}$ remains small indicating that the spontaneous breaking of $S_{R,+} \times S_{R,-}$ symmetry just like in the limit of $\lambda^{-1}\rightarrow 0$. As explained above, the spontaneously broken $S_{R,+} \times S_{R,-}$ symmetry results in volume-law entanglement entropy scaling.

With the broken $S_{R,+} \times S_{R,-}$ symmetry, the remaining massless fluctuations are the compact bosons in the diagonal elements of $\hat{X}$. Conceptually, one should stop following the RG generated by the beta functions Eq.~\eqref{eq:1loop} at the scale when $M a^2$ becomes $O(1)$. Next, one should extract the stiffness and vortex fugacity of these massless compact bosons at this scale. Depending on the stiffness and the vortex fugacity, the compact bosons can either have quasi-long-range order or become disordered, which respectively corresponds to the charge-fuzzy and the charge-sharpened phases of the interacting monitored dynamics. Based on our current result on the 1-loop RG Eq.~\eqref{eq:1loop}, both charge-fuzzy and charge-sharpened phases are plausible in the weak coupling limit. That is to say, both scenario 1 and scenario 2 discussed above remains possible. More refined RG analysis might be needed to understand the conditions for each scenario (if possible) to exist. Including a finite $\bar{\Gamma}_U$ is expected not to qualitatively change the analysis above.

\section{Conclusions}

In this work, we study the charge-conserving monitored dynamics of interacting fermions in 1+1D. We obtain a unifying replica Keldysh field theory Eq.~\eqref{eq:ZFull} that provides us a comprehensive view of the phases and transitions of the relevant dynamical systems. This field theory features the average density sector $S_{\textrm{U(1)}}$ [Eq.~(\ref{eq:U(1)})] and the entanglement sector $S_{\textrm{SU($R$)}} + S_I$ [Eqs.~(\ref{eq:PCM}) and (\ref{eq:SI})]. The average density sector controls the dynamics of the average charge density, which turns out to be always diffusive. The entanglement sector governs the dynamics of the entanglement, higher-moment
charge fluctuations that probe the charge-sharpening physics, and other observables that depend non-linearly on the system's density matrix. As we show, this entanglement sector leads to a rich set of phases with different entanglement scalings and charge fuzziness (or sharpness), as well as the entanglement transition and the charge-sharpening transition between them.

We obtain this replica Keldysh field theory Eq.~\eqref{eq:ZFull} of the charge-conserving interacting dynamics using complementary methods. We first obtain it by applying a simple and exact ``gauging trick" (see Sec. \ref{sec:Results}) to the NLSM that was previously shown by Refs. \cite{Mirlin23,Mirlin2024_Above1D,Chahine23,FavaU1} to capture the monitored dynamics of non-interacting fermions in symmetry class AIII \cite{CMJian2022}. This replica Keldysh field theory is corroborated by an independent microscopic Keldysh calculation using a saddle point analysis and a higher-order trace-log expansion (see Sec. \ref{sec:KDeriv}). A similar Keldysh theory is also obtained using the $G-\Sigma$ formulation applied to a variant of the interacting monitored dynamics (see App. \ref{app:Brownian}).

We also show that this replica Keldysh field theory is equivalent to a path integral that captures the time evolution of (replicas of) the fermionic system's density matrix mapped to the (replicated) doubled Hilbert space using the Choi-Jamio{\l}kowski isomorphism
(see Sec.~\ref{sec:doublehilbert}). The latter is the path integral representation of the statistical mechanics model of the monitored dynamics of interacting fermions, analogous to the established statistical mechanics model \cite{Jian20,Bao20} for the monitored dynamics of qudit systems. From the Keldysh field theory viewpoint, the interpretation of it as a statistical mechanics model that describes the interacting dynamics in the bulk of spacetime is due to the measurement-induced heating effects that trivialize the fermion distribution function (see Sec.~\ref{sec:KDeriv}). In other words, the microscopic details of the initial state become insignificant, after repeated measurements, for the universal behavior of the dynamics in the long-time limit. Another important physical consequence of the trivialized distribution functions is that the system decouples from external electric potentials, and therefore the system is expected to be insensitive to disorders or inhomogeneities.

Compared to the charge-conserving monitored dynamics of non-interacting fermions in 1+1D, the interacting fermion dynamics exhibits a richer phase diagram. The entanglement sector of the non-interacting dynamics is given by the NLSM whose action is just $S_{\textrm{SU($R$)}}$, which has a large $\rSU(R)_+ \times \rSU(R)_-$ replica symmetry \cite{Mirlin23,Mirlin2024_Above1D,Chahine23,FavaU1}. It only admits one phase where the matrix field $\hat{X}$ is fully disordered, and consequently, the entanglement entropy follows an area-law scaling. Once the interaction is turned on, we find that $S_I$ [Eq.~\eqref{eq:SI}] reduces the symmetry of the field theory to $\calG_+ \times \calG_-$ (see Sec.~\ref{sec:phasesintmodel}), which combines the discrete $S_{R,+} \times S_{R,-}$ replica permutation symmetry and the $[\rU(1)_+]^{(R-1)}\times [\rU(1)_-]^{(R-1)} $ symmetry associated with the charge conservation within each replica. From a weak-coupling RG analysis applicable to the weak-interaction and low-measurement-rate regime (see Sec.~\ref{sec:RG}), we can already identify a separatrix in the RG flow that separates the dynamical phases with volume-law entanglement scaling and the phase with area-law scaling.

A more global perspective of the phase diagram (see Fig.~\ref{fig:PhaseDiagram_AIII}) of the interacting dynamics can be obtained following the general symmetry principle. The entanglement entropy scaling is dictated by the
$S_{R,+} \times S_{R,-}$ replica permutation symmetry. The volume-law entanglement scaling emerges from the spontaneous breaking of $S_{R,+} \times S_{R,-}$ into its diagonal subgroup characterized by the expectation value of $\langle |X|^2_{j\bar{k}}\rangle$. When $S_{R,+} \times S_{R,-}$ is unbroken, the resulting dynamics exhibits an area-law entanglement scaling. In the presence of the volume-law entanglement scaling, the fluctuations in the off-diagonal elements $\hat{X}$ become massive. Only the diagonal elements $\hat{X}_{jj} = e^{\frac{\ii}{\sqrt{\lambda}}Y_{jj}}$ remain massless. They transform under the axial subgroup of the $[\rU(1)_+]^{(R-1)}\times [\rU(1)_-]^{(R-1)} $ symmetry and are equivalent to $(R-1)$ compact bosons in 1+1D. The charge-fuzzy phase corresponds to the quasi-long-range order of these compact bosons $Y_{jj}$, while the charge-sharpened phase occurs when
the
$Y_{jj}$'s are disordered. The charge-sharpening transition between them is of the Kosterlitz-Thouless type driven by the vortices of the boson fields $Y_{jj}$, which is consistent with the theory of the charge-sharpening transition previously obtained in Ref.~\cite{Barratt_U1_FT} for a microscopically very different dynamical quantum system based on
qudits.

We observe that there is no symmetry-breaking pattern of $\calG_+ \times \calG_-$ (that can be captured by the field $\hat{X}$) which can simultaneously lead to an area-law entanglement scaling and the charge-fuzzy dynamics. In other words, a charge-fuzzy area-law phase might not exist in the charge-conserving monitored dynamics. The absence of such a phase also implies that the charge-sharpening transition has to happen in the presence of the volume-law entanglement scaling. Equivalently, as we increase the measurement rate (at a given interaction strength), the charge-sharpening transition should occur before the entanglement transition (between the volume-law and area-law entanglement scaling) if the two transitions are separated. Arguments for the order of these two phase transitions were previously given for monitored qudit systems with large qudit dimensions \cite{Barratt_U1_FT}. Here, we obtain a similar conclusion for the monitored interacting fermionic systems using a general symmetry-based analysis.

\acknowledgments

We thank Romain Vasseur, Igor Poboiko, Igor Gornyi, and Sasha Mirlin for discussions.
This work was supported by the Welch Foundation Grant No.~C-1809 (M.S.F.) and the Alfred P. Sloan Foundation through a Sloan Research Fellowship (C.-M.J.).

{\it Note added:} As we were completing this paper, we became aware of contemporaneous and independent related work on the monitored dynamics of interacting fermions \cite{Poboiko2024}.

\bibliography{refCombined}

\begin{thebibliography}{75}%
\makeatletter
\providecommand \@ifxundefined [1]{%
 \@ifx{#1\undefined}
}%
\providecommand \@ifnum [1]{%
 \ifnum #1\expandafter \@firstoftwo
 \else \expandafter \@secondoftwo
 \fi
}%
\providecommand \@ifx [1]{%
 \ifx #1\expandafter \@firstoftwo
 \else \expandafter \@secondoftwo
 \fi
}%
\providecommand \natexlab [1]{#1}%
\providecommand \enquote  [1]{``#1''}%
\providecommand \bibnamefont  [1]{#1}%
\providecommand \bibfnamefont [1]{#1}%
\providecommand \citenamefont [1]{#1}%
\providecommand \href@noop [0]{\@secondoftwo}%
\providecommand \href [0]{\begingroup \@sanitize@url \@href}%
\providecommand \@href[1]{\@@startlink{#1}\@@href}%
\providecommand \@@href[1]{\endgroup#1\@@endlink}%
\providecommand \@sanitize@url [0]{\catcode `\\12\catcode `\$12\catcode
  `\&12\catcode `\#12\catcode `\^12\catcode `\_12\catcode `\%12\relax}%
\providecommand \@@startlink[1]{}%
\providecommand \@@endlink[0]{}%
\providecommand \url  [0]{\begingroup\@sanitize@url \@url }%
\providecommand \@url [1]{\endgroup\@href {#1}{\urlprefix }}%
\providecommand \urlprefix  [0]{URL }%
\providecommand \Eprint [0]{\href }%
\providecommand \doibase [0]{https://doi.org/}%
\providecommand \selectlanguage [0]{\@gobble}%
\providecommand \bibinfo  [0]{\@secondoftwo}%
\providecommand \bibfield  [0]{\@secondoftwo}%
\providecommand \translation [1]{[#1]}%
\providecommand \BibitemOpen [0]{}%
\providecommand \bibitemStop [0]{}%
\providecommand \bibitemNoStop [0]{.\EOS\space}%
\providecommand \EOS [0]{\spacefactor3000\relax}%
\providecommand \BibitemShut  [1]{\csname bibitem#1\endcsname}%
\let\auto@bib@innerbib\@empty
\bibitem [{\citenamefont {Li}\ \emph {et~al.}(2018)\citenamefont {Li},
  \citenamefont {Chen},\ and\ \citenamefont {Fisher}}]{li2018quantum}%
  \BibitemOpen
  \bibfield  {author} {\bibinfo {author} {\bibfnamefont {Y.}~\bibnamefont
  {Li}}, \bibinfo {author} {\bibfnamefont {X.}~\bibnamefont {Chen}},\ and\
  \bibinfo {author} {\bibfnamefont {M.~P.}\ \bibnamefont {Fisher}},\ }\bibfield
   {title} {\bibinfo {title} {Quantum zeno effect and the many-body
  entanglement transition},\ }\href@noop {} {\bibfield  {journal} {\bibinfo
  {journal} {Physical Review B}\ }\textbf {\bibinfo {volume} {98}},\ \bibinfo
  {pages} {205136} (\bibinfo {year} {2018})}\BibitemShut {NoStop}%
\bibitem [{\citenamefont {Li}\ \emph {et~al.}(2019)\citenamefont {Li},
  \citenamefont {Chen},\ and\ \citenamefont {Fisher}}]{li2019measurement}%
  \BibitemOpen
  \bibfield  {author} {\bibinfo {author} {\bibfnamefont {Y.}~\bibnamefont
  {Li}}, \bibinfo {author} {\bibfnamefont {X.}~\bibnamefont {Chen}},\ and\
  \bibinfo {author} {\bibfnamefont {M.~P.}\ \bibnamefont {Fisher}},\ }\bibfield
   {title} {\bibinfo {title} {Measurement-driven entanglement transition in
  hybrid quantum circuits},\ }\href@noop {} {\bibfield  {journal} {\bibinfo
  {journal} {Physical Review B}\ }\textbf {\bibinfo {volume} {100}},\ \bibinfo
  {pages} {134306} (\bibinfo {year} {2019})}\BibitemShut {NoStop}%
\bibitem [{\citenamefont {Skinner}\ \emph {et~al.}(2019)\citenamefont
  {Skinner}, \citenamefont {Ruhman},\ and\ \citenamefont
  {Nahum}}]{skinner2019measurement}%
  \BibitemOpen
  \bibfield  {author} {\bibinfo {author} {\bibfnamefont {B.}~\bibnamefont
  {Skinner}}, \bibinfo {author} {\bibfnamefont {J.}~\bibnamefont {Ruhman}},\
  and\ \bibinfo {author} {\bibfnamefont {A.}~\bibnamefont {Nahum}},\ }\bibfield
   {title} {\bibinfo {title} {Measurement-induced phase transitions in the
  dynamics of entanglement},\ }\href
  {https://doi.org/10.1103/PhysRevX.9.031009} {\bibfield  {journal} {\bibinfo
  {journal} {Phys. Rev. X}\ }\textbf {\bibinfo {volume} {9}},\ \bibinfo {pages}
  {031009} (\bibinfo {year} {2019})}\BibitemShut {NoStop}%
\bibitem [{\citenamefont {Chan}\ \emph {et~al.}(2019)\citenamefont {Chan},
  \citenamefont {Nandkishore}, \citenamefont {Pretko},\ and\ \citenamefont
  {Smith}}]{chan2019unitary}%
  \BibitemOpen
  \bibfield  {author} {\bibinfo {author} {\bibfnamefont {A.}~\bibnamefont
  {Chan}}, \bibinfo {author} {\bibfnamefont {R.~M.}\ \bibnamefont
  {Nandkishore}}, \bibinfo {author} {\bibfnamefont {M.}~\bibnamefont
  {Pretko}},\ and\ \bibinfo {author} {\bibfnamefont {G.}~\bibnamefont
  {Smith}},\ }\bibfield  {title} {\bibinfo {title} {Unitary-projective
  entanglement dynamics},\ }\href@noop {} {\bibfield  {journal} {\bibinfo
  {journal} {Physical Review B}\ }\textbf {\bibinfo {volume} {99}},\ \bibinfo
  {pages} {224307} (\bibinfo {year} {2019})}\BibitemShut {NoStop}%
\bibitem [{\citenamefont {Fisher}\ \emph {et~al.}(2023)\citenamefont {Fisher},
  \citenamefont {Khemani}, \citenamefont {Nahum},\ and\ \citenamefont
  {Vijay}}]{fisher2023random}%
  \BibitemOpen
  \bibfield  {author} {\bibinfo {author} {\bibfnamefont {M.~P.~A.}\
  \bibnamefont {Fisher}}, \bibinfo {author} {\bibfnamefont {V.}~\bibnamefont
  {Khemani}}, \bibinfo {author} {\bibfnamefont {A.}~\bibnamefont {Nahum}},\
  and\ \bibinfo {author} {\bibfnamefont {S.}~\bibnamefont {Vijay}},\ }\bibfield
   {title} {\bibinfo {title} {Random quantum circuits},\ }\href@noop {}
  {\bibfield  {journal} {\bibinfo  {journal} {Annual Review of Condensed Matter
  Physics}\ }\textbf {\bibinfo {volume} {14}},\ \bibinfo {pages} {335}
  (\bibinfo {year} {2023})}\BibitemShut {NoStop}%
\bibitem [{\citenamefont {{Potter}}\ and\ \citenamefont
  {{Vasseur}}(2021)}]{PotterVasseurReview2021}%
  \BibitemOpen
  \bibfield  {author} {\bibinfo {author} {\bibfnamefont {A.~C.}\ \bibnamefont
  {{Potter}}}\ and\ \bibinfo {author} {\bibfnamefont {R.}~\bibnamefont
  {{Vasseur}}},\ }\bibfield  {title} {\bibinfo {title} {{Entanglement dynamics
  in hybrid quantum circuits}},\ }\href
  {https://doi.org/10.48550/arXiv.2111.08018} {\bibfield  {journal} {\bibinfo
  {journal} {arXiv e-prints}\ ,\ \bibinfo {eid} {arXiv:2111.08018}} (\bibinfo
  {year} {2021})},\ \Eprint {https://arxiv.org/abs/2111.08018}
  {arXiv:2111.08018 [quant-ph]} \BibitemShut {NoStop}%
\bibitem [{\citenamefont {Zhou}\ and\ \citenamefont
  {Nahum}(2019{\natexlab{a}})}]{zhou2019emergent}%
  \BibitemOpen
  \bibfield  {author} {\bibinfo {author} {\bibfnamefont {T.}~\bibnamefont
  {Zhou}}\ and\ \bibinfo {author} {\bibfnamefont {A.}~\bibnamefont {Nahum}},\
  }\bibfield  {title} {\bibinfo {title} {Emergent statistical mechanics of
  entanglement in random unitary circuits},\ }\href
  {https://doi.org/10.1103/PhysRevB.99.174205} {\bibfield  {journal} {\bibinfo
  {journal} {Physical Review B}\ }\textbf {\bibinfo {volume} {99}},\ \bibinfo
  {pages} {174205} (\bibinfo {year} {2019}{\natexlab{a}})}\BibitemShut
  {NoStop}%
\bibitem [{\citenamefont {Cao}\ \emph {et~al.}(2019)\citenamefont {Cao},
  \citenamefont {Tilloy},\ and\ \citenamefont {De~Luca}}]{cao2019entanglement}%
  \BibitemOpen
  \bibfield  {author} {\bibinfo {author} {\bibfnamefont {X.}~\bibnamefont
  {Cao}}, \bibinfo {author} {\bibfnamefont {A.}~\bibnamefont {Tilloy}},\ and\
  \bibinfo {author} {\bibfnamefont {A.}~\bibnamefont {De~Luca}},\ }\bibfield
  {title} {\bibinfo {title} {Entanglement in a fermion chain under continuous
  monitoring},\ }\href@noop {} {\bibfield  {journal} {\bibinfo  {journal}
  {SciPost Physics}\ }\textbf {\bibinfo {volume} {7}},\ \bibinfo {pages} {024}
  (\bibinfo {year} {2019})}\BibitemShut {NoStop}%
\bibitem [{\citenamefont {Vasseur}\ \emph {et~al.}(2019)\citenamefont
  {Vasseur}, \citenamefont {Potter}, \citenamefont {You},\ and\ \citenamefont
  {Ludwig}}]{VasseurRTN2019}%
  \BibitemOpen
  \bibfield  {author} {\bibinfo {author} {\bibfnamefont {R.}~\bibnamefont
  {Vasseur}}, \bibinfo {author} {\bibfnamefont {A.~C.}\ \bibnamefont {Potter}},
  \bibinfo {author} {\bibfnamefont {Y.-Z.}\ \bibnamefont {You}},\ and\ \bibinfo
  {author} {\bibfnamefont {A.~W.~W.}\ \bibnamefont {Ludwig}},\ }\bibfield
  {title} {\bibinfo {title} {Entanglement transitions from holographic random
  tensor networks},\ }\href {https://doi.org/10.1103/PhysRevB.100.134203}
  {\bibfield  {journal} {\bibinfo  {journal} {Phys. Rev. B}\ }\textbf {\bibinfo
  {volume} {100}},\ \bibinfo {pages} {134203} (\bibinfo {year}
  {2019})}\BibitemShut {NoStop}%
\bibitem [{\citenamefont {Zabalo}\ \emph {et~al.}(2020)\citenamefont {Zabalo},
  \citenamefont {Gullans}, \citenamefont {Wilson}, \citenamefont
  {Gopalakrishnan}, \citenamefont {Huse},\ and\ \citenamefont
  {Pixley}}]{zabalo2020critical}%
  \BibitemOpen
  \bibfield  {author} {\bibinfo {author} {\bibfnamefont {A.}~\bibnamefont
  {Zabalo}}, \bibinfo {author} {\bibfnamefont {M.~J.}\ \bibnamefont {Gullans}},
  \bibinfo {author} {\bibfnamefont {J.~H.}\ \bibnamefont {Wilson}}, \bibinfo
  {author} {\bibfnamefont {S.}~\bibnamefont {Gopalakrishnan}}, \bibinfo
  {author} {\bibfnamefont {D.~A.}\ \bibnamefont {Huse}},\ and\ \bibinfo
  {author} {\bibfnamefont {J.}~\bibnamefont {Pixley}},\ }\bibfield  {title}
  {\bibinfo {title} {Critical properties of the measurement-induced transition
  in random quantum circuits},\ }\href@noop {} {\bibfield  {journal} {\bibinfo
  {journal} {Physical Review B}\ }\textbf {\bibinfo {volume} {101}},\ \bibinfo
  {pages} {060301} (\bibinfo {year} {2020})}\BibitemShut {NoStop}%
\bibitem [{\citenamefont {{Choi}}\ \emph {et~al.}(2020)\citenamefont {{Choi}},
  \citenamefont {{Bao}}, \citenamefont {{Qi}},\ and\ \citenamefont
  {{Altman}}}]{choi_2020}%
  \BibitemOpen
  \bibfield  {author} {\bibinfo {author} {\bibfnamefont {S.}~\bibnamefont
  {{Choi}}}, \bibinfo {author} {\bibfnamefont {Y.}~\bibnamefont {{Bao}}},
  \bibinfo {author} {\bibfnamefont {X.-L.}\ \bibnamefont {{Qi}}},\ and\
  \bibinfo {author} {\bibfnamefont {E.}~\bibnamefont {{Altman}}},\ }\bibfield
  {title} {\bibinfo {title} {{Quantum Error Correction in Scrambling Dynamics
  and Measurement-Induced Phase Transition}},\ }\href
  {https://doi.org/10.1103/PhysRevLett.125.030505} {\bibfield  {journal}
  {\bibinfo  {journal} {\prl}\ }\textbf {\bibinfo {volume} {125}},\ \bibinfo
  {eid} {030505} (\bibinfo {year} {2020})},\ \Eprint
  {https://arxiv.org/abs/1903.05124} {arXiv:1903.05124 [quant-ph]} \BibitemShut
  {NoStop}%
\bibitem [{\citenamefont {Jian}\ \emph {et~al.}(2020)\citenamefont {Jian},
  \citenamefont {You}, \citenamefont {Vasseur},\ and\ \citenamefont
  {Ludwig}}]{Jian20}%
  \BibitemOpen
  \bibfield  {author} {\bibinfo {author} {\bibfnamefont {C.-M.}\ \bibnamefont
  {Jian}}, \bibinfo {author} {\bibfnamefont {Y.-Z.}\ \bibnamefont {You}},
  \bibinfo {author} {\bibfnamefont {R.}~\bibnamefont {Vasseur}},\ and\ \bibinfo
  {author} {\bibfnamefont {A.~W.~W.}\ \bibnamefont {Ludwig}},\ }\bibfield
  {title} {\bibinfo {title} {Measurement-induced criticality in random quantum
  circuits},\ }\href {https://doi.org/10.1103/PhysRevB.101.104302} {\bibfield
  {journal} {\bibinfo  {journal} {Phys. Rev. B}\ }\textbf {\bibinfo {volume}
  {101}},\ \bibinfo {pages} {104302} (\bibinfo {year} {2020})}\BibitemShut
  {NoStop}%
\bibitem [{\citenamefont {{Bao}}\ \emph {et~al.}(2020)\citenamefont {{Bao}},
  \citenamefont {{Choi}},\ and\ \citenamefont {{Altman}}}]{Bao20}%
  \BibitemOpen
  \bibfield  {author} {\bibinfo {author} {\bibfnamefont {Y.}~\bibnamefont
  {{Bao}}}, \bibinfo {author} {\bibfnamefont {S.}~\bibnamefont {{Choi}}},\ and\
  \bibinfo {author} {\bibfnamefont {E.}~\bibnamefont {{Altman}}},\ }\bibfield
  {title} {\bibinfo {title} {{Theory of the phase transition in random unitary
  circuits with measurements}},\ }\href
  {https://doi.org/10.1103/PhysRevB.101.104301} {\bibfield  {journal} {\bibinfo
   {journal} {\prb}\ }\textbf {\bibinfo {volume} {101}},\ \bibinfo {eid}
  {104301} (\bibinfo {year} {2020})},\ \Eprint
  {https://arxiv.org/abs/1908.04305} {arXiv:1908.04305 [cond-mat.stat-mech]}
  \BibitemShut {NoStop}%
\bibitem [{\citenamefont {Nahum}\ and\ \citenamefont
  {Skinner}(2020)}]{nahum2020entanglement}%
  \BibitemOpen
  \bibfield  {author} {\bibinfo {author} {\bibfnamefont {A.}~\bibnamefont
  {Nahum}}\ and\ \bibinfo {author} {\bibfnamefont {B.}~\bibnamefont
  {Skinner}},\ }\bibfield  {title} {\bibinfo {title} {Entanglement and dynamics
  of diffusion-annihilation processes with majorana defects},\ }\href@noop {}
  {\bibfield  {journal} {\bibinfo  {journal} {Physical Review Research}\
  }\textbf {\bibinfo {volume} {2}},\ \bibinfo {pages} {023288} (\bibinfo {year}
  {2020})}\BibitemShut {NoStop}%
\bibitem [{\citenamefont {Lang}\ and\ \citenamefont
  {B\"uchler}(2020)}]{lang2020entanglement}%
  \BibitemOpen
  \bibfield  {author} {\bibinfo {author} {\bibfnamefont {N.}~\bibnamefont
  {Lang}}\ and\ \bibinfo {author} {\bibfnamefont {H.~P.}\ \bibnamefont
  {B\"uchler}},\ }\bibfield  {title} {\bibinfo {title} {Entanglement transition
  in the projective transverse field ising model},\ }\href
  {https://doi.org/10.1103/PhysRevB.102.094204} {\bibfield  {journal} {\bibinfo
   {journal} {Phys. Rev. B}\ }\textbf {\bibinfo {volume} {102}},\ \bibinfo
  {pages} {094204} (\bibinfo {year} {2020})}\BibitemShut {NoStop}%
\bibitem [{\citenamefont {Gullans}\ and\ \citenamefont
  {Huse}(2020{\natexlab{a}})}]{gullans2020dynamical}%
  \BibitemOpen
  \bibfield  {author} {\bibinfo {author} {\bibfnamefont {M.~J.}\ \bibnamefont
  {Gullans}}\ and\ \bibinfo {author} {\bibfnamefont {D.~A.}\ \bibnamefont
  {Huse}},\ }\bibfield  {title} {\bibinfo {title} {Dynamical purification phase
  transition induced by quantum measurements},\ }\href
  {https://doi.org/10.1103/PhysRevX.10.041020} {\bibfield  {journal} {\bibinfo
  {journal} {Phys. Rev. X}\ }\textbf {\bibinfo {volume} {10}},\ \bibinfo
  {pages} {041020} (\bibinfo {year} {2020}{\natexlab{a}})}\BibitemShut
  {NoStop}%
\bibitem [{\citenamefont {Gullans}\ and\ \citenamefont
  {Huse}(2020{\natexlab{b}})}]{gullans2020scalable}%
  \BibitemOpen
  \bibfield  {author} {\bibinfo {author} {\bibfnamefont {M.~J.}\ \bibnamefont
  {Gullans}}\ and\ \bibinfo {author} {\bibfnamefont {D.~A.}\ \bibnamefont
  {Huse}},\ }\bibfield  {title} {\bibinfo {title} {Scalable probes of
  measurement-induced criticality},\ }\href
  {https://doi.org/10.1103/PhysRevLett.125.070606} {\bibfield  {journal}
  {\bibinfo  {journal} {Phys. Rev. Lett.}\ }\textbf {\bibinfo {volume} {125}},\
  \bibinfo {pages} {070606} (\bibinfo {year} {2020}{\natexlab{b}})}\BibitemShut
  {NoStop}%
\bibitem [{\citenamefont {Bao}\ \emph {et~al.}(2020)\citenamefont {Bao},
  \citenamefont {Choi},\ and\ \citenamefont {Altman}}]{BaoChoiAltman2019}%
  \BibitemOpen
  \bibfield  {author} {\bibinfo {author} {\bibfnamefont {Y.}~\bibnamefont
  {Bao}}, \bibinfo {author} {\bibfnamefont {S.}~\bibnamefont {Choi}},\ and\
  \bibinfo {author} {\bibfnamefont {E.}~\bibnamefont {Altman}},\ }\bibfield
  {title} {\bibinfo {title} {Theory of the phase transition in random unitary
  circuits with measurements},\ }\href
  {https://doi.org/10.1103/PhysRevB.101.104301} {\bibfield  {journal} {\bibinfo
   {journal} {Phys. Rev. B}\ }\textbf {\bibinfo {volume} {101}},\ \bibinfo
  {pages} {104301} (\bibinfo {year} {2020})}\BibitemShut {NoStop}%
\bibitem [{\citenamefont {Fidkowski}\ \emph {et~al.}(2021)\citenamefont
  {Fidkowski}, \citenamefont {Haah},\ and\ \citenamefont
  {Hastings}}]{fidkowskiHaahHastingsHowDynamicalMemoriesForget-arXiv2008.10611}%
  \BibitemOpen
  \bibfield  {author} {\bibinfo {author} {\bibfnamefont {L.}~\bibnamefont
  {Fidkowski}}, \bibinfo {author} {\bibfnamefont {J.}~\bibnamefont {Haah}},\
  and\ \bibinfo {author} {\bibfnamefont {M.~B.}\ \bibnamefont {Hastings}},\
  }\bibfield  {title} {\bibinfo {title} {How {D}ynamical {Q}uantum {M}emories
  {F}orget},\ }\href {https://doi.org/10.22331/q-2021-01-17-382} {\bibfield
  {journal} {\bibinfo  {journal} {{Quantum}}\ }\textbf {\bibinfo {volume}
  {5}},\ \bibinfo {pages} {382} (\bibinfo {year} {2021})}\BibitemShut {NoStop}%
\bibitem [{\citenamefont {Lavasani}\ \emph {et~al.}(2021)\citenamefont
  {Lavasani}, \citenamefont {Alavirad},\ and\ \citenamefont
  {Barkeshli}}]{lavasani2021measurement}%
  \BibitemOpen
  \bibfield  {author} {\bibinfo {author} {\bibfnamefont {A.}~\bibnamefont
  {Lavasani}}, \bibinfo {author} {\bibfnamefont {Y.}~\bibnamefont {Alavirad}},\
  and\ \bibinfo {author} {\bibfnamefont {M.}~\bibnamefont {Barkeshli}},\
  }\bibfield  {title} {\bibinfo {title} {Measurement-induced topological
  entanglement transitions in symmetric random quantum circuits},\ }\href@noop
  {} {\bibfield  {journal} {\bibinfo  {journal} {Nature Physics}\ }\textbf
  {\bibinfo {volume} {17}},\ \bibinfo {pages} {342} (\bibinfo {year}
  {2021})}\BibitemShut {NoStop}%
\bibitem [{\citenamefont {{Sang}}\ \emph {et~al.}(2021)\citenamefont {{Sang}},
  \citenamefont {{Li}}, \citenamefont {{Zhou}}, \citenamefont {{Chen}},
  \citenamefont {{Hsieh}},\ and\ \citenamefont {{Fisher}}}]{Sang2021LoopModel}%
  \BibitemOpen
  \bibfield  {author} {\bibinfo {author} {\bibfnamefont {S.}~\bibnamefont
  {{Sang}}}, \bibinfo {author} {\bibfnamefont {Y.}~\bibnamefont {{Li}}},
  \bibinfo {author} {\bibfnamefont {T.}~\bibnamefont {{Zhou}}}, \bibinfo
  {author} {\bibfnamefont {X.}~\bibnamefont {{Chen}}}, \bibinfo {author}
  {\bibfnamefont {T.~H.}\ \bibnamefont {{Hsieh}}},\ and\ \bibinfo {author}
  {\bibfnamefont {M.~P.~A.}\ \bibnamefont {{Fisher}}},\ }\bibfield  {title}
  {\bibinfo {title} {{Entanglement Negativity at Measurement-Induced
  Criticality}},\ }\href {https://doi.org/10.1103/PRXQuantum.2.030313}
  {\bibfield  {journal} {\bibinfo  {journal} {PRX Quantum}\ }\textbf {\bibinfo
  {volume} {2}},\ \bibinfo {eid} {030313} (\bibinfo {year} {2021})},\ \Eprint
  {https://arxiv.org/abs/2012.00031} {arXiv:2012.00031 [cond-mat.stat-mech]}
  \BibitemShut {NoStop}%
\bibitem [{\citenamefont {Turkeshi}\ \emph {et~al.}(2020)\citenamefont
  {Turkeshi}, \citenamefont {Fazio},\ and\ \citenamefont
  {Dalmonte}}]{Turkeshi2020MIPT2D}%
  \BibitemOpen
  \bibfield  {author} {\bibinfo {author} {\bibfnamefont {X.}~\bibnamefont
  {Turkeshi}}, \bibinfo {author} {\bibfnamefont {R.}~\bibnamefont {Fazio}},\
  and\ \bibinfo {author} {\bibfnamefont {M.}~\bibnamefont {Dalmonte}},\
  }\bibfield  {title} {\bibinfo {title} {Measurement-induced criticality in
  $(2+1)$-dimensional hybrid quantum circuits},\ }\href
  {https://doi.org/10.1103/PhysRevB.102.014315} {\bibfield  {journal} {\bibinfo
   {journal} {Phys. Rev. B}\ }\textbf {\bibinfo {volume} {102}},\ \bibinfo
  {pages} {014315} (\bibinfo {year} {2020})}\BibitemShut {NoStop}%
\bibitem [{\citenamefont {{Zabalo}}\ \emph {et~al.}(2022)\citenamefont
  {{Zabalo}}, \citenamefont {{Gullans}}, \citenamefont {{Wilson}},
  \citenamefont {{Vasseur}}, \citenamefont {{Ludwig}}, \citenamefont
  {{Gopalakrishnan}}, \citenamefont {{Huse}},\ and\ \citenamefont
  {{Pixley}}}]{Zabalo2021}%
  \BibitemOpen
  \bibfield  {author} {\bibinfo {author} {\bibfnamefont {A.}~\bibnamefont
  {{Zabalo}}}, \bibinfo {author} {\bibfnamefont {M.~J.}\ \bibnamefont
  {{Gullans}}}, \bibinfo {author} {\bibfnamefont {J.~H.}\ \bibnamefont
  {{Wilson}}}, \bibinfo {author} {\bibfnamefont {R.}~\bibnamefont {{Vasseur}}},
  \bibinfo {author} {\bibfnamefont {A.~W.~W.}\ \bibnamefont {{Ludwig}}},
  \bibinfo {author} {\bibfnamefont {S.}~\bibnamefont {{Gopalakrishnan}}},
  \bibinfo {author} {\bibfnamefont {D.~A.}\ \bibnamefont {{Huse}}},\ and\
  \bibinfo {author} {\bibfnamefont {J.~H.}\ \bibnamefont {{Pixley}}},\
  }\bibfield  {title} {\bibinfo {title} {{Operator Scaling Dimensions and
  Multifractality at Measurement-Induced Transitions}},\ }\href
  {https://doi.org/10.1103/PhysRevLett.128.050602} {\bibfield  {journal}
  {\bibinfo  {journal} {\prl}\ }\textbf {\bibinfo {volume} {128}},\ \bibinfo
  {eid} {050602} (\bibinfo {year} {2022})},\ \Eprint
  {https://arxiv.org/abs/2107.03393} {arXiv:2107.03393 [cond-mat.dis-nn]}
  \BibitemShut {NoStop}%
\bibitem [{\citenamefont {Li}\ \emph {et~al.}(2021)\citenamefont {Li},
  \citenamefont {Chen}, \citenamefont {Ludwig},\ and\ \citenamefont
  {Fisher}}]{LiChenLudwigFisher2021}%
  \BibitemOpen
  \bibfield  {author} {\bibinfo {author} {\bibfnamefont {Y.}~\bibnamefont
  {Li}}, \bibinfo {author} {\bibfnamefont {X.}~\bibnamefont {Chen}}, \bibinfo
  {author} {\bibfnamefont {A.~W.~W.}\ \bibnamefont {Ludwig}},\ and\ \bibinfo
  {author} {\bibfnamefont {M.~P.~A.}\ \bibnamefont {Fisher}},\ }\bibfield
  {title} {\bibinfo {title} {Conformal invariance and quantum nonlocality in
  critical hybrid circuits},\ }\href
  {https://doi.org/10.1103/PhysRevB.104.104305} {\bibfield  {journal} {\bibinfo
   {journal} {Phys. Rev. B}\ }\textbf {\bibinfo {volume} {104}},\ \bibinfo
  {pages} {104305} (\bibinfo {year} {2021})}\BibitemShut {NoStop}%
\bibitem [{\citenamefont {Nahum}\ \emph {et~al.}(2021)\citenamefont {Nahum},
  \citenamefont {Roy}, \citenamefont {Skinner},\ and\ \citenamefont
  {Ruhman}}]{NahumRoySkinnerRuhmanAlltoAll2021}%
  \BibitemOpen
  \bibfield  {author} {\bibinfo {author} {\bibfnamefont {A.}~\bibnamefont
  {Nahum}}, \bibinfo {author} {\bibfnamefont {S.}~\bibnamefont {Roy}}, \bibinfo
  {author} {\bibfnamefont {B.}~\bibnamefont {Skinner}},\ and\ \bibinfo {author}
  {\bibfnamefont {J.}~\bibnamefont {Ruhman}},\ }\bibfield  {title} {\bibinfo
  {title} {Measurement and entanglement phase transitions in all-to-all quantum
  circuits, on quantum trees, and in landau-ginsburg theory},\ }\href
  {https://doi.org/10.1103/PRXQuantum.2.010352} {\bibfield  {journal} {\bibinfo
   {journal} {PRX Quantum}\ }\textbf {\bibinfo {volume} {2}},\ \bibinfo {pages}
  {010352} (\bibinfo {year} {2021})}\BibitemShut {NoStop}%
\bibitem [{\citenamefont {Alberton}\ \emph {et~al.}(2021)\citenamefont
  {Alberton}, \citenamefont {Buchhold},\ and\ \citenamefont
  {Diehl}}]{AlbertonBuchholdDiehlFermionPRL2021}%
  \BibitemOpen
  \bibfield  {author} {\bibinfo {author} {\bibfnamefont {O.}~\bibnamefont
  {Alberton}}, \bibinfo {author} {\bibfnamefont {M.}~\bibnamefont {Buchhold}},\
  and\ \bibinfo {author} {\bibfnamefont {S.}~\bibnamefont {Diehl}},\ }\bibfield
   {title} {\bibinfo {title} {Entanglement transition in a monitored
  free-fermion chain: From extended criticality to area law},\ }\href
  {https://doi.org/10.1103/PhysRevLett.126.170602} {\bibfield  {journal}
  {\bibinfo  {journal} {Phys. Rev. Lett.}\ }\textbf {\bibinfo {volume} {126}},\
  \bibinfo {pages} {170602} (\bibinfo {year} {2021})}\BibitemShut {NoStop}%
\bibitem [{\citenamefont {Buchhold}\ \emph {et~al.}(2021)\citenamefont
  {Buchhold}, \citenamefont {Minoguchi}, \citenamefont {Altland},\ and\
  \citenamefont {Diehl}}]{MBuchhold2021}%
  \BibitemOpen
  \bibfield  {author} {\bibinfo {author} {\bibfnamefont {M.}~\bibnamefont
  {Buchhold}}, \bibinfo {author} {\bibfnamefont {Y.}~\bibnamefont {Minoguchi}},
  \bibinfo {author} {\bibfnamefont {A.}~\bibnamefont {Altland}},\ and\ \bibinfo
  {author} {\bibfnamefont {S.}~\bibnamefont {Diehl}},\ }\bibfield  {title}
  {\bibinfo {title} {Effective {{Theory}} for the {{Measurement-Induced Phase
  Transition}} of {{Dirac Fermions}}},\ }\href
  {https://doi.org/10.1103/PhysRevX.11.041004} {\bibfield  {journal} {\bibinfo
  {journal} {Phys. Rev. X}\ }\textbf {\bibinfo {volume} {11}},\ \bibinfo
  {pages} {041004} (\bibinfo {year} {2021})}\BibitemShut {NoStop}%
\bibitem [{\citenamefont {Zabalo}\ \emph {et~al.}(2022)\citenamefont {Zabalo},
  \citenamefont {Gullans}, \citenamefont {Wilson}, \citenamefont {Vasseur},
  \citenamefont {Ludwig}, \citenamefont {Gopalakrishnan}, \citenamefont
  {Huse},\ and\ \citenamefont
  {Pixley}}]{ZabaloGullansWilsonVasseurLudwigGopalakrishnanHusePixley}%
  \BibitemOpen
  \bibfield  {author} {\bibinfo {author} {\bibfnamefont {A.}~\bibnamefont
  {Zabalo}}, \bibinfo {author} {\bibfnamefont {M.~J.}\ \bibnamefont {Gullans}},
  \bibinfo {author} {\bibfnamefont {J.~H.}\ \bibnamefont {Wilson}}, \bibinfo
  {author} {\bibfnamefont {R.}~\bibnamefont {Vasseur}}, \bibinfo {author}
  {\bibfnamefont {A.~W.~W.}\ \bibnamefont {Ludwig}}, \bibinfo {author}
  {\bibfnamefont {S.}~\bibnamefont {Gopalakrishnan}}, \bibinfo {author}
  {\bibfnamefont {D.~A.}\ \bibnamefont {Huse}},\ and\ \bibinfo {author}
  {\bibfnamefont {J.~H.}\ \bibnamefont {Pixley}},\ }\bibfield  {title}
  {\bibinfo {title} {Operator scaling dimensions and multifractality at
  measurement-induced transitions},\ }\href
  {https://doi.org/10.1103/PhysRevLett.128.050602} {\bibfield  {journal}
  {\bibinfo  {journal} {Phys. Rev. Lett.}\ }\textbf {\bibinfo {volume} {128}},\
  \bibinfo {pages} {050602} (\bibinfo {year} {2022})}\BibitemShut {NoStop}%
\bibitem [{\citenamefont {Jian}\ \emph {et~al.}(2022)\citenamefont {Jian},
  \citenamefont {Bauer}, \citenamefont {Keselman},\ and\ \citenamefont
  {Ludwig}}]{CMJian2022}%
  \BibitemOpen
  \bibfield  {author} {\bibinfo {author} {\bibfnamefont {C.-M.}\ \bibnamefont
  {Jian}}, \bibinfo {author} {\bibfnamefont {B.}~\bibnamefont {Bauer}},
  \bibinfo {author} {\bibfnamefont {A.}~\bibnamefont {Keselman}},\ and\
  \bibinfo {author} {\bibfnamefont {A.~W.~W.}\ \bibnamefont {Ludwig}},\
  }\bibfield  {title} {\bibinfo {title} {Criticality and entanglement in
  nonunitary quantum circuits and tensor networks of noninteracting fermions},\
  }\href {https://doi.org/10.1103/PhysRevB.106.134206} {\bibfield  {journal}
  {\bibinfo  {journal} {Phys. Rev. B}\ }\textbf {\bibinfo {volume} {106}},\
  \bibinfo {pages} {134206} (\bibinfo {year} {2022})}\BibitemShut {NoStop}%
\bibitem [{\citenamefont {Agrawal}\ \emph {et~al.}(2022)\citenamefont
  {Agrawal}, \citenamefont {Zabalo}, \citenamefont {Chen}, \citenamefont
  {Wilson}, \citenamefont {Potter}, \citenamefont {Pixley}, \citenamefont
  {Gopalakrishnan},\ and\ \citenamefont {Vasseur}}]{Agrawal22}%
  \BibitemOpen
  \bibfield  {author} {\bibinfo {author} {\bibfnamefont {U.}~\bibnamefont
  {Agrawal}}, \bibinfo {author} {\bibfnamefont {A.}~\bibnamefont {Zabalo}},
  \bibinfo {author} {\bibfnamefont {K.}~\bibnamefont {Chen}}, \bibinfo {author}
  {\bibfnamefont {J.~H.}\ \bibnamefont {Wilson}}, \bibinfo {author}
  {\bibfnamefont {A.~C.}\ \bibnamefont {Potter}}, \bibinfo {author}
  {\bibfnamefont {J.~H.}\ \bibnamefont {Pixley}}, \bibinfo {author}
  {\bibfnamefont {S.}~\bibnamefont {Gopalakrishnan}},\ and\ \bibinfo {author}
  {\bibfnamefont {R.}~\bibnamefont {Vasseur}},\ }\bibfield  {title} {\bibinfo
  {title} {Entanglement and charge-sharpening transitions in u(1) symmetric
  monitored quantum circuits},\ }\href
  {https://doi.org/10.1103/PhysRevX.12.041002} {\bibfield  {journal} {\bibinfo
  {journal} {Phys. Rev. X}\ }\textbf {\bibinfo {volume} {12}},\ \bibinfo
  {pages} {041002} (\bibinfo {year} {2022})}\BibitemShut {NoStop}%
\bibitem [{\citenamefont {{Barratt}}\ \emph {et~al.}(2022)\citenamefont
  {{Barratt}}, \citenamefont {{Agrawal}}, \citenamefont {{Gopalakrishnan}},
  \citenamefont {{Huse}}, \citenamefont {{Vasseur}},\ and\ \citenamefont
  {{Potter}}}]{Barratt_U1_FT}%
  \BibitemOpen
  \bibfield  {author} {\bibinfo {author} {\bibfnamefont {F.}~\bibnamefont
  {{Barratt}}}, \bibinfo {author} {\bibfnamefont {U.}~\bibnamefont
  {{Agrawal}}}, \bibinfo {author} {\bibfnamefont {S.}~\bibnamefont
  {{Gopalakrishnan}}}, \bibinfo {author} {\bibfnamefont {D.~A.}\ \bibnamefont
  {{Huse}}}, \bibinfo {author} {\bibfnamefont {R.}~\bibnamefont {{Vasseur}}},\
  and\ \bibinfo {author} {\bibfnamefont {A.~C.}\ \bibnamefont {{Potter}}},\
  }\bibfield  {title} {\bibinfo {title} {{Field Theory of Charge Sharpening in
  Symmetric Monitored Quantum Circuits}},\ }\href
  {https://doi.org/10.1103/PhysRevLett.129.120604} {\bibfield  {journal}
  {\bibinfo  {journal} {\prl}\ }\textbf {\bibinfo {volume} {129}},\ \bibinfo
  {eid} {120604} (\bibinfo {year} {2022})},\ \Eprint
  {https://arxiv.org/abs/2111.09336} {arXiv:2111.09336 [quant-ph]} \BibitemShut
  {NoStop}%
\bibitem [{\citenamefont {Barratt}\ \emph {et~al.}(2022)\citenamefont
  {Barratt}, \citenamefont {Agrawal}, \citenamefont {Potter}, \citenamefont
  {Gopalakrishnan},\ and\ \citenamefont
  {Vasseur}}]{LearnabilityPhysRevLett.129.200602}%
  \BibitemOpen
  \bibfield  {author} {\bibinfo {author} {\bibfnamefont {F.}~\bibnamefont
  {Barratt}}, \bibinfo {author} {\bibfnamefont {U.}~\bibnamefont {Agrawal}},
  \bibinfo {author} {\bibfnamefont {A.~C.}\ \bibnamefont {Potter}}, \bibinfo
  {author} {\bibfnamefont {S.}~\bibnamefont {Gopalakrishnan}},\ and\ \bibinfo
  {author} {\bibfnamefont {R.}~\bibnamefont {Vasseur}},\ }\bibfield  {title}
  {\bibinfo {title} {Transitions in the learnability of global charges from
  local measurements},\ }\href {https://doi.org/10.1103/PhysRevLett.129.200602}
  {\bibfield  {journal} {\bibinfo  {journal} {Phys. Rev. Lett.}\ }\textbf
  {\bibinfo {volume} {129}},\ \bibinfo {pages} {200602} (\bibinfo {year}
  {2022})}\BibitemShut {NoStop}%
\bibitem [{\citenamefont {Li}\ \emph {et~al.}(2023{\natexlab{a}})\citenamefont
  {Li}, \citenamefont {Zou}, \citenamefont {Glorioso}, \citenamefont {Altman},\
  and\ \citenamefont {Fisher}}]{li2023cross}%
  \BibitemOpen
  \bibfield  {author} {\bibinfo {author} {\bibfnamefont {Y.}~\bibnamefont
  {Li}}, \bibinfo {author} {\bibfnamefont {Y.}~\bibnamefont {Zou}}, \bibinfo
  {author} {\bibfnamefont {P.}~\bibnamefont {Glorioso}}, \bibinfo {author}
  {\bibfnamefont {E.}~\bibnamefont {Altman}},\ and\ \bibinfo {author}
  {\bibfnamefont {M.~P.~A.}\ \bibnamefont {Fisher}},\ }\bibfield  {title}
  {\bibinfo {title} {Cross entropy benchmark for measurement-induced phase
  transitions},\ }\href@noop {} {\bibfield  {journal} {\bibinfo  {journal}
  {Physical Review Letters}\ }\textbf {\bibinfo {volume} {130}},\ \bibinfo
  {pages} {220404} (\bibinfo {year} {2023}{\natexlab{a}})}\BibitemShut
  {NoStop}%
\bibitem [{\citenamefont {Majidy}\ \emph {et~al.}(2023)\citenamefont {Majidy},
  \citenamefont {Agrawal}, \citenamefont {Gopalakrishnan}, \citenamefont
  {Potter}, \citenamefont {Vasseur},\ and\ \citenamefont
  {Halpern}}]{MajidyAgrawalGopalakrishnanPotterVasseurHalpern2023}%
  \BibitemOpen
  \bibfield  {author} {\bibinfo {author} {\bibfnamefont {S.}~\bibnamefont
  {Majidy}}, \bibinfo {author} {\bibfnamefont {U.}~\bibnamefont {Agrawal}},
  \bibinfo {author} {\bibfnamefont {S.}~\bibnamefont {Gopalakrishnan}},
  \bibinfo {author} {\bibfnamefont {A.~C.}\ \bibnamefont {Potter}}, \bibinfo
  {author} {\bibfnamefont {R.}~\bibnamefont {Vasseur}},\ and\ \bibinfo {author}
  {\bibfnamefont {N.~Y.}\ \bibnamefont {Halpern}},\ }\bibfield  {title}
  {\bibinfo {title} {Critical phase and spin sharpening in su(2)-symmetric
  monitored quantum circuits},\ }\href
  {https://doi.org/10.1103/PhysRevB.108.054307} {\bibfield  {journal} {\bibinfo
   {journal} {Phys. Rev. B}\ }\textbf {\bibinfo {volume} {108}},\ \bibinfo
  {pages} {054307} (\bibinfo {year} {2023})}\BibitemShut {NoStop}%
\bibitem [{\citenamefont {Li}\ \emph {et~al.}(2023{\natexlab{b}})\citenamefont
  {Li}, \citenamefont {Vijay},\ and\ \citenamefont
  {Fisher}}]{LiVijayPolymer2023entanglement}%
  \BibitemOpen
  \bibfield  {author} {\bibinfo {author} {\bibfnamefont {Y.}~\bibnamefont
  {Li}}, \bibinfo {author} {\bibfnamefont {S.}~\bibnamefont {Vijay}},\ and\
  \bibinfo {author} {\bibfnamefont {M.~P.~A.}\ \bibnamefont {Fisher}},\
  }\bibfield  {title} {\bibinfo {title} {Entanglement domain walls in monitored
  quantum circuits and the directed polymer in a random environment},\
  }\href@noop {} {\bibfield  {journal} {\bibinfo  {journal} {PRX Quantum}\
  }\textbf {\bibinfo {volume} {4}},\ \bibinfo {pages} {010331} (\bibinfo {year}
  {2023}{\natexlab{b}})}\BibitemShut {NoStop}%
\bibitem [{\citenamefont {Jian}\ \emph {et~al.}(2023)\citenamefont {Jian},
  \citenamefont {Shapourian}, \citenamefont {Bauer},\ and\ \citenamefont
  {Ludwig}}]{CMJian2023}%
  \BibitemOpen
  \bibfield  {author} {\bibinfo {author} {\bibfnamefont {C.-M.}\ \bibnamefont
  {Jian}}, \bibinfo {author} {\bibfnamefont {H.}~\bibnamefont {Shapourian}},
  \bibinfo {author} {\bibfnamefont {B.}~\bibnamefont {Bauer}},\ and\ \bibinfo
  {author} {\bibfnamefont {A.~W.~W.}\ \bibnamefont {Ludwig}},\ }\href
  {https://doi.org/10.48550/arXiv.2302.09094} {\bibinfo {title}
  {Measurement-induced entanglement transitions in quantum circuits of
  non-interacting fermions: {{Born-rule}} versus forced measurements}}
  (\bibinfo {year} {2023}),\ \Eprint {https://arxiv.org/abs/2302.09094}
  {arXiv:2302.09094 [cond-mat, physics:quant-ph]} \BibitemShut {NoStop}%
\bibitem [{\citenamefont {Fava}\ \emph {et~al.}(2023)\citenamefont {Fava},
  \citenamefont {Piroli}, \citenamefont {Swann}, \citenamefont {Bernard},\ and\
  \citenamefont {Nahum}}]{MFava2023}%
  \BibitemOpen
  \bibfield  {author} {\bibinfo {author} {\bibfnamefont {M.}~\bibnamefont
  {Fava}}, \bibinfo {author} {\bibfnamefont {L.}~\bibnamefont {Piroli}},
  \bibinfo {author} {\bibfnamefont {T.}~\bibnamefont {Swann}}, \bibinfo
  {author} {\bibfnamefont {D.}~\bibnamefont {Bernard}},\ and\ \bibinfo {author}
  {\bibfnamefont {A.}~\bibnamefont {Nahum}},\ }\href
  {https://doi.org/10.48550/arXiv.2302.12820} {\bibinfo {title} {Nonlinear
  sigma models for monitored dynamics of free fermions}} (\bibinfo {year}
  {2023}),\ \Eprint {https://arxiv.org/abs/2302.12820} {arXiv:2302.12820
  [cond-mat, physics:quant-ph]} \BibitemShut {NoStop}%
\bibitem [{\citenamefont {Poboiko}\ \emph {et~al.}(2023)\citenamefont
  {Poboiko}, \citenamefont {P\"opperl}, \citenamefont {Gornyi},\ and\
  \citenamefont {Mirlin}}]{Mirlin23}%
  \BibitemOpen
  \bibfield  {author} {\bibinfo {author} {\bibfnamefont {I.}~\bibnamefont
  {Poboiko}}, \bibinfo {author} {\bibfnamefont {P.}~\bibnamefont {P\"opperl}},
  \bibinfo {author} {\bibfnamefont {I.~V.}\ \bibnamefont {Gornyi}},\ and\
  \bibinfo {author} {\bibfnamefont {A.~D.}\ \bibnamefont {Mirlin}},\ }\bibfield
   {title} {\bibinfo {title} {Theory of free fermions under random projective
  measurements},\ }\href {https://doi.org/10.1103/PhysRevX.13.041046}
  {\bibfield  {journal} {\bibinfo  {journal} {Phys. Rev. X}\ }\textbf {\bibinfo
  {volume} {13}},\ \bibinfo {pages} {041046} (\bibinfo {year}
  {2023})}\BibitemShut {NoStop}%
\bibitem [{\citenamefont {{Chahine}}\ and\ \citenamefont
  {{Buchhold}}(2024)}]{Chahine23}%
  \BibitemOpen
  \bibfield  {author} {\bibinfo {author} {\bibfnamefont {K.}~\bibnamefont
  {{Chahine}}}\ and\ \bibinfo {author} {\bibfnamefont {M.}~\bibnamefont
  {{Buchhold}}},\ }\bibfield  {title} {\bibinfo {title} {{Entanglement phases,
  localization, and multifractality of monitored free fermions in two
  dimensions}},\ }\href {https://doi.org/10.1103/PhysRevB.110.054313}
  {\bibfield  {journal} {\bibinfo  {journal} {\prb}\ }\textbf {\bibinfo
  {volume} {110}},\ \bibinfo {eid} {054313} (\bibinfo {year} {2024})},\ \Eprint
  {https://arxiv.org/abs/2309.12391} {arXiv:2309.12391 [cond-mat.str-el]}
  \BibitemShut {NoStop}%
\bibitem [{\citenamefont {Nahum}\ and\ \citenamefont
  {Wiese}(2023)}]{NahumWiese}%
  \BibitemOpen
  \bibfield  {author} {\bibinfo {author} {\bibfnamefont {A.}~\bibnamefont
  {Nahum}}\ and\ \bibinfo {author} {\bibfnamefont {K.~J.}\ \bibnamefont
  {Wiese}},\ }\bibfield  {title} {\bibinfo {title} {Renormalization group for
  measurement and entanglement phase transitions},\ }\href
  {https://doi.org/10.1103/PhysRevB.108.104203} {\bibfield  {journal} {\bibinfo
   {journal} {Phys. Rev. B}\ }\textbf {\bibinfo {volume} {108}},\ \bibinfo
  {pages} {104203} (\bibinfo {year} {2023})}\BibitemShut {NoStop}%
\bibitem [{\citenamefont {Piroli}\ \emph {et~al.}(2023)\citenamefont {Piroli},
  \citenamefont {Li}, \citenamefont {Vasseur},\ and\ \citenamefont
  {Nahum}}]{PiroliLiVasseurNahumControl2023}%
  \BibitemOpen
  \bibfield  {author} {\bibinfo {author} {\bibfnamefont {L.}~\bibnamefont
  {Piroli}}, \bibinfo {author} {\bibfnamefont {Y.}~\bibnamefont {Li}}, \bibinfo
  {author} {\bibfnamefont {R.}~\bibnamefont {Vasseur}},\ and\ \bibinfo {author}
  {\bibfnamefont {A.}~\bibnamefont {Nahum}},\ }\bibfield  {title} {\bibinfo
  {title} {Triviality of quantum trajectories close to a directed percolation
  transition},\ }\href {https://doi.org/10.1103/PhysRevB.107.224303} {\bibfield
   {journal} {\bibinfo  {journal} {Phys. Rev. B}\ }\textbf {\bibinfo {volume}
  {107}},\ \bibinfo {pages} {224303} (\bibinfo {year} {2023})}\BibitemShut
  {NoStop}%
\bibitem [{\citenamefont {Li}\ \emph {et~al.}(2024)\citenamefont {Li},
  \citenamefont {Vasseur}, \citenamefont {Fisher},\ and\ \citenamefont
  {Ludwig}}]{LiVasseurFisherLudwig}%
  \BibitemOpen
  \bibfield  {author} {\bibinfo {author} {\bibfnamefont {Y.}~\bibnamefont
  {Li}}, \bibinfo {author} {\bibfnamefont {R.}~\bibnamefont {Vasseur}},
  \bibinfo {author} {\bibfnamefont {M.~P.~A.}\ \bibnamefont {Fisher}},\ and\
  \bibinfo {author} {\bibfnamefont {A.~W.~W.}\ \bibnamefont {Ludwig}},\
  }\bibfield  {title} {\bibinfo {title} {Statistical mechanics model for
  clifford random tensor networks and monitored quantum circuits},\ }\href
  {https://doi.org/10.1103/PhysRevB.109.174307} {\bibfield  {journal} {\bibinfo
   {journal} {Phys. Rev. B}\ }\textbf {\bibinfo {volume} {109}},\ \bibinfo
  {pages} {174307} (\bibinfo {year} {2024})}\BibitemShut {NoStop}%
\bibitem [{\citenamefont {Kumar}\ \emph {et~al.}(2024)\citenamefont {Kumar},
  \citenamefont {Aziz}, \citenamefont {Chakraborty}, \citenamefont {Ludwig},
  \citenamefont {Gopalakrishnan}, \citenamefont {Pixley},\ and\ \citenamefont
  {Vasseur}}]{KumarKemalChakrabortyLudwigGopalakrishnanPixleyVasseur}%
  \BibitemOpen
  \bibfield  {author} {\bibinfo {author} {\bibfnamefont {A.}~\bibnamefont
  {Kumar}}, \bibinfo {author} {\bibfnamefont {K.}~\bibnamefont {Aziz}},
  \bibinfo {author} {\bibfnamefont {A.}~\bibnamefont {Chakraborty}}, \bibinfo
  {author} {\bibfnamefont {A.~W.~W.}\ \bibnamefont {Ludwig}}, \bibinfo {author}
  {\bibfnamefont {S.}~\bibnamefont {Gopalakrishnan}}, \bibinfo {author}
  {\bibfnamefont {J.~H.}\ \bibnamefont {Pixley}},\ and\ \bibinfo {author}
  {\bibfnamefont {R.}~\bibnamefont {Vasseur}},\ }\bibfield  {title} {\bibinfo
  {title} {Boundary transfer matrix spectrum of measurement-induced
  transitions},\ }\href {https://doi.org/10.1103/PhysRevB.109.014303}
  {\bibfield  {journal} {\bibinfo  {journal} {Phys. Rev. B}\ }\textbf {\bibinfo
  {volume} {109}},\ \bibinfo {pages} {014303} (\bibinfo {year}
  {2024})}\BibitemShut {NoStop}%
\bibitem [{\citenamefont {Gonz\'alez-Garc\'{\i}a}\ \emph
  {et~al.}(2024)\citenamefont {Gonz\'alez-Garc\'{\i}a}, \citenamefont {Sang},
  \citenamefont {Hsieh}, \citenamefont {Boixo}, \citenamefont {Vidal},
  \citenamefont {Potter},\ and\ \citenamefont
  {Vasseur}}]{VidalPotterVasseur2024}%
  \BibitemOpen
  \bibfield  {author} {\bibinfo {author} {\bibfnamefont {S.}~\bibnamefont
  {Gonz\'alez-Garc\'{\i}a}}, \bibinfo {author} {\bibfnamefont {S.}~\bibnamefont
  {Sang}}, \bibinfo {author} {\bibfnamefont {T.~H.}\ \bibnamefont {Hsieh}},
  \bibinfo {author} {\bibfnamefont {S.}~\bibnamefont {Boixo}}, \bibinfo
  {author} {\bibfnamefont {G.}~\bibnamefont {Vidal}}, \bibinfo {author}
  {\bibfnamefont {A.~C.}\ \bibnamefont {Potter}},\ and\ \bibinfo {author}
  {\bibfnamefont {R.}~\bibnamefont {Vasseur}},\ }\bibfield  {title} {\bibinfo
  {title} {Random insights into the complexity of two-dimensional tensor
  network calculations},\ }\href {https://doi.org/10.1103/PhysRevB.109.235102}
  {\bibfield  {journal} {\bibinfo  {journal} {Phys. Rev. B}\ }\textbf {\bibinfo
  {volume} {109}},\ \bibinfo {pages} {235102} (\bibinfo {year}
  {2024})}\BibitemShut {NoStop}%
\bibitem [{\citenamefont {Lovas}\ \emph {et~al.}(2024)\citenamefont {Lovas},
  \citenamefont {Agrawal},\ and\ \citenamefont
  {Vijay}}]{LovasAgrawalVijayBoundaryDiss2024}%
  \BibitemOpen
  \bibfield  {author} {\bibinfo {author} {\bibfnamefont {I.}~\bibnamefont
  {Lovas}}, \bibinfo {author} {\bibfnamefont {U.}~\bibnamefont {Agrawal}},\
  and\ \bibinfo {author} {\bibfnamefont {S.}~\bibnamefont {Vijay}},\ }\bibfield
   {title} {\bibinfo {title} {Quantum coding transitions in the presence of
  boundary dissipation},\ }\href {https://doi.org/10.1103/PRXQuantum.5.030327}
  {\bibfield  {journal} {\bibinfo  {journal} {PRX Quantum}\ }\textbf {\bibinfo
  {volume} {5}},\ \bibinfo {pages} {030327} (\bibinfo {year}
  {2024})}\BibitemShut {NoStop}%
\bibitem [{\citenamefont {{Poboiko}}\ \emph {et~al.}(2024)\citenamefont
  {{Poboiko}}, \citenamefont {{Gornyi}},\ and\ \citenamefont
  {{Mirlin}}}]{Mirlin2024_Above1D}%
  \BibitemOpen
  \bibfield  {author} {\bibinfo {author} {\bibfnamefont {I.}~\bibnamefont
  {{Poboiko}}}, \bibinfo {author} {\bibfnamefont {I.~V.}\ \bibnamefont
  {{Gornyi}}},\ and\ \bibinfo {author} {\bibfnamefont {A.~D.}\ \bibnamefont
  {{Mirlin}}},\ }\bibfield  {title} {\bibinfo {title} {{Measurement-Induced
  Phase Transition for Free Fermions above One Dimension}},\ }\href
  {https://doi.org/10.1103/PhysRevLett.132.110403} {\bibfield  {journal}
  {\bibinfo  {journal} {\prl}\ }\textbf {\bibinfo {volume} {132}},\ \bibinfo
  {eid} {110403} (\bibinfo {year} {2024})},\ \Eprint
  {https://arxiv.org/abs/2309.12405} {arXiv:2309.12405 [quant-ph]} \BibitemShut
  {NoStop}%
\bibitem [{\citenamefont {{Fava}}\ \emph {et~al.}(2024)\citenamefont {{Fava}},
  \citenamefont {{Piroli}}, \citenamefont {{Bernard}},\ and\ \citenamefont
  {{Nahum}}}]{FavaU1}%
  \BibitemOpen
  \bibfield  {author} {\bibinfo {author} {\bibfnamefont {M.}~\bibnamefont
  {{Fava}}}, \bibinfo {author} {\bibfnamefont {L.}~\bibnamefont {{Piroli}}},
  \bibinfo {author} {\bibfnamefont {D.}~\bibnamefont {{Bernard}}},\ and\
  \bibinfo {author} {\bibfnamefont {A.}~\bibnamefont {{Nahum}}},\ }\bibfield
  {title} {\bibinfo {title} {{A tractable model of monitored fermions with
  conserved $\mathrm{U}(1)$ charge}},\ }\href
  {https://doi.org/10.48550/arXiv.2407.08045} {\bibfield  {journal} {\bibinfo
  {journal} {arXiv e-prints}\ ,\ \bibinfo {eid} {arXiv:2407.08045}} (\bibinfo
  {year} {2024})},\ \Eprint {https://arxiv.org/abs/2407.08045}
  {arXiv:2407.08045 [cond-mat.stat-mech]} \BibitemShut {NoStop}%
\bibitem [{\citenamefont {{Merritt}}\ and\ \citenamefont
  {{Fidkowski}}(2023)}]{MerrittFidkowski2023}%
  \BibitemOpen
  \bibfield  {author} {\bibinfo {author} {\bibfnamefont {J.}~\bibnamefont
  {{Merritt}}}\ and\ \bibinfo {author} {\bibfnamefont {L.}~\bibnamefont
  {{Fidkowski}}},\ }\bibfield  {title} {\bibinfo {title} {{Entanglement
  transitions with free fermions}},\ }\href
  {https://doi.org/10.1103/PhysRevB.107.064303} {\bibfield  {journal} {\bibinfo
   {journal} {\prb}\ }\textbf {\bibinfo {volume} {107}},\ \bibinfo {eid}
  {064303} (\bibinfo {year} {2023})},\ \Eprint
  {https://arxiv.org/abs/2210.05681} {arXiv:2210.05681 [cond-mat.str-el]}
  \BibitemShut {NoStop}%
\bibitem [{\citenamefont {Lumia}\ \emph {et~al.}(2024)\citenamefont {Lumia},
  \citenamefont {Tirrito}, \citenamefont {Fazio},\ and\ \citenamefont
  {Collura}}]{Lumia2024}%
  \BibitemOpen
  \bibfield  {author} {\bibinfo {author} {\bibfnamefont {L.}~\bibnamefont
  {Lumia}}, \bibinfo {author} {\bibfnamefont {E.}~\bibnamefont {Tirrito}},
  \bibinfo {author} {\bibfnamefont {R.}~\bibnamefont {Fazio}},\ and\ \bibinfo
  {author} {\bibfnamefont {M.}~\bibnamefont {Collura}},\ }\bibfield  {title}
  {\bibinfo {title} {Measurement-induced transitions beyond gaussianity: A
  single particle description},\ }\href
  {https://doi.org/10.1103/PhysRevResearch.6.023176} {\bibfield  {journal}
  {\bibinfo  {journal} {Phys. Rev. Res.}\ }\textbf {\bibinfo {volume} {6}},\
  \bibinfo {pages} {023176} (\bibinfo {year} {2024})}\BibitemShut {NoStop}%
\bibitem [{\citenamefont {Fuji}\ and\ \citenamefont
  {Ashida}(2020)}]{YoheiAshida2020}%
  \BibitemOpen
  \bibfield  {author} {\bibinfo {author} {\bibfnamefont {Y.}~\bibnamefont
  {Fuji}}\ and\ \bibinfo {author} {\bibfnamefont {Y.}~\bibnamefont {Ashida}},\
  }\bibfield  {title} {\bibinfo {title} {Measurement-induced quantum
  criticality under continuous monitoring},\ }\href
  {https://doi.org/10.1103/PhysRevB.102.054302} {\bibfield  {journal} {\bibinfo
   {journal} {Phys. Rev. B}\ }\textbf {\bibinfo {volume} {102}},\ \bibinfo
  {pages} {054302} (\bibinfo {year} {2020})}\BibitemShut {NoStop}%
\bibitem [{Note1()}]{Note1}%
  \BibitemOpen
  \bibinfo {note} {Certainly, special features can enable other types of
  entanglement scaling in non-interacting fermion systems; see, for example,
  the spacetime dual a unitary circuit in studied Ref. \cite
  {Tarun}.}\BibitemShut {Stop}%
\bibitem [{\citenamefont {Basko}\ \emph {et~al.}(2006)\citenamefont {Basko},
  \citenamefont {Aleiner},\ and\ \citenamefont {Altshuler}}]{BAA}%
  \BibitemOpen
  \bibfield  {author} {\bibinfo {author} {\bibfnamefont {D.~M.}\ \bibnamefont
  {Basko}}, \bibinfo {author} {\bibfnamefont {I.~L.}\ \bibnamefont {Aleiner}},\
  and\ \bibinfo {author} {\bibfnamefont {B.~L.}\ \bibnamefont {Altshuler}},\
  }\bibfield  {title} {\bibinfo {title} {Metal-insulator transition in a weakly
  interacting many-electron system with localized single-particle states},\
  }\href {https://doi.org/10.1016/j.aop.2005.11.014} {\bibfield  {journal}
  {\bibinfo  {journal} {Annals of Physics (Amsterdam)}\ }\textbf {\bibinfo
  {volume} {321}},\ \bibinfo {pages} {1126} (\bibinfo {year}
  {2006})}\BibitemShut {NoStop}%
\bibitem [{\citenamefont {{Kamenev}}(2023)}]{Kamenev23}%
  \BibitemOpen
  \bibfield  {author} {\bibinfo {author} {\bibfnamefont {A.}~\bibnamefont
  {{Kamenev}}},\ }\href@noop {} {\emph {\bibinfo {title} {Field Theory of
  Non-Equilibrium Systems}}},\ \bibinfo {edition} {2nd}\ ed.\ (\bibinfo
  {publisher} {Cambridge University Press},\ \bibinfo {address} {Cambridge,
  England},\ \bibinfo {year} {2023})\BibitemShut {NoStop}%
\bibitem [{\citenamefont {McKane}\ and\ \citenamefont
  {Stone}(1979)}]{McKaneStone79}%
  \BibitemOpen
  \bibfield  {author} {\bibinfo {author} {\bibfnamefont {A.}~\bibnamefont
  {McKane}}\ and\ \bibinfo {author} {\bibfnamefont {M.}~\bibnamefont {Stone}},\
  }\bibfield  {title} {\bibinfo {title} {Non-linear $\sigma$ models: A
  perturbative approach to symmetry restoration},\ }\href
  {https://doi.org/10.1016/0550-3213(80)90396-X} {\bibfield  {journal}
  {\bibinfo  {journal} {Nuclear Physics B}\ }\textbf {\bibinfo {volume}
  {163}},\ \bibinfo {pages} {169} (\bibinfo {year} {1979})}\BibitemShut
  {NoStop}%
\bibitem [{Note2()}]{Note2}%
  \BibitemOpen
  \bibinfo {note} {These can naturally be thought of as arising from a
  space-time constant interaction by coarse-graining, given the other sources
  of space-time randomness in the circuit, i.e. the measurement outcomes, and
  they thus are expected to yield equivalent results.}\BibitemShut {Stop}%
\bibitem [{\citenamefont {Altshuler}\ \emph {et~al.}(1982)\citenamefont
  {Altshuler}, \citenamefont {Aronov},\ and\ \citenamefont
  {Khmelnitsky}}]{AAK82}%
  \BibitemOpen
  \bibfield  {author} {\bibinfo {author} {\bibfnamefont {B.~L.}\ \bibnamefont
  {Altshuler}}, \bibinfo {author} {\bibfnamefont {A.~G.}\ \bibnamefont
  {Aronov}},\ and\ \bibinfo {author} {\bibfnamefont {D.~E.}\ \bibnamefont
  {Khmelnitsky}},\ }\bibfield  {title} {\bibinfo {title} {Effects of
  electron-electron collisions with small energy transfers on quantum
  localisation},\ }\href {https://doi.org/10.1088/0022-3719/15/36/018}
  {\bibfield  {journal} {\bibinfo  {journal} {Journal of Physics C: Solid State
  Physics}\ }\textbf {\bibinfo {volume} {15}},\ \bibinfo {pages} {7367}
  (\bibinfo {year} {1982})}\BibitemShut {NoStop}%
\bibitem [{\citenamefont {{Liao}}\ \emph {et~al.}(2017)\citenamefont {{Liao}},
  \citenamefont {{Levchenko}},\ and\ \citenamefont {{Foster}}}]{Liao17}%
  \BibitemOpen
  \bibfield  {author} {\bibinfo {author} {\bibfnamefont {Y.}~\bibnamefont
  {{Liao}}}, \bibinfo {author} {\bibfnamefont {A.}~\bibnamefont
  {{Levchenko}}},\ and\ \bibinfo {author} {\bibfnamefont {M.~S.}\ \bibnamefont
  {{Foster}}},\ }\bibfield  {title} {\bibinfo {title} {{Response theory of the
  ergodic many-body delocalized phase: Keldysh Finkel'stein sigma models and
  the 10-fold way}},\ }\href {https://doi.org/10.1016/j.aop.2017.08.020}
  {\bibfield  {journal} {\bibinfo  {journal} {Annals of Physics}\ }\textbf
  {\bibinfo {volume} {386}},\ \bibinfo {pages} {97} (\bibinfo {year} {2017})},\
  \Eprint {https://arxiv.org/abs/1706.07066} {arXiv:1706.07066
  [cond-mat.dis-nn]} \BibitemShut {NoStop}%
\bibitem [{\citenamefont {Finkel'stein}(1983)}]{Finkelstein83}%
  \BibitemOpen
  \bibfield  {author} {\bibinfo {author} {\bibfnamefont {A.~M.}\ \bibnamefont
  {Finkel'stein}},\ }\bibfield  {title} {\bibinfo {title} {Influence of
  {Coulomb} interaction on the properties of disordered metals},\ }\href@noop
  {} {\bibfield  {journal} {\bibinfo  {journal} {Zh. Eksp. Teor. Fiz.}\
  }\textbf {\bibinfo {volume} {84}},\ \bibinfo {pages} {168} (\bibinfo {year}
  {1983})},\ \bibinfo {note} {[Sov. Phys. JETP {\bf 57}, 97
  (1983)]}\BibitemShut {NoStop}%
\bibitem [{\citenamefont {Belitz}\ and\ \citenamefont
  {Kirkpatrick}(1994)}]{BK94}%
  \BibitemOpen
  \bibfield  {author} {\bibinfo {author} {\bibfnamefont {D.}~\bibnamefont
  {Belitz}}\ and\ \bibinfo {author} {\bibfnamefont {T.}~\bibnamefont
  {Kirkpatrick}},\ }\bibfield  {title} {\bibinfo {title} {The {Anderson}-{Mott}
  transition},\ }\href {https://doi.org/10.1103/RevModPhys.66.261} {\bibfield
  {journal} {\bibinfo  {journal} {Rev. Mod. Phys}\ }\textbf {\bibinfo {volume}
  {66}},\ \bibinfo {pages} {261} (\bibinfo {year} {1994})}\BibitemShut
  {NoStop}%
\bibitem [{\citenamefont {Burmistrov}(2019)}]{Burmistrov19}%
  \BibitemOpen
  \bibfield  {author} {\bibinfo {author} {\bibfnamefont {I.}~\bibnamefont
  {Burmistrov}},\ }\bibfield  {title} {\bibinfo {title} {Finkel'stein nonlinear
  sigma model: Interplay of disorder and interaction in 2d electron systems},\
  }\href@noop {} {\bibfield  {journal} {\bibinfo  {journal} {J. Exp. Theor.
  Phys.}\ }\textbf {\bibinfo {volume} {129}},\ \bibinfo {pages} {669} (\bibinfo
  {year} {2019})}\BibitemShut {NoStop}%
\bibitem [{\citenamefont {Ha}\ \emph {et~al.}(2024)\citenamefont {Ha},
  \citenamefont {Pandey}, \citenamefont {Gopalakrishnan},\ and\ \citenamefont
  {Huse}}]{Ha2024}%
  \BibitemOpen
  \bibfield  {author} {\bibinfo {author} {\bibfnamefont {H.}~\bibnamefont
  {Ha}}, \bibinfo {author} {\bibfnamefont {A.}~\bibnamefont {Pandey}}, \bibinfo
  {author} {\bibfnamefont {S.}~\bibnamefont {Gopalakrishnan}},\ and\ \bibinfo
  {author} {\bibfnamefont {D.~A.}\ \bibnamefont {Huse}},\ }\bibfield  {title}
  {\bibinfo {title} {Measurement-induced phase transitions in systems with
  diffusive dynamics},\ }\href {https://doi.org/10.1103/PhysRevB.110.L140301}
  {\bibfield  {journal} {\bibinfo  {journal} {Phys. Rev. B}\ }\textbf {\bibinfo
  {volume} {110}},\ \bibinfo {pages} {L140301} (\bibinfo {year}
  {2024})}\BibitemShut {NoStop}%
\bibitem [{Note3()}]{Note3}%
  \BibitemOpen
  \bibinfo {note} {Note that this form of the expectation value $\langle
  |X_{j\protect \bar {k}}|^2 \rangle = B \protect \, \delta _{j\protect \bar
  {k}} + (1-B) R^{-1}$ is due to constraint $\protect \hat {X}^\dagger \protect
  \hat {X}=\protect \hat {1}_R$. $S_R \times S_R$ is spontaneously broken to
  its diagonal subgroup for any $B>0$. More details of the order parameter of
  the spontaneous breaking of $S_R \times S_R$ is provided in Sec. \ref
  {sec:CFVLphase}.}\BibitemShut {Stop}%
\bibitem [{Note4()}]{Note4}%
  \BibitemOpen
  \bibinfo {note} {For an equilibrium $F(\omega ) = \tanh (\omega /2T)$,
  $F_B(\omega ) = \coth (\omega /2T)$ and the first term in Eq.~(\ref {eq:St})
  vanishes due to the identity $ \coth (x - y)\left [\tanh (x) - \tanh
  (y)\right ] = 1 - \tanh (x) \protect \, \tanh (y) $.}\BibitemShut {Stop}%
\bibitem [{Note5()}]{Note5}%
  \BibitemOpen
  \bibinfo {note} {One-loop anomalous dimension function.}\BibitemShut {Stop}%
\bibitem [{\citenamefont {Altshuler}\ and\ \citenamefont
  {Aronov}(1985)}]{AA85}%
  \BibitemOpen
  \bibfield  {author} {\bibinfo {author} {\bibfnamefont {B.~L.}\ \bibnamefont
  {Altshuler}}\ and\ \bibinfo {author} {\bibfnamefont {A.~G.}\ \bibnamefont
  {Aronov}},\ }\bibfield  {title} {\bibinfo {title} {Electron-electron
  interaction in disordered conductors},\ }in\ \href@noop {} {\emph {\bibinfo
  {booktitle} {Electron-Electron Interactions in Disordered Systems}}},\
  \bibinfo {editor} {edited by\ \bibinfo {editor} {\bibfnamefont {A.~L.}\
  \bibnamefont {Efros}}\ and\ \bibinfo {editor} {\bibfnamefont
  {M.}~\bibnamefont {Pollak}}}\ (\bibinfo  {publisher} {North-Holland},\
  \bibinfo {address} {Amsterdam},\ \bibinfo {year} {1985})\ Chap.~\bibinfo
  {chapter} {1}, p.~\bibinfo {pages} {1}\BibitemShut {NoStop}%
\bibitem [{\citenamefont {Nielsen}\ and\ \citenamefont
  {Chuang}(2010)}]{NielsenChuang2010}%
  \BibitemOpen
  \bibfield  {author} {\bibinfo {author} {\bibfnamefont {M.~A.}\ \bibnamefont
  {Nielsen}}\ and\ \bibinfo {author} {\bibfnamefont {I.~L.}\ \bibnamefont
  {Chuang}},\ }\href@noop {} {\emph {\bibinfo {title} {Computation and Quantum
  Information}}}\ (\bibinfo  {publisher} {Cambridge University Press},\
  \bibinfo {address} {Cambridge, UK},\ \bibinfo {year} {2010})\BibitemShut
  {NoStop}%
\bibitem [{\citenamefont {Zhou}\ and\ \citenamefont
  {Nahum}(2019{\natexlab{b}})}]{ZhouRUC2019}%
  \BibitemOpen
  \bibfield  {author} {\bibinfo {author} {\bibfnamefont {T.}~\bibnamefont
  {Zhou}}\ and\ \bibinfo {author} {\bibfnamefont {A.}~\bibnamefont {Nahum}},\
  }\bibfield  {title} {\bibinfo {title} {Emergent statistical mechanics of
  entanglement in random unitary circuits},\ }\href
  {https://doi.org/10.1103/PhysRevB.99.174205} {\bibfield  {journal} {\bibinfo
  {journal} {Phys. Rev. B}\ }\textbf {\bibinfo {volume} {99}},\ \bibinfo
  {pages} {174205} (\bibinfo {year} {2019}{\natexlab{b}})}\BibitemShut
  {NoStop}%
\bibitem [{\citenamefont {Klich}\ and\ \citenamefont
  {Levitov}(2009)}]{Klich09}%
  \BibitemOpen
  \bibfield  {author} {\bibinfo {author} {\bibfnamefont {I.}~\bibnamefont
  {Klich}}\ and\ \bibinfo {author} {\bibfnamefont {L.}~\bibnamefont
  {Levitov}},\ }\bibfield  {title} {\bibinfo {title} {Quantum noise as an
  entanglement meter},\ }\href {https://doi.org/10.1103/PhysRevLett.102.100502}
  {\bibfield  {journal} {\bibinfo  {journal} {Phys. Rev. Lett.}\ }\textbf
  {\bibinfo {volume} {102}},\ \bibinfo {pages} {100502} (\bibinfo {year}
  {2009})}\BibitemShut {NoStop}%
\bibitem [{Note6()}]{Note6}%
  \BibitemOpen
  \bibinfo {note} {The $S_R$-symmetric saddle is $ \protect \hat {Q}_{{\protect
  \mathsf {sp}}} = \protect \mathrm {i}\protect \, \gamma \protect \, \protect
  \hat {\tau }^3 \protect \, \protect \hat {\protect \mathcal {S}}_R \otimes
  \protect \hat {1}_2, $ where $(\protect \hat {\protect \mathcal {S}}_R)_{i j}
  = (2/R) - \delta _{i j}$. Parameterizing fluctuations around this saddle
  gives an identical effective theory for non-interacting fermions [Eq.~(\ref
  {eq:PCM})]; however, the resulting interacting theory possesses additional,
  unphysical gapless modes.}\BibitemShut {Stop}%
\bibitem [{Note7()}]{Note7}%
  \BibitemOpen
  \bibinfo {note} {The notion of charge-fuzzy phase is first introduced in
  Ref.~\cite {Agrawal22} in a different setting with dynamical qubit
  systems.}\BibitemShut {Stop}%
\bibitem [{\citenamefont {Greenblatt}\ \emph {et~al.}(2009)\citenamefont
  {Greenblatt}, \citenamefont {Aizenman},\ and\ \citenamefont
  {Lebowitz}}]{Greenblatt2009RoundingQuantum}%
  \BibitemOpen
  \bibfield  {author} {\bibinfo {author} {\bibfnamefont {R.~L.}\ \bibnamefont
  {Greenblatt}}, \bibinfo {author} {\bibfnamefont {M.}~\bibnamefont
  {Aizenman}},\ and\ \bibinfo {author} {\bibfnamefont {J.~L.}\ \bibnamefont
  {Lebowitz}},\ }\bibfield  {title} {\bibinfo {title} {Rounding of first order
  transitions in low-dimensional quantum systems with quenched disorder},\
  }\href {https://doi.org/10.1103/PhysRevLett.103.197201} {\bibfield  {journal}
  {\bibinfo  {journal} {Phys. Rev. Lett.}\ }\textbf {\bibinfo {volume} {103}},\
  \bibinfo {pages} {197201} (\bibinfo {year} {2009})}\BibitemShut {NoStop}%
\bibitem [{\citenamefont {Poboiko}\ \emph {et~al.}(2025)\citenamefont
  {Poboiko}, \citenamefont {P{\"o}pperl}, \citenamefont {Gornyi},\ and\
  \citenamefont {Mirlin}}]{Poboiko2024}%
  \BibitemOpen
  \bibfield  {author} {\bibinfo {author} {\bibfnamefont {I.}~\bibnamefont
  {Poboiko}}, \bibinfo {author} {\bibfnamefont {P.}~\bibnamefont
  {P{\"o}pperl}}, \bibinfo {author} {\bibfnamefont {I.~V.}\ \bibnamefont
  {Gornyi}},\ and\ \bibinfo {author} {\bibfnamefont {A.~D.}\ \bibnamefont
  {Mirlin}},\ }\bibfield  {title} {\bibinfo {title} {Measurement-induced
  transitions for interacting fermions},\ }\href
  {https://doi.org/10.1103/PhysRevB.111.024204} {\bibfield  {journal} {\bibinfo
   {journal} {Phys. Rev. B}\ }\textbf {\bibinfo {volume} {111}},\ \bibinfo
  {pages} {024204} (\bibinfo {year} {2025})},\ \Eprint
  {https://arxiv.org/abs/2410.07334} {arXiv:2410.07334 [quant-ph]} \BibitemShut
  {NoStop}%
\bibitem [{\citenamefont {Lu}\ and\ \citenamefont {Grover}(2021)}]{Tarun}%
  \BibitemOpen
  \bibfield  {author} {\bibinfo {author} {\bibfnamefont {T.-C.}\ \bibnamefont
  {Lu}}\ and\ \bibinfo {author} {\bibfnamefont {T.}~\bibnamefont {Grover}},\
  }\bibfield  {title} {\bibinfo {title} {Spacetime duality between localization
  transitions and measurement-induced transitions},\ }\href
  {https://doi.org/10.1103/PRXQuantum.2.040319} {\bibfield  {journal} {\bibinfo
   {journal} {PRX Quantum}\ }\textbf {\bibinfo {volume} {2}},\ \bibinfo {pages}
  {040319} (\bibinfo {year} {2021})}\BibitemShut {NoStop}%
\bibitem [{\citenamefont {{Foster}}\ \emph {et~al.}(2014)\citenamefont
  {{Foster}}, \citenamefont {{Xie}},\ and\ \citenamefont {{Chou}}}]{Foster14}%
  \BibitemOpen
  \bibfield  {author} {\bibinfo {author} {\bibfnamefont {M.~S.}\ \bibnamefont
  {{Foster}}}, \bibinfo {author} {\bibfnamefont {H.-Y.}\ \bibnamefont
  {{Xie}}},\ and\ \bibinfo {author} {\bibfnamefont {Y.-Z.}\ \bibnamefont
  {{Chou}}},\ }\bibfield  {title} {\bibinfo {title} {{Topological protection,
  disorder, and interactions: Survival at the surface of three-dimensional
  topological superconductors}},\ }\href
  {https://doi.org/10.1103/PhysRevB.89.155140} {\bibfield  {journal} {\bibinfo
  {journal} {Phys. Rev. B}\ }\textbf {\bibinfo {volume} {89}},\ \bibinfo
  {pages} {155140} (\bibinfo {year} {2014})}\BibitemShut {NoStop}%
\bibitem [{\citenamefont {Amit}\ and\ \citenamefont
  {Martin-mayor}(2005)}]{AmitBook}%
  \BibitemOpen
  \bibfield  {author} {\bibinfo {author} {\bibfnamefont {D.~J.}\ \bibnamefont
  {Amit}}\ and\ \bibinfo {author} {\bibfnamefont {V.}~\bibnamefont
  {Martin-mayor}},\ }\href@noop {} {\emph {\bibinfo {title} {Field Theory, The
  Renormalization Group, And Critical Phenomena: Graphs To Computers (3rd
  Edition)}}}\ (\bibinfo  {publisher} {World Scientific Publishing},\ \bibinfo
  {address} {Singapore},\ \bibinfo {year} {2005})\BibitemShut {NoStop}%
\end{thebibliography}%

\appendix

\section{Away from half filling} \label{app:ahf}

In this appendix, we derive the form of the non-linear sigma model for monitored, non-interacting fermions at arbitrary filling. After replicating and averaging over quantum trajectories, the effective action is 
given by Eq.~(\ref{eq:barS}). The saddle-point equation can be phrased in terms of the fermion self-energies in the LO basis $\hat{\Sigma}^{R,A,K}$, defined via
\begin{align}
    \hat{\Sigma}
    \equiv
    \begin{bmatrix}
    \hat{\Sigma}^R & \hat{\Sigma}^K \\
    0 & \hat{\Sigma}^A
    \end{bmatrix}_{\mathsf{LO}}
    \equiv
    -\hat{Q}
    +
    f_\alpha
    \,
    \tauh^1
    \,
    \tr\left[\hat{Q} \, \tauh^1\right].
\end{align}
The saddle-point equation is
\begin{align}
    -
    \begin{bmatrix}
    \hat{\Sigma}^A & 0 \\
    \hat{\Sigma}^K & \hat{\Sigma}^R
    \end{bmatrix}_{\mathsf{LO}}
    =
    \frac{4}{\lambda}
    \intl{\omega,k}
    \left\{
    \begin{aligned}
    &\,
        \hat{G}_\spp(\omega,k)
    \\&\,
        -        \,
        f_\alpha
        \,
        \tauh^1
        \,
        \tr\left[
        \hat{G}^K_{\spp}(\omega,k)
        \right]
    \end{aligned}
    \right\},
\end{align}
where 
\begin{align}
\begin{aligned}
    \hat{G}_{\spp}^{R,A}(\omega,k) 
    =&\,
    \left[
        \hat{G}_0^{-1}(\omega,k) - \hat{\Sigma}^{R,A}
    \right]^{-1},
\\[2pt]
    \hat{G}^K_{\spp}(\omega,k)
    =&\,
    \hat{G}^R_{\spp}(\omega,k)
    \,
    \hat{\Sigma}^K
    \,
    \hat{G}^A_{\spp}(\omega,k).
\end{aligned}
\end{align}
With the choice $f_\alpha = 1 / 2 R$ (as discussed in Sec.~\ref{sec:KDeriv} and further explicated in Appendix \ref{app:WardU(1)}), 
the U($R$)-symmetric saddle-point is 
\begin{align}
    \hat{Q}_{\spp}
    =
    \hat{1}_R
    \otimes
    \hat{1}_2
    \begin{bmatrix}
        \ii \, \gamma 
    & 
        0 
    \\
        \Sigma^K 
    & 
        - \ii \, \gamma
    \end{bmatrix}_{\mathsf{LO}},
\end{align}
which is an identity matrix in replica and left/right-mover spaces, 
$\gamma = (2/\lambda)(\Lambda/2\pi)$ is the measurement decay rate, 
and $\Sigma^K$ is not determined by the saddle-point equation. 
Its value can be fixed by computing 
\begin{align}
    \left\langle n_\cl \right\rangle
    =
    \ii 
    \,
    \frac{\delta Z}{\delta V_q(t,x)}
    =
    2 R 
    \left(\frac{\Lambda}{2 \pi}\right)
    \mathcal{N}_0,
\quad
    \mathcal{N}_0
    =
    -
    \frac{\ii \, \Sigma^K}{2 \gamma}.
\end{align}
The parameter $\mathcal{N}_0$ is the dimensionless filling per channel relative to half-filling, with $-1 \leq \mathcal{N}_0 \leq 1$; $\Lambda/2\pi$ is the maximum particle density per channel (i.e., a filled band). 

As discussed in Sec.~\ref{sec:KDeriv}, the microscopic Keldysh action possesses U($R$) $\times$ U($R)$ symmetry. This is generated by the transformation
\begin{align}
    \psi \rightarrow \hat{\mathcal{U}} \, \psi,
    \quad
    \hat{\mathcal{U}}
    =
    \frac{1}{2}
    \begin{bmatrix}
    \hat{U}_1 + \hat{U}_2 & \hat{U}_1 - \hat{U}_2 \\
    \hat{U}_1 - \hat{U}_2 & \hat{U}_1 + \hat{U}_2
    \end{bmatrix}_{\mathsf{LO}},
\end{align}
where $\hat{U}_{1,2}$ are independent U($R$) matrices. 
Then, the fluctuating field matrix is given by 
$\hat{Q}(t,x) = \hat{\mathcal{U}}^\dagger(t,x) \, \hat{Q}_{\spp} \, \hat{\mathcal{U}}(t,x)$.
Transforming back to the Keldysh-branch $T/\bar{T}$ basis, one
obtains the parameterization
\begin{align}
    \hat{Q}
    =
    \gamma
    \begin{bmatrix}
    0 & \ii \left(1 + \mathcal{N}_0\right) \hxc
    \\
    \ii \left(1 - \mathcal{N}_0\right) \hxc^\dagger
    & 0
    \end{bmatrix}_{T,\bar{T}}
    +
    2 \, \mathcal{N}_0 \, \gamma \, \hat{1},
\end{align}
where 
\begin{align}
    \hxc = \hat{U}_1^\dagger \, \hat{U}_2 \equiv e^{i \theta/R} \, \hx,
\end{align}
and $\hx$ is an SU($R$) matrix.
Then we can generalize the identification in Eq.~(\ref{eq:XFieldIdentify}) to generic filling:\begin{align}\label{eq:XFieldIdentify2}
\begin{aligned}
    \psi_{+,j} \, \bar{\psi}_{-,\bar{k}}
    \Leftrightarrow&\,
    \left(1 + \mathcal{N}_0\right) e^{i \theta/R} \, X_{j \bar{k}},
\\
    \psi_{-,\bar{j}} \, \bar{\psi}_{+,k}
    \Leftrightarrow&\,
    \left(1 - \mathcal{N}_0\right) e^{-i \theta/R} \, X^\dagger_{\bar{j} k}.
\end{aligned}
\end{align}
These make sense on the Keldysh contour, with the anti-time-ordered (-) branch occuring \emph{after} the time-ordered (+) branch. In particular, the limit $\mathcal{N}_0 \rightarrow 1$ requires that we annihilate before we create on the contour, whereas $\mathcal{N}_0 \rightarrow -1$ requires that we create before we annihilate. 

We incorporate fluctuations of the average density by replacing 
\begin{align}
    \mathcal{N}_0 \rightarrow \mathcal{N}_0 + \delta \mathcal{N}(t,x).
\end{align}
Calculating the 
third-order
trace-log expansion (details omitted), we find that each component of the action in Eqs.~(\ref{eq:PCM}), (\ref{eq:U(1)}), and (\ref{eq:SV}) is multiplied by the prefactor $\mathcal{I}_0 \equiv 1 - \mathcal{N}_0^2$, where we identify 
\begin{align}
    \phi(t,x) \equiv \ii \, \frac{R \sqrt{\lambda}}{\mathcal{I}_0} \, \delta \mathcal{N}(t,x)
\end{align}
after performing a shift $\theta \rightarrow \theta + \phi$.
We also find a new term (anticipated in \cite{Ha2024}) that encodes the modulation of the entanglement sector stiffness by the average density:
\begin{align}
    S_{\delta \mathcal{N}}
    =
    -
    2
    \lambda
    &\,
    \mathcal{N}_0
    \int
    \frac{dt d x}{8}
    \delta \mathcal{N}(t,x)
\nonumber\\
    &\,
    \times
    \trr_R\left[
        \parr_t\hx^\dagger \parr_t\hx + v^2 \parr_x \hx^\dagger \parr_x \hx
    \right].
\end{align}
The same result obtains (up to irrelevant operators) from the action at fixed $\mathcal{N}_0$ simply by varying 
$\delta \mathcal{I}_0 \rightarrow -2 \mathcal{N}_0 \, \delta \mathcal{N}$.
The SU($R$) and U(1) sectors decouple at half filling ($\mathcal{N}_0 = 0$),
Eqs.~(\ref{eq:PCM}) and (\ref{eq:U(1)}).

\section{Kinetic equation} \label{app:KE}

In this appendix, we derive the kinetic equation (\ref{eq:KE}) in 
Sec.~\ref{sec:MIH} that describes the continuous heating of the 
interacting, monitored fermions by the measurements. 
The starting point is the microscopic Keldysh theory in Sec.~\ref{sec:KDeriv}, 
given by Eqs.~(\ref{eq:Zadecoup})--(\ref{Sa}). 
We work in 1+1-D and at half-filling in this Appendix. 

In order to treat interaction effects at the saddle-point (semiclassical) level,
we employ the bilocal $G$-$\Sigma$-$D$-$\Pi$ decoupling scheme for the fermion and boson
Green's functions and self energies; the same scheme is used to treat Brownian interactions in Appendix~\ref{app:Brownian}.
We replicate the fermion, boson, and bilocal fields, 
and average over the measurements. We again decouple the latter via the $\hat{Q}$ matrix. 
The theory is then described by the path integral 
\begin{widetext}
\begin{align}
    \bar{Z}
    =
    \int
    \calD{\hqq}
    \,
    \calD{a} 
    \,
    \calD\bar{\psi} \, \calD \psi 
    \,
    \calD\hat{G}
    \,
    \calD\hat{\Sigma}
    \,
    \calD\hat{D}
    \,
    \calD\hat{\Pi}
    \,
    e^{-\bar{S}},
\end{align}
and the action is
\begin{align}
    \bar{S}
    =&
    \frac{\lambda }{8}
    \intl{t,x}
        \left\{
        \trr\left[\left(\hqq \, \tauh^1\right)^2\right]
        -
        f_\alpha
        \left[\trr\left(\hqq \, \tauh^1\right)\right]^2
        \right\}
\nonumber\\
&\,
    +
    \frac{1}{2 \ii}
    \intl{\omega,\omega',k,k'}
    a_{s,i}(-\omega,-k)
    \left[
        \delta_{\omega,\omega'}
        \,
        \delta_{k,k'}
        \,
        \delta_{i j}
        \,
        \frac{2}{U}
        (\hat{s}^1)^{s,s'}
        -
        \Pi_{i j}^{s,s'}(\omega,k;-\omega',-k')
    \right]
    a_{s',j}(\omega',k')
\nonumber\\
&\,
    +
    \frac{1}{\ii}
    \intl{\omega,\omega',k,k'}
    \bar{\psi}_{i}^\tau(\omega,k)
    \left\{
    \begin{aligned}
    &\,
        \delta_{\omega,\omega'} \, \delta_{k,k'} \, \delta_{i,j} \left[\hat{G}_0^{-1}(\omega,k)\right]^{\tau,\tau'}
        -
        \Sigma_{i j}^{\tau,\tau'}(\omega,k;\omega',k')
    \\&\,
        + Q_{i j}^{\tau,\tau'}(\omega - \omega',k - k')
        -
        \delta_{i j} \, f_\alpha \, \trr\left[\hqq(\omega-\omega',k - k') \, \tauh^1\right] (\tauh^1)^{\tau,\tau'}
    \end{aligned}
    \right\}
    \psi_{j}^{\tau'}(\omega',k')
\nonumber\\
&\,
    -
    \intl{\omega,\omega',k,k'}
    \left[
        \Sigma_{i j}^{\tau,\tau'}(\omega,k;\omega',k')
        \,
        G_{j i}^{\tau',\tau}(\omega',k';\omega,k)
        -
        \frac{1}{2}
        \Pi^{s,s'}_{i j}(\omega,k;-\omega',-k')
        \,
        D^{s',s}_{j i}(\omega',k';-\omega,-k)
    \right]
\nonumber\\
&\,
-
    \frac{1}{\ii}
    \sum_{j}
    \intl{\omega,\omega',k,k'}
    \bar{\psi}_j^\tau(\omega,k) 
    \,
    (\hat{\Gamma}^s)^{\tau,\tau'}
    \,
    \psi_{j}^{\tau'}(\omega',k')
    \,
    a_{s,j}(\omega - \omega',k-k'),
\end{align}
\end{widetext}
where doubly repeated indices are summed,
$s \in \{\cl,\q\}$, and $\hat{\Gamma}^{\cl,\q}=\{\hat{1}_\tau,\hat{\tau}^1\}$.
The matrix $\hat{s}^1$ acts on the $s \in \{\cl,\q\}$ space of the bosons. 
Integrating out the fermions and bosons leads to 
\begin{align}
    \bar{Z}
    =
    \int
    \calD{\hqq}
    \,
    \calD\hat{G}
    \,
    \calD\hat{\Sigma}
    \,
    \calD\hat{D}
    \,
    \calD\hat{\Pi}
    \,
    e^{-\bar{S}},
\end{align}
with 
\begin{align}
    \bar{S}
    =&
    \frac{\lambda }{8}
    \intl{t,x}
        \left\{
        \trr\left[\left(\hqq \, \tauh^1\right)^2\right]
        -
        f_\alpha
        \left[\trr\left(\hqq \, \tauh^1\right)\right]^2
        \right\}
\nonumber\\
&\,
    +
    \frac{1}{2}
    \Tr\log\left[\frac{2}{U} \hat{s}^1 - \hat{\Pi}\right]
\nonumber\\
&\,
    -
    \Tr\log\left[\hat{G}_0^{-1}+\hat{Q} - f_\alpha \, \trr\left(\hqq \, \tauh^1\right) \tauh^1 - \hat{\Sigma}\right]
\nonumber\\
&\,
    -
    \Tr\left[\hat{\Sigma} \, \hat{G}\right]
    +
    \frac{1}{2}
    \Tr\left[\hat{\Pi} \, \hat{D}\right]
\nonumber\\
&\,
    +
    \frac{\ii}{2}
    \intl{1,2,3,4}
    (\hat{\Gamma}^s)^{\tau_1, \tau_2}
    (\hat{\Gamma}^{s'})^{\tau_3, \tau_4}
    G_{j,i}^{\tau_2,\tau_3}(2,3)
    \,
    G_{i,j}^{\tau_4,\tau_1}(4,1)
\nonumber\\
&\,
\qquad\quad\times
    \,
    D_{j,i}^{s,s'}(1-2;3-4),
\end{align}
where in the last term we have abbreviated $(\omega_n,k_n) \equiv n$, 
$n \in \{1,2,3,4\}$.

Assuming spacetime-translational invariance, we can derive saddle-point equations 
for $\{\hat{Q},\hat{G},\hat{\Sigma},\hat{D},\hat{\Pi}\}$. 
The equation for $\hat{Q}$ is unchanged;
again we adopt the U($R$)-symmetric saddle point 
$\hat{Q} = \hat{Q}_{\spp} = \ii \, \gamma \, \tauh^3 \, \hat{1}_R$, Eq.~(\ref{eq:qSP}).
The saddle-point Dyson equations for $\hat{G}$ and $\hat{D}$ are
\begin{align}
\begin{aligned}
    \hat{G}^{-1}
    =&\,
    \hat{G}_0^{-1}
    +
    \ii \, \gamma \, \tauh^3
    -
    \hat{\Sigma},
\\
    \hat{D}^{-1}
    =&\
    \frac{2}{U} 
    \hat{s}^1
    -
    \hat{\Pi}.
\end{aligned}
\end{align}
Finally, the equations for $\hat{\Sigma}$ and $\hat{\Pi}$ are self-consistently determined. 
All quantities can be taken to be diagonal in replica space, and we therefore drop replica indices to lighten the notation. Decomposing the Green's functions in Keldysh space gives the standard Eliashberg-type equations
\begin{align}
\label{eq:SigmaR}
    \Sigma^R(\omega,k)
    =&\,
    \ii
    \intl{\Omega,q}
    \left[
    \begin{aligned}
    &\,
    G^R(\Omega,q)
    \,
    D^K(\Omega - \omega,q - k)
    \\&\,
    +
    G^K(\Omega,q)
    \,
    D^A(\Omega - \omega,q - k)
    \end{aligned}
    \right],
\\
\label{eq:SigmaK}
    \Sigma^K(\omega,k)
    =&\,
    \ii
    \intl{\Omega,q}
    \left\{
    \begin{aligned}
    &\,
    G^K(\Omega,q)
    \,
    D^K(\Omega - \omega,q - k)
    \\&\,
    -
    \left[G^R - G^A\right](\Omega,q)
    \\&\,
    \times
    \left[D^R - D^A\right](\Omega - \omega,q - k)
    \end{aligned}
    \right\},    
\end{align}
\begin{align}\label{eq:PiR}
    \Pi^R(\Omega,q)
    =&\,
    -
    \ii
    \intl{\omega,k}
    \left[
    \begin{aligned}
    &\,
    G^R(\omega+\Omega,k+q) \, G^K(\omega,k)
    \\&\,
    +
    G^K(\omega+\Omega,k+q) \, G^A(\omega,k)
    \end{aligned}
    \right],
\\
\label{eq:PiK}
    \Pi^K(\Omega,q)
    =&\,
    -
    \ii
    \intl{\omega,k}
    \left\{
    \begin{aligned}
    &\,
    G^K(\omega-\Omega,k-q) \, G^K(\omega,k)
    \\&\,
    -
    \left[G^R - G^A\right](\omega - \Omega,k-q)
    \\&\,
    \times
    \left[G^R - G^A\right](\omega,k)
    \end{aligned}
    \right\}.
\end{align}
We further assume that the self-energies can be taken as functions of frequency only; we will see that this is a consistent choice. 
Without loss of generality, we then define
\begin{align}\label{eq:SigmaKF}
    \Sigma^K(\omega)
    \equiv
    -2 \ii 
    \left[ \gamma + |\Im(\Sigma^R)(\omega)| \right] F(\omega),
\end{align}
for \emph{some} fermion distribution function $F(\omega)$. 
Then 
\begin{align}
    G^K(\omega,k) = \left[G^R(\omega,k) - G^A(\omega,k)\right] F(\omega). 
\end{align}
In equilibrium at temperature $T$ with $F(\omega) = \tanh(\omega/ 2 T)$,
the last equation is the fluctuation-dissipation theorem \cite{Kamenev23}. 

We restrict to energies much smaller than the measurement rate $\gamma$. 
Then Eq.~(\ref{eq:PiR}) gives $\Pi^R(\Omega) \sim - 2 \kappa - \ii \Omega/\pi v \gamma$,
where $\kappa$ is the effective charge compressibility. The retarded boson propagator 
can be approximated as
\begin{align}
    D^R(\Omega)
    \sim
    \frac{U}{2}
    \left[1 + \kappa \, U + \ii \frac{U \, \Omega}{2 \pi v \gamma}\right]^{-1}.
\end{align}
To obtain the above, we assume that $F(\omega = \infty) - F(-\infty) = 2$, as in equilibrium. 
Evaluating Eq.~(\ref{eq:PiK}), we find that the Keldysh boson propagator also satisfies a non-equilibrium generalization of the fluctuation-dissipation theorem,
\begin{align}
    D^K(\Omega)
    \sim
    \left[D^R(\Omega) - D^A(\Omega)\right]F_B(\Omega).
\end{align}
Here the boson distribution function $F_B(\Omega)$ is not independent, but instead
determined by the fermion one $F(\Omega)$ via Eq.~(\ref{eq:FB}).
Eqs.~(\ref{eq:SigmaR}) and (\ref{eq:SigmaK}) reduce to 
\begin{align}\label{eq:SigmaRE}
    \Im\Sigma^R(\omega)
    =&\,
    -
    \frac{1}{2 v}
    \intl{\Omega}
    \varrho(\Omega - \omega)
    \left[
        F_B(\Omega - \omega) - F(\Omega) 
    \right],
\\\label{eq:SigmaKE}
    \Sigma^K(\omega)
    =&\,
    -
    \frac{\ii}{v}
    \intl{\Omega}
    \varrho(\Omega - \omega)
    \left[
        F(\Omega) \, F_B(\Omega - \omega) - 1
    \right],
\end{align}
where
\begin{align}
    \varrho(\Omega)
    \equiv
    -2 \Im D^R(\Omega)
    =
    \frac{U^2 \, \Omega}{2 \pi v \gamma}
    \!\!
    \left[
        (1 + \kappa \, U)^2 \!+\! \left(\frac{U \, \Omega}{2 \pi v \gamma}\right)^2
    \right]^{-1}
\end{align}    
is the spectral function of the bosons in the Eliashberg approximation.

Using Eqs.~(\ref{eq:SigmaRE}) and (\ref{eq:SigmaKE}) in Eq.~(\ref{eq:SigmaKF}) gives
\begin{align}\label{eq:Stapp}
    \gamma \, F(\omega) 
    =&\,
    \frac{\ii}{2}
    \Sigma^K(\omega)
    - 
    |\Im(\Sigma^R)| F(\omega)
\nonumber\\
    =&\,
    \intl{\Omega}
    \frac{\varrho(\Omega - \omega)}{2 v}
\left\{
\begin{aligned}
&\,
        F_B(\Omega - \omega)
        \left[F(\Omega) - F(\omega)\right] 
\\&\,
        - 
        \left[1
            -
            F(\Omega) \, F(\omega) 
        \right]
\end{aligned}
\right\}.
\end{align}
This assumed steady-state equation has only the trivial solution $F(\Omega) = 0$. 
In fact, the two terms on either side of Eq.~(\ref{eq:Stapp}) are easily identified as the balance of the measurement-induced decay rate (left-hand side) with the energy-conserving, inelastic collision integral (right-hand side).
In the more general time-dependent setting, we get the kinetic equation in Eq.~(\ref{eq:KE}), with the total 
collision integral determined by the difference of these, Eq.~(\ref{eq:St}).

\section{1-loop RG \label{app:RG}}

For the renormalization group, we rewrite the theory in Eqs.~(\ref{eq:PCM}) and (\ref{eq:SI}) as
\begin{widetext}
\begin{align}\label{eq:SRG}
    S_1
    =
    \int
    \rd t \, \rd x
    \left\{
        \frac{\lambda}{8}
        \trr_R
        \left[
            \vex{\nabla}
            \hx^\dagger
            \cdot
            \vex{\nabla}
            \hx
        \right]
        -
        \ii
        \,
        \Gamma_U
        \sum_{j = 1}^R
        \sum_{\mu}
        \eta^{\mu \mu}
        \left[
            \left(\hat{J}_\mu\right)_{jj}^2
            -
            \left(\hat{\overline{J}}_\mu\right)_{\bar{j}\bar{j}}^2
        \right]
    -
        \frac{M}{16}
        \sum_{j,k = 1}^R
        |X_{j,\bar{k}}|^4
    \right\}.
\end{align}
\end{widetext}
Here we have set the velocity $v = 1$ and $\vex{\nabla} = (\parr_t,\parr_x)$ is the ``Euclidean'' spacetime gradient operator. The mass for off-diagonal SU($R$) fluctuations is encoded in the parameter $M \equiv \lambda \, m^2$. 

The replica Noether currents that appear in the interaction term were defined above in Eq.~(\ref{eq:Noether}). By comparison, we have introduced a diagonal ``metric''
$\eta^{\mu \mu}$ into the interaction term in Eq.~(\ref{eq:SRG}). The current-current interactions in Eq.~(\ref{eq:SI}) obtain from microscopic density-density interactions in the fermion theory. In 1+1D, these can alternatively be encoded in the \emph{Minkowski} current-current interaction
\begin{align}
    (J_0)^2 - (J_1)^2 = 4 \bar{\psi}_R \, \psi_R \, \bar{\psi}_L \, \psi_L,
\end{align}
where $\psi_{L,R}$ annihilates a left- (L) or right- (R) moving fermion.
In what follows we therefore take $\eta^{\mu \mu} = \diag(1,-1)$. 

In addition to the 
those
presented in Eq.~\eqref{eq:SRG}, there is another term which parameterizes the deformation of the target manifold due to 
the
interaction. It is an anisotropic stiffness term,
\begin{equation}
    S_2=\int \rd t \, \rd x \, \frac{\lambda_d}{16}\sum_{j = 1}^R
        \sum_{\mu}
        \left[
            \left(\hat{J}_\mu\right)_{jj}^2
            +
        \left(\hat{\overline{J}}_\mu\right)_{\bar{j}\bar{j}}^2
        \right]\,.
\end{equation}  Here the signature is Euclidean. The term is called anisotropic stiffness because after expanding the $\hx$ fields into $\hy$, to leading order it is a stiffness term for the diagonal components of $\hy$. This term is not generated at the leading order of the gauging trick, but will be generated along the RG flow. From the effective field theory perspective, the total action $S_1+S_2$ can be constrained from the $S_R\times S_R$ permutation symmetry and the  anti-unitary symmetry [which descends from Eq.\eqref{eq:antiU_Sym}]
\begin{align}
    \hx \leftrightarrow \hx^\mathsf{T},
    \quad
    \ii \rightarrow - \ii.
\end{align}

We perform a field-theoretic RG calculation by expanding the field $\hat{X}=\exp(i\hy)$ in powers of $\hy$, and compute the beta functions perturbatively in powers of $1/\lambda$, assuming that $\lambda_d/\lambda,M/\lambda$ and $\Gamma_U/\lambda$ are of order one.

We obtained the anomalous dimension $\gamma_X$ of the $\hx$ field by renormalizing the $\braket{\hx\hx}$ correlator, which reads
\begin{equation}
    \gamma_X=\frac{2(R-1)\left[\lambda+R(\lambda+\lambda_d)\right]}{\pi R \lambda (\lambda+\lambda_d)}\,.
\end{equation} When $\lambda_d=0$, this reduces back to the well-known result of $\gamma_X=2(R^2-1)/(\pi R\lambda)$. From the same calculation, we also obtain the beta functions of $\lambda,\lambda_d,M$ to be 
\begin{widetext}
\begin{equation}\label{}
  \frac{\rd \lambda}{\rd \ln L} = \frac{128 \Gamma_U ^2+\lambda _d \left(\lambda _d+2 \lambda \right)-\lambda  R \left(\lambda _d+\lambda \right)}{\pi  \lambda  \left(\lambda _d+\lambda \right)}\,,
\end{equation}
\begin{equation}\label{}
  \frac{\rd \lambda_d}{\rd \ln L} = \frac{-\lambda  \lambda _d \left[\lambda _d+2 R \lambda _d+2 \lambda  (R+1)\right]+128 \Gamma_U ^2 \left[2 R \lambda _d+\lambda  (2 R-1)\right]}{\pi  \lambda ^2 \left(\lambda _d+\lambda \right)}\,,
\end{equation}
\begin{equation}
    \frac{\rd M}{\rd \ln L}= \left[2-\frac{4(R+1)\lambda(\lambda+\lambda_d)+\lambda_d^2}{\pi\lambda^2(\lambda+\lambda_d)}-\frac{128\Gamma_U^2}{\pi\lambda^2(\lambda+\lambda_d)}\right]M+\Lambda^2\frac{128\Gamma_U^2}{\pi \lambda^2}\,.
\end{equation}

Finally, the beta function for the interaction $\Gamma_U$ is obtained by renormalizing the $\braket{\hx\hx\hx}$ correlator, with the result 
\begin{equation}\label{}
  \frac{\rd \ln \Gamma_U}{\rd \ln L}=\frac{3}{2}\gamma_X\Gamma_U-\frac{\Gamma_U  \left[3 R^2 \left(9 \lambda  \lambda _d+2 \lambda _d^2+7 \lambda ^2\right)+R \left(768 \Gamma_U ^2+6 \lambda _d^2-24 \lambda  \lambda _d-2 \lambda ^2\right)-18 \lambda ^2\right]}{6 \pi  \lambda ^2 R \left(\lambda _d+\lambda \right)}\,.
\end{equation}

The anisotropic stiffness term does not change the RG flow qualitatively, so we choose to present it in the main text with $\lambda_d=0$ for clarity. 
The results above reduce to Eqs.~\eqref{eq:1loop} and \eqref{eq:betaGamma} after setting $\lambda_d=0$, $R=1$, and redefining $\bar{\Gamma}_U=\Gamma_U/\lambda$.  
\end{widetext}

We performed several additional, independent calculations to verify aspects of these flow equations.
We separately obtained the scaling dimension of the mass in Eq.~(\ref{eq:DeltaM}) from the large-level $k \equiv \lambda \pi \gg 1$ limit of the SU($R$) Wess-Zumino-Novikov-Witten (WZNW) model \cite{Foster14}. In the WZNW model, the holomorphic half of the mass operator belongs to the traceless bisymmetrized tensor representation of SU($R$) with highest weight $\Lambda = 2 \theta$, where $\theta$ is the highest root.

For the case $\lambda_d = 0$, we also verified all $R$-independent interaction corrections via a hard cutoff, field-theory renormalization scheme applied to the SU(2) case. Because this theory is a modified O(4)/O(3) sigma model, we exploited the multiplicative renormalizability of ``pi''-field correlators to extract all beta functions \cite{AmitBook}. Eqs.~(\ref{eq:betalambda}) and (\ref{eq:betaM}) were obtained from the two-point correlator, while a three-point correlator was needed for Eq.~(\ref{eq:betaGamma}).

\section{Path integral of the double-Hilbert space formalism and its equivalence with Keldysh formalism} \label{app:pathintegral}

In this appendix, we provide a proper path integral treatment of the double-Hilbert space formulation introduced in Sec.~\ref{sec:doublehilbert}, and demonstrate that it is equivalent to the Keldysh formalism.
For simplicity, we first focus on the unitary part of the system, and the generalization to the measurement part is straightforward. Our notation here will be slightly different from Sec.~\ref{sec:doublehilbert}. We will use $c^\dagger,c$ to denote fermion operators in the physical Hilbert space (instead of $\psi^\dagger,\psi$), and $d^\dagger,d$ to denote fermion operators in the auxiliary Hilbert space (instead of $\chi^\dagger,\chi$). To lighten the notation we will also drop the lattice site subscripts.

\subsection{The partition function}

The primary object that the path integral computes is the partition of the system, which is
\begin{equation}\label{}
  Z= \Tr(U(T){\rho} U(T)^\dagger)=\Tr\left(U(T)^\dagger U(T){\rho}\right)\,.
\end{equation} Here $U(T)=\exp(-\ii HT)$.

We transform the above object to the double-Hilbert space formalism
\begin{equation}\label{}
\begin{split}
  &Z=\sum_n \braket{n|U(T)^\dagger U(T) \hat{\rho}|n}\\
  &=\sum_{n,m} \bra{n}\otimes \bra{\bar{n}} U(T)^\dagger U(T)\hat{\rho} \ket{m}\otimes \ket{\bar{m}}\,.
\end{split}
\end{equation} In the second line, we have decoupled the trace by  introducing the auxiliary Hilbert space with basis $\ket{\bar{n}}$. Identifying $\sum_n \ket{n}\otimes \ket{\bar{n}}=\ket{\Phi\rangle}$ with the EPR state introduced in Sec.\ref{sec:doublehilbert}, we arrive at
\begin{equation}\label{eq:Z=PhiUUrho}
  Z=\braket{\langle\Phi|U(T)^\dagger U(T)\hat{\rho}|\Phi\rangle}=\braket{\langle\Phi|U(T)\otimes U^\mathsf{T}(T)^\dagger|\rho\rangle}\,,
\end{equation} where in the second equality we have used the CJ isomorphism.

Our task is therefore to construct a path-integral expression for Eq.~\eqref{eq:Z=PhiUUrho}.

\subsection{A quick recap on conventional fermion coherent state}

We first consider $U(T)$ in the physical Hilbert space. Its path integral is given by the conventional fermion coherent state, which we now review.

Following textbooks, we trotterize $U(T)=U_N U_{N-1}\cdots U_1$ into small segments by dividing $[0,T]=[t_0,t_1]\cup [t_1,t_2]\cup\dots \cup[t_{N_1},t_N]$, where $t_i=T(i/N)$, and
\begin{equation}\label{}
  U_i\approx \exp(-\ii\delta t H(c^\dagger, c))\,,\quad \delta t=T/N\,.
\end{equation}

Since by construction $H(c^\dagger,c)$ is normal ordered, its matrix element can be evaluated on the coherent states. The coherent states are the eigenstates of the fermion operator $c$. By definition
\begin{equation}\label{}
  c \ket{\eta}= \eta \ket{\eta}\,,
\end{equation} where $\eta$ is a Grassman number. The conjugate bra is
\begin{equation}\label{}
  \bra{\bar{\eta}}c^\dagger = \bra{\bar{\eta}}\bar{\eta}\,.
\end{equation} We note that $\bar{\eta}$ is independent of $\eta$.

Expanding $\ket{\eta}$ in the Fock basis $\ket{0},\ket{1}$, we obtain
\begin{equation}\label{}
  \ket{\eta}=\ket{0}-\eta\ket{1}= e^{-\eta c^\dagger}\ket{0}\,,
\end{equation} accordingly
\begin{equation}\label{}
  \bra{\bar{\eta}}=\bra{0}+\bar{\eta}\bra{1}= \bra{0} e^{-c\bar{\eta}}\,.
\end{equation}

The inner product between the coherent states are given by
\begin{equation}\label{}
  \braket{\bar{\eta}|\eta}=e^{\bar{\eta}\eta}=1+\bar{\eta}\eta\,.
\end{equation} The resolution of identity is
\begin{equation}\label{}
  \int \rd \bar{\eta} \rd \eta e^{-\bar{\eta}\eta} \ket{\eta}\bra{\bar{\eta}}=\hat{1}\,.
\end{equation}

The normal ordered $H(c^\dagger,c)$ therefore satisfies
\begin{equation}\label{}
  \braket{\bar{\eta}|H(c^\dagger,c)|\eta}=H(\bar{\eta},\eta)\,.
\end{equation}

Inserting the resolution of identity between every trotterization factors, we obtain the path integral as
\begin{align}\label{}
  &\,
  \braket{\bar{\eta}|U|\eta}
\nonumber\\
&\,
\quad
  =
  \int_{\psi(0)=\eta}^{\bar{\psi}(t)=\bar{\eta}} \calD \bar{\psi} \calD \psi \exp\left(\ii \int \rd t'\left( \ii \bar{\psi}\partial_{t'}\psi-H(\bar{\psi},\psi)\right)\right)\!.
\end{align}

\subsection{Path integral in the auxiliary Hilbert space}

Next, we treat the time evolution in the auxiliary Hilbert space. As discussed in Sec.~\ref{sec:doublehilbert}, the Hamiltonian $H^{\mathsf{T}}$ of $U^{\dagger \mathsf{T}}$ is anti-normal ordered, so we write it as $H^{\mathsf{T}}(d,d^\dagger)$.

To construct the path integral, we need a new set of coherent states which satisfies
\begin{equation}\label{}
  \braket{\xi|H^{\mathsf{T}}|\bar{\xi}}=H^{\mathsf{T}}(\xi,\bar{\xi})\,.
\end{equation} Therefore, we consider the coherent states
\begin{equation}\label{}
  d^\dagger \ket{\bar{\xi}}=\bar{\xi}\ket{\bar{\xi}}\,,\quad \bra{\xi}d =\bra{\xi}\xi\,.
\end{equation}

    Here we note a small presentational difference with the main text, that the definition of the coherent state $\bra{\xi}$ differs by a minus sign. In this appendix, we have defined the second set of coherent states as above, but later in the next subsection, we will perform a change of integration variables on $\bra{\xi}\to\bra{-\xi}$. The exposition in the main text combines these two steps into one, causing a slightly different definition but with equivalent results.

Through simple computations, we find
\begin{equation}\label{}
  \ket{\bar{\xi}}=\ket{1}-\bar{\xi}\ket{0}= e^{-\bar{\xi}d}\ket{1}\,,
\end{equation} and
\begin{equation}\label{}
  \bra{\xi}=\bra{1}-\xi\bra{0}=\bra{1}e^{-d^\dagger \xi}\,.
\end{equation}

The inner product is
\begin{equation}\label{}
  \braket{\xi|\bar{\xi}}=e^{\xi\bar{\xi}}=1+\xi\bar{\xi}\,,
\end{equation} and the resolution of identity is
\begin{equation}\label{}
  \int \rd \xi \rd \bar{\xi} e^{-\xi\bar{\xi}}\ket{\bar{\xi}}\bra{\xi}=\hat{1}\,.
\end{equation}

The path integral of $U^{\dagger \mathsf{T}}$ is therefore
\begin{align}\label{eq:DUT}
&\,
    \braket{\xi|U^{\dagger \mathsf{T}}|\bar{\xi}}
\nonumber\\
&\,
    \quad\,
  =
  \int_{\bar{\zeta}(0)=\bar{\xi}}^{\zeta(t)=\xi}\calD \zeta \calD \bar{\zeta}\exp\left(\ii\int \rd t' \left(i\bar{\zeta}\partial_{t'} \zeta+H^{\mathsf{T}}(\zeta,\bar{\zeta})\right)\right).
\end{align}

\subsection{Boundary condition and equivalence with Keldysh path integral}

Finally, we take care of the boundary conditions $\bra{\langle\Phi}$ and $\ket{\rho\rangle}$ in Eq.~\eqref{eq:Z=PhiUUrho}. By direct computation, we can show that with EPR state $\ket{\Phi_{AB}\rangle}=\ket{0_A 1_B}+\ket{1_A 0_B}$, we have
\begin{equation}\label{eq:Phisimplify1}
  \braket{\langle\Phi_{AB}|\eta_A}=\bra{-\eta_B}\,,\quad \braket{\bar{\eta}_A|\Phi_{AB}\rangle}=\ket{-\bar{\eta}_B}\,,
\end{equation}
\begin{equation}\label{eq:Phisimplify2}
  \braket{\langle\Phi_{AB}|\bar{\eta}_B}=\bra{\bar{\eta}_A}\,,\quad \braket{\eta_B|\Phi_{AB}\rangle}= \ket{\eta_A}\,.
\end{equation}

 Inserting resolution of identity at the initial and the final states of Eq.~\eqref{eq:Z=PhiUUrho}, we obtain
\begin{equation}\label{eq:Zglue}
\begin{split}
   Z & = \int \rd \bar{\eta}_f \rd \eta_f \rd \xi_f \rd \bar{\xi}_f e^{-\bar{\eta}_f \eta_f-\xi_f\bar{\xi}_f-\eta_f\bar{\xi}_f} \\
   &\times\braket{\bar{\eta}_f|U|\eta_i} \braket{\xi_f|U^{\dagger \mathsf{T}}|\bar{\xi}_i}\\
   &\times \int \rd\bar{\eta}_i \rd \eta_i \rd \xi_i \rd \bar{\xi}_i e^{-\bar{\eta}_i \eta_i} e^{-\xi_i \bar{\xi}_i} \braket{\langle\bar{\eta_i}\xi_i|\rho\rangle}\,.
\end{split}
\end{equation} Eq.~\eqref{eq:Zglue} can be evaluated in several steps. First, we perform the integral over $\xi_f$ and $\bar{\eta_f}$, and the first two lines of Eq.~\eqref{eq:Zglue} becomes
\begin{equation}\label{eq:UU}
  \int \rd \xi_f \int \rd \bar{\eta}_f e^{\bar{\eta}_f \xi_f}\braket{\xi_f|U^{\dagger \mathsf{T}}|\bar{\xi}_i}\braket{\bar{\eta}_f|U|\eta_i}\,.
\end{equation}  The crucial step  now is to perform a change of variables by flipping $\xi_f\to -\xi_f$. Accordingly, we also flip the sign of $\zeta$ in Eq.~\eqref{eq:DUT}:
\begin{equation}\label{}
\begin{split}
  &\braket{-\xi_f|U^{\dagger \mathsf{T}}|\bar{\xi}_i}\\
  &=\int_{\bar{\zeta}(0)=\bar{\xi}_i}^{\zeta(t)=\xi_f} D\bar{\zeta} D\zeta \exp\left(\ii\int \rd t' \left(-i\bar{\zeta}\partial_{t'}\zeta+H^{\mathsf{T}}(-\zeta,\bar{\zeta})\right)\right)
\end{split}
\end{equation} Recall from Sec.~\ref{sec:doublehilbert}, that $H^{\mathsf{T}}$ is obtained from $H$ by changing the normal order to anti-normal order, and flip the sign of the quadratic term. When we replace fermion operators by Grassman numbers, the ordering doesn't matter, and flipping the sign of the quadratic term can exactly be cancelled by flipping $\zeta$, so $H^{\mathsf{T}}(-\zeta,\bar{\zeta})=H(\bar{\zeta},\zeta)$. Therefore, Eq.~\eqref{eq:UU} can be exactly written as
\begin{equation}\label{}
\begin{split}
  &\braket{\bar{\xi}_i|U^\dagger(T) U(T)|\eta_i}=\int_{\psi(0)=\eta_i,\bar{\zeta}(0)=\bar{\xi}_i}
   \calD \bar{\psi}\calD \psi \calD \bar{\zeta} \calD \zeta\\
   &\times \exp\left(\ii S[\psi]-\ii S[\zeta]\right)\,,
\end{split}
\end{equation} where
\begin{equation}\label{}
  S[\psi]=\int \rd t \left[\ii \bar{\psi}\partial_t \psi-H\left(\bar{\psi},\psi\right)\right]\,.
\end{equation} This is exactly the form of Keldysh path integral.

Finally, we simplify the last line of Eq.~\eqref{eq:Zglue}. Using $\ket{\rho\rangle}={\rho}\otimes \hat{1}\ket{\Phi_{AB}\rangle}$ and Eq.~\eqref{eq:Phisimplify2}, we obtain
\begin{equation}\label{}
  \braket{\langle\bar{\eta}_i \xi_i|\rho\rangle}=\braket{\bar{\eta}_i|{\rho}|\xi_i}\,.
\end{equation} Further making a change of variable $\xi_i\to-\xi_i$, we obtain the full path integral to be
\begin{equation}\label{eq:Zresult}
\begin{split}
  Z=\int \rd\bar{\eta_i} \rd \eta_i \rd \bar{\xi}_i \rd \bar{\xi} e^{-\bar{\xi}_i \xi_i}\braket{\bar{\xi}_i|U^\dagger U|\eta_i} e^{-\bar{\eta}_i \eta_i} \braket{\bar{\eta}_i|{\rho}|-\xi_i}\,.
\end{split}
\end{equation} Eq.~\eqref{eq:Zresult} is fully consistent with the Keldysh field theory. First, the dynamics part Eq.~\eqref{eq:UU} takes the same form as the Keldysh action, which is conventionally formulated as a path-integral of $U^\dagger U$. Second, the fermion path integral over the density matrix has anti-periodic boundary condition, as it should.

Finally, the measurement terms can be included following a similar exercise. According to Eq.~\eqref{eq:H_otimes_HT} and \eqref{eq:K_otimes_KT}, we see that the measurement terms has the opposite relative sign between the two time branches compared to the unitary part. Therefore, the measurement terms will appear as $S_+^\textrm{m}+S_-^\textrm{m}$.

\section{Brownian Interaction and $G$-$\Sigma$ formulation}\label{app:Brownian}

In the main text, our discussion focused on the theory with an interaction term that is invariant under spacetime translations. However, in random unitary circuits, the generated interaction tends to be random as well. In this appendix, we model these effects using the Brownian interaction and the $G$-$\Sigma$ formalism, and the resulting theory is qualitatively similar to the one presented in the main text.
\subsection{Model}
The action we consider is $S=S_\text{free}+S_\text{int}$, where the free fermion part is
\begin{equation}\label{eq:Sfree}
\begin{split}
  S_\text{free} &=\sum_{i} \int_{t,\vec{x}} \! \sum_{\alpha}
  \bar{\psi}_{i\alpha}(\vec{x},t)
  \!\left[\ii\partial_t-\left(\frac{-\nabla_{\vec{x}}^2}{2m}-\mu\right)\right]\!
  \psi_{i\alpha}(\vec{x},t)\\
  &-\ii \upsilon(\vec{x},t)\sum_{\alpha\beta}\bar{\psi}_{i\alpha} \hat{\tau}^1_{\alpha\beta} \psi_{i\beta}\,.
\end{split}
\end{equation}
Here $i=1,\dots,R$ denote the replica indices and $\alpha=1,2$ denotes the Keldysh index in the Larkin-Ovchinnikov basis (see Eq.~\eqref{eq:LO}). The measurement term $\upsilon(\vec{x},t)$ is random in spacetime. Here we model it by a Gaussian random variable with zero mean and variance $\overline{\upsilon(\vec{x},t)\upsilon(\vec{x}',t')}=g^2\delta(\vec{x}-\vec{x}')\delta(t-t')$

The brownian interaction is described by an auxiliary bosonic field $\phi$:
\begin{equation}\label{eq:Sint}
\begin{split}
   S_\text{int} &= \sum_{i}\int_{t,\vec{x}} \frac{1}{2}\sum_{\alpha\beta} \phi_{i\alpha} \hat{\tau}^1_{\alpha\beta} (-1/U) \phi_{i\beta}\\
   &+u(\vec{x},t)\sum_{\alpha\beta\gamma} \bar{\psi}_{i\alpha} \hat{\Gamma}^{\gamma}_{\alpha\beta} \psi_{i\beta} \phi_{i\gamma}\,.
\end{split}
\end{equation} Here, the boson carries replica index $i=1,\dots,R$ and Keldysh index $\alpha=\text{cl},\text{q}$.
The coupling constant $u$ is a Gaussian random variable in spacetime, with zero mean and unit variance $\overline{u(\vec{x},t)u(\vec{x}',t')}=\delta(\vec{x}-\vec{x}')\delta(t-t')$. $U$ is the strength of the interaction. The tensor $\hat{\Gamma}$ encodes the Keldysh structure of the interaction:
\begin{equation}\label{}
  \hat{\Gamma}^{\rm cl,q}=\frac{1}{\sqrt{2}}(1,\hat{\tau}^1)\,.
\end{equation}

We can rewrite the theory as a path-integral over bi-local variables $G$-$\Sigma$-$D$-$\Pi$
\begin{equation}\label{}
  Z=\int\calD G\calD \Sigma \calD D \calD \Pi \exp(\ii S)\,,
\end{equation} where
\begin{equation}\label{eq:SGS}
\begin{split}
   &\ii S  =\ln\det(G_0^{-1}-\Sigma)-\frac{1}{2}\ln\det(D_0^{-1}-\Pi)\\
   &+\int_{\vec{x},t}\tr_{RK}(\Sigma\cdot G)-\frac{1}{2}\tr_{RK}(\Pi\cdot D) \\
     & +\int_{\vec{x},t}\frac{g^2}{2}\left[\tr_{RK}(G\hat{\tau}^1G_{}\hat{\tau}^1)-f_\upsilon\tr_{RK}^2(G\hat{\tau}^1)\right]\\
     &+\int_{\vec{x},t}\frac{\ii }{2} \sum_{\alpha\beta,ij}\left[\tr_K\left(\hat{\gamma}^{\alpha}G_{ij}\hat{\gamma}^{\beta}G_{ji}\right)\right.\\
     &-\left.f_U\tr_K(\hat{\gamma}^{\alpha}G_{ii})\tr_K(\hat{\gamma}^{\beta}G_{jj})\right]\times D_{ab}^{\alpha\beta}\,.
\end{split}
\end{equation} The action is derived by rewriting $\psi,\phi$ into bi-local variables with the definitions $\ii G^{\alpha\beta}_{ij}(\vec{x},t)=\psi_{i\alpha}(\vec{x},t)\bar{\psi}_{j\beta}(\vec{x},t)$ and $\ii D^{\alpha\beta}_{ij}(\vec{x},t)=-\phi_{i\alpha}(\vec{x},t)\phi_{j\beta}(\vec{x},t)$, which are enforced via the Lagrangian multiplier $\Sigma$ and $\Pi$ respectively. When the indices are not explicitly written out, matrix multiplication is implied. In Eq.~\eqref{eq:SGS}, $\tr_{R(K)}$ means tracing out the replica (Keldysh) indices, and the $\ln\det$ computes the functional determinant over Keldysh, replica and spacetime indices. Due to the fact that measurement and interaction induces space-time local couplings, there is an ambiguity in the rewriting of fermion quartic terms $(\bar{\psi} \hat{\tau}^1 \psi)^2=-\psi\hat{\tau}^1 \bar{\psi}\psi \hat{\tau}^1 \bar{\psi}$. We therefore keep both possibilities and denote their relative weights by $f_\upsilon$ and $f_U$ for the measurement channel and the interaction channel respectively. The value of $f_\upsilon$ and $f_U$ does not affect the saddle point but has an important effect on the fluctuations. In particular, we need to choose $f_\upsilon=1/R$ and $f_U=1$ to ensure the existence of diffusion mode. Note here the normalization of $f_\upsilon$ appears to be different from the main text because here we do not introduce left/right movers. Finally, $G_0=(\omega-\xi_k)^{-1}$ ($\xi_k=k^2/(2m)-\mu$) and $D_0=\hat{\tau}^1/m_b^2$ are the free propagators of the fermion and the boson respectively.

\subsection{Saddle point}

We can obtain the saddle point of the system by extremizing Eq.~\eqref{eq:SGS} over $G,\Sigma,D,\Pi$. We obtain the Schwinger-Dyson equation which relates the Green's functions $G,D$ to the self-energies $\Sigma,\Pi$. Assuming the fermion band is half filled, the fermion self-energy only has the retarded and the advanced components $\Sigma_{R/A}$ and the boson only has a Keldysh component $\Pi_K$, both of which are momentum and frequency independent:
  \begin{eqnarray}
    \Sigma_R &=& \Sigma_A^*=-\frac{\ii}{2}g^2 V_\text{BZ}\left(1+\frac{U^2 V_\text{BZ}^3\Lambda}{4g^2}\right)\,, \\
    \Pi_K &=& \frac{\ii  V_\text{BZ}^2}{2}\,.
  \end{eqnarray} Here $V_\text{BZ}$ is the volume of the Brillouin zone and $\Lambda$ is the cutoff of the  interaction. We note that the solutions are consistent with the interpretation that the system is heated to infinite temperature in the steady state.

\subsection{Gaussian fluctuations}

  We continue the analyze the Gaussian fluctuations of the theory. This is carried out by expanding Eq.~\eqref{eq:SGS} to second order in $\delta G,\delta \Sigma,\delta D,\delta \Pi$, and then integrating out all but $\delta \Sigma$.  The result can be written into the form
  \begin{equation}\label{}
  \begin{split}
  & iS[\delta\Sigma]=-\frac{1}{2}\sum_{ijkl,\alpha\beta\gamma\delta}\int_{\vec{x},t,\vec{x}',t'} \delta \Sigma_{lk}^{\delta\gamma}(\vec{x},t) \\
  &\times K_{kl,ij}^{\gamma\delta,\alpha\beta}(\vec{x}-\vec{x}',t-t')\delta \Sigma_{ij}^{\alpha\beta}(\vec{x}',t')\,.
  \end{split}
\end{equation} Here the fluctuations  $\delta\Sigma$ is a two-index tensor in both the replica and the Keldysh indices, whereas the kernel $K$ is a four-index tensor which generates the Bethe-Salpeter ladder. Schematically, it can be written as the sum over several Feynmann diagrams

\begin{equation}
\begin{split}
K = &-\begin{tikzpicture}[baseline={([yshift=4pt]current bounding box.center)}]
                     \draw[thick, mid arrow] (40pt,12pt)--(0pt,12pt);
                     \draw[thick, mid arrow] (0pt,-12pt)--(40pt,-12pt);
                     \node[below] at (20pt,-12pt) {Self-energy};
\end{tikzpicture}+
\begin{tikzpicture}[baseline={([yshift=4pt]current bounding box.center)}]
                     \draw[thick, mid arrow] (40pt,12pt)--(0pt,12pt);
                     \draw[thick, mid arrow] (0pt,-12pt)--(40pt,-12pt);
                     \draw[thick, dashed] (20pt,12pt)--(20pt,-12pt);
                     \node[below] at (20pt,-12pt) {Measurement};
\end{tikzpicture}+
\begin{tikzpicture}[baseline={([yshift=4pt]current bounding box.center)}]
                     \draw[thick, mid arrow] (40pt,12pt)--(0pt,12pt);
                     \draw[thick, mid arrow] (0pt,-12pt)--(40pt,-12pt);
                     \draw[thick, boson] (20pt,12pt)--(20pt,-12pt);
                     \node[below] at (20pt,-12pt) {Interaction};
\end{tikzpicture}\\
&+\begin{tikzpicture}[baseline={([yshift=4pt]current bounding box.center)}]
                     \draw[thick, mid arrow] (40pt,12pt)--(30pt,12pt)--(30pt,-12pt)--(40pt,-12pt);
                     \draw[thick, mid arrow] (0pt,-12pt)--(10pt,-12pt)--(10pt,12pt)--(0pt,12pt);
                     \draw[thick, boson] (30pt,12pt)--(10pt,12pt);
                     \draw[thick, boson] (30pt,-12pt)--(10pt,-12pt);
                     \node[below] at (20pt,-12pt) {AL1};
                     \end{tikzpicture}+
                     \begin{tikzpicture}[baseline={([yshift=4pt]current bounding box.center)}]
                     \draw[thick, mid arrow] (40pt,12pt)--(30pt,12pt)--(30pt,-12pt)--(40pt,-12pt);
                     \draw[thick, mid arrow] (0pt,-12pt)--(10pt,-12pt)--(10pt,12pt)--(0pt,12pt);
                     \draw[thick, boson] (30pt,12pt)--(10pt,-12pt);
                     \draw[thick, boson] (30pt,-12pt)--(10pt,12pt);
                     \node[below] at (20pt,-12pt) {AL2};
                     \end{tikzpicture}\,.
\end{split}
\end{equation} Here, the solid line is the full fermion propagator that incorporates the self-energy from the saddle point. The dashed line denotes measurement, and the wavy line denotes the interaction boson. The second and the third diagrams in the first line are the so called Maki-Thompson diagrams, and the diagrams in the second line are the Aslamazov-Larkin diagrams. As discussed in the main text, the Maki-Thompson diagram of the interaction will gap out fluctuations that are off-diagonal in the replica indices, and keep the diagonal fluctuations gapless. It turns out that the Aslamazov-Larkin diagrams does not affect the low-energy modes, because the Keldysh index structure is orthogonal ($K_\text{AL1+AL2}^{\gamma\delta,\alpha\beta}\delta\Sigma^{\alpha\beta}$=0). The structure of $\delta\Sigma$ is given in Eq.~\eqref{eq:sigma_ansatz} below.

    We present the action for the low-energy modes, i.e. the modes that are gapless in the free limit, and some of them can acquire a gap in the presence of interaction.  The low-energy fluctuations take the form
  \begin{equation}\label{eq:sigma_ansatz}
    \delta\Sigma_{ij}^{\alpha\beta}=\theta_x (\hat{\tau}^1)^{\alpha\beta} \delta_{ij}+(\theta_y\delta_{ij}+{Y}^{ij}) (\hat{\tau}^2)^{\alpha\beta}\,.
  \end{equation}
  Here ${Y}$ is traceless in the replica indices. The action can be decomposed into two parts $S=S[Y]+S[\theta_x,\theta_y]$. The traceless part is
\begin{equation}\label{eq:SY}
\begin{split}
  &iS[Y_y]=-\frac{1}{2g^2} \sum_{ij}\int_P \frac{4}{4+\alpha_U \alpha_\Lambda}Y^{ij}(-P)Y^{ji}(P) \\
  &\times\left[ \frac{C_{+}(P)+C_{-}(P)}{2}-\frac{4+\alpha_U \alpha_\Lambda \delta_{ij}}{4+\alpha_U \alpha_\Lambda}\right] \,.
\end{split}
\end{equation} Here we have transformed  to momentum space with 3-momentum $P=(\Omega,\vec{p})$. The two functions $C_\pm(P)$ arise from fermion bubbles, and their long-wavelength limits are given by
\begin{equation}\label{eq:Cpm}
     C_{\pm}(P) = 1 \pm \frac{i\Omega}{\Gamma}-\frac{1}{d}\frac{v_F^2 \vn{p}^2}{\Gamma^2}-\frac{\Omega^2}{\Gamma^2}+\calO\left(\Omega^3,\vn{p}^4\right)\,,
\end{equation}
and $\Gamma=g^2V_\text{BZ}(1+\alpha_U\alpha_\Lambda/4)$ is the elastic scattering rate of the fermion extracted from the self-energy. $d$ is the spatial dimension of the system. We have introduced two  parameters $\alpha_U=U^2V_\text{BZ}^4$ and $\alpha_\Lambda=\Lambda/(g^2V_\text{BZ})$. Expanding $C_\pm(P)$ in small $\Omega$ and $\vec{p}$, we find that the diagonal fluctuations $Y^{ii}$ are gapless with relativistic dispersion in Euclidean time with velocity $v=v_F/\sqrt{d}$, and the off-diagonal modes $Y^{i\neq j}$ are gapped with gap $\Delta=g^2 V_\text{BZ} \sqrt{\alpha_U \alpha_\Lambda(4+\alpha_U \alpha_\Lambda)}$. As a sanity check, in the absence of interaction $\alpha_U=0$, all the $R^2-1$ modes are gapless, consistent with the $\rm{SU(R)}$ PCM.

The traceful part is (when $f_U=1$)
\begin{widetext}
\begin{equation}\label{}
  iS[\theta_x,\theta_y]=-\frac{1}{2g^2} R \int_P \begin{pmatrix}
                                     \theta_x(-P) & \theta_y(-P)
                                   \end{pmatrix}
                                   \begin{pmatrix}
                                    \frac{C_{+}(P)+C_{-}(P)}{2}-\frac{4+\alpha_U \alpha_\Lambda}{\alpha_U \alpha_\Lambda+8f_\upsilon R-4} & -\frac{i}{2}\left(C_{+}(P)-C_{-}(P)\right) \\
                                     \frac{i}{2}\left(C_{+}(P)-C_{-}(P)\right) & \frac{C_{+}(P)+C_{-}(P)}{2}-1
                                   \end{pmatrix}
                                   \begin{pmatrix}
                                     \theta_x(P) \\
                                     \theta_y(P)
                                   \end{pmatrix}\,,
\end{equation}
\end{widetext} We notice that for generic $f_\upsilon$, the $\theta_x$ mode is gapped. However, for $f_\upsilon R=1$, it becomes gapless and can hybridize with $\theta_y$. We reparameterize the fluctuations with $\sqrt{2}\theta_{1,2}=\theta_x\pm i\theta_y$, and we obtain
\begin{equation}\label{eq:Stheta}
\begin{split}
  &iS[\theta_x,\theta_y]=-\frac{R}{2\gamma^2}\int_P (C_{+}(P)-1)\theta_1(-P)\theta_2(P)\\
  &+\left(C_{-}(P)-1\right)\theta_2(-P)\theta_1(P)\,,
\end{split}
\end{equation} The poles are therefore located at $C_{\pm}(P)=1$, which becomes exactly the diffusion mode when we expand $C_\pm(P)$ in small $\Omega,\vec{p}$.


To summarize, Eq.~\eqref{eq:SY} reproduces the PCM (Eq.~\eqref{eq:PCM}) + the mass term deformation in Eq.~\eqref{eq:SIY}, and Eq.~\eqref{eq:Stheta} reproduces the U(1) part (Eq.~\eqref{eq:U(1)}). To further reproduce the Yukawa term in Eq.~\eqref{eq:SIY}, we can further expand the fermion determinant term in Eq.~\eqref{eq:SGS} to cubic order, and substitute the parameterization \eqref{eq:sigma_ansatz}.

We note that the above derivations are insensitive to spatial dimensions except in Eq.~\eqref{eq:Cpm}, which alters some numerical prefactors. Therefore we expect the structure of the theory can be generalized to higher dimensions.

\subsection{Away from half-filling}

We also briefly comment on the situation when the system is away from half-filling. For simplicity we will restrict ourselves to non-interacting fermions. 
The fermion self-energy reads
\begin{equation}
    \hat{\Sigma}=\begin{pmatrix}
        \Sigma_R & \Sigma_K \\
        0 & \Sigma_A
    \end{pmatrix} = -\frac{\ii}{2}g^2 V_\text{BZ} \begin{pmatrix}
        1 & -2\calQ \\
        0 & -1
    \end{pmatrix} \,,
\end{equation} where $\calQ\in [-1,1]$ is related to the filling fraction of the fermion via $n=(1+\calQ)/2$. We note that this solution is only possible because of the choice $f_\upsilon R=1$.

As for the Gaussian fluctuation, we are still able to identify the diffusive and the relativistic fluctuations. However, their form is more complicated than Eq.~\eqref{eq:sigma_ansatz} due to nonzero $\calQ$, so we do not present it here. Nevertheless, we found that the charge diffusion constant is not renormalized by $\calQ$, but the velocity of the relativistic mode is renormalized by $v(\calQ)=v(\calQ=0)\sqrt{1-\calQ^2}$.

\section{Hubbard-Stratonovich and global U(1) Ward identity\label{app:WardU(1)}}

In this Appendix, we demonstrate the generalized Hubbard-Stratonovich decoupling used in Eq.~(\ref{eq:barS}), and show how one particular choice for the parameter $\alpha$ is required in order for the theory to give the diffusive response for the average density.

Eq.~(\ref{eq:barS}) obtains from the following identity.
For a Hermitian matrix-valued field $\hqq$, under the shift
\begin{align}
    \hqq \rightarrow \hqq - \ii \, g^2 \alpha \left(\tauh^1 \, \psi\bar{\psi} \, \tauh^1\right),
\end{align}
one gets
\begin{widetext}
\begin{multline}
    -
    \frac{1}{2 g^2 \alpha}
    \,
    \trr\left[\left(\hqq \, \tauh^1\right)^2\right]
    -
    \ii
    \,
    \tr\left(\hqq \, \psi \bar{\psi}\right)
    +
    \frac{f_\alpha}{2 g^2 \alpha}
    \left[\trr\left(\hqq \, \tauh^1\right)\right]^2
    +
    \ii
    \,
    f_\alpha
    \,
    \tr\left(\hqq \, \tauh^1\right)
    \,
    \tr\left(\psi\bar{\psi} \, \tauh^1\right)
\\
    \rightarrow
    -
    \frac{1}{2 g^2 \alpha}
    \,
    \trr\left[\left(\hqq \, \tauh^1\right)^2\right]
    +
    \frac{f_\alpha}{2 g^2 \alpha}
    \left[\trr\left(\hqq \, \tauh^1\right)\right]^2
    +
    \frac{g^2}{2}
    \left(\bar{\psi} \, \tauh^1 \, \psi\right)^2,
\end{multline}
where $f_\alpha = (1 - \alpha)/\alpha$.

Using the parameterization of Gaussian fluctuations in Eq.~(\ref{YDef}),
we expand the action $\bar{S}$ in Eq.~(\ref{eq:barS}) to quadratic order in the fluctuations.
Ignoring the $\tilde{a}_{\cl,\q}$ fields that incorporate density-density interactions and source fields $V_{\cl,\q}$, one finds in 1+1D
\begin{align}
    \bar{S}
    =&\,
    \frac{\gamma^2}{2}
    \intl{\omega,k}
    \left[
    \begin{aligned}
    &\,
        \Xi(\omega,k)
        \,
        \left\{
                \tr_R\left[\hat{Y}(-\omega,-k) \,\hat{Y}(\omega,k)\right]
                +
                \frac{1}{R}
                \,
                \theta(-\omega,-k) \, \theta(\omega,k)
        \right\}
   \\&\,
        +
        p_\alpha
        \left\{
                1
                +
                R \, p_\alpha
                \left[
                    1
                    -
                    \Xi(\omega,k)
                \right]
        \right\}
        \phi(-\omega,-k) \, \phi(\omega,k)
        -
        \ii \, p_\alpha
        \,
        \frac{\omega}{\gamma}
        \,
        \phi(-\omega,-k) \, \theta(\omega,k)
    \end{aligned}
    \right].
\end{align}
\end{widetext}
Here $\gamma$ is the saddle-point decay rate induced by the measurements for the fermions
[Eq.~(\ref{eq:qSP})], and
\begin{align}
    \Xi(\omega,k)
    \equiv
    \frac{v^2 \, k^2  + \omega^2}{4 \gamma^2},
\quad p_\alpha
    \equiv
    \frac{1}{R} - 4 f_\alpha.
\end{align}
The mode $\phi(\omega,k)$ is generically massive, leading to an unphysical \emph{Euclidean} relativistic dispersion for $\theta(\omega,k)$
and the Keldysh density-density correlation function.
For the special choice $p_\alpha = -1/R$
[$f_\alpha = 1/2R$], however, $\phi$ is made massless, leading to Eq.~(\ref{eq:U(1)}). This choice gives the diffusive Keldysh correlator for the average density that satisfies the global U(1) Ward identity in Eq.~(\ref{Ward1+1}).


\end{document}